\documentclass[12pt]{iopart}
\usepackage{iopams}
\usepackage{color}
\usepackage{graphicx}
\usepackage[normalem]{ulem}

\def\ra{\rangle}
\def\la{\langle}

\def\sgn{\mathrm{sgn}}

\def\q{\bi{q}}
\def\k{\bi{k}}

\def\rr{\bi{r}}

\def\Q{\bi{Q}}
\def\S{\bi{S}}
\newcommand{\ii}{{\mathrm{i}}}
\newcommand{\LOFA}{LaFeAsO}
\newcommand{\LOFFA}{LaFeAsO$_{1-x}$F$_x$}

\newcommand{\SrKFeAs}{Sr$_{1-x}$K$_x$Fe$_2$As$_2$}

\newcommand{\BFCA}{Ba(Fe$_{1-x}$Co$_x$)$_2$As$_2$}
\newcommand{\BFA}{BaFe$_2$As$_2$}
\newcommand{\BFNA}{Ba(Fe$_{1-x}$Ni$_x$)$_2$As$_2$}
\newcommand{\BFKA}{{Ba$_{1-x}$K$_x$Fe$_2$As$_2$}}
\newcommand{\LFPO}{LaFePO}
\newcommand{\KFA}{KFe$_2$As$_2$}
\newcommand{\KFS}{KFe$_2$Se$_2$}
\newcommand{\BFAP}{BaFe$_2$(As$_{1-x}$P$_x$)$_2$}
\newcommand{\BFAR}{BaFe$_2$(As$_{1-x}$Ru$_x$)$_2$}
\newcommand{\LFA}{LiFeAs}
\newcommand{\LFP}{LiFeP}

\begin{document}

\review{Gap symmetry and structure of Fe-based superconductors}

\author{P.J. Hirschfeld}
\address{Department of Physics, University of Florida, Gainesville, Florida 32611, USA}
\ead{pjh@ufl.edu}

\author{M.M. Korshunov}
\address{Department of Physics, University of Florida, Gainesville, Florida 32611, USA}
\address{L.V. Kirensky Institute of Physics, Siberian Branch of Russian Academy of Sciences, 660036 Krasnoyarsk, Russia}
\ead{korshunov@phys.ufl.edu}

\author{I.I. Mazin}
\address{Code 6393, Naval Research Laboratory, Washington, D.C. 20375, USA}
\ead{mazin@nrl.navy.mil}

\begin{abstract}

The recently discovered Fe pnictide and chalcogenide superconductors display low temperature properties suggesting superconducting gap structures which appear to vary substantially from family to family, and even within family as a function of doping or pressure.  We propose that this apparent nonuniversality can actually be understood by considering the predictions of spin fluctuation theory and accounting for the peculiar electronic structure of these systems, coupled with the likely ``sign-changing $s$-wave'' ($s_\pm$) symmetry.  We review theoretical aspects, materials properties, and experimental evidence relevant to this suggestion, and discuss which further measurements would be useful to settle these issues.

\end{abstract}

\tableofcontents

\maketitle

\date{\today}

\begin{indented}
\scriptsize
\item[] \textit{
Satisfactoriness has to be measured by a multitude of standards, of which some, 
for aught we know, may fail in any given case; and what is more satisfactory 
than any alternative in sight, may to the end be a sum of \textbf{pluses} and \textbf{minuses}, 
concerning which we can only trust that by ulterior corrections and improvements 
a maximum of the one and a minimum of the other may some day be approached.\newline
William James, Meaning of Truth}
\end{indented}

\section{Introduction.}
\label{sec:intro}

\subsection{Aim and scope of this article.}

The iron arsenide superconductor LaFeAsO with critical temperature 26K  was discovered in 2008 by Hideo Hosono and collaborators \cite{Hosono}. Within two months, materials based on substitution of La with other rare earths had been synthesized,  raising the critical temperature of Fe-based superconductors (FeBSs) to 55K.  This rapid sequence of discoveries captured the attention of the  high-temperature superconductivity community.  The following three years saw the discovery of several related families of materials, the rapid calculation
of their electronic structure within density functional theory (DFT), and the development of microscopic models for superconductivity largely based on these DFT calculations. The existence of a second class of high-temperature superconductors is generally agreed to be important not only for the possible existence of materials with even higher $T_c$'s within the same  class of Fe-based materials, but because the comparison with the cuprates can allow one to potentially understand the essential ingredients of high-temperature superconductivity. Because of the extremely rapidly advancing nature of the field, the perils of writing a review are obvious.  Several  authors have nevertheless recently attempted to summarize the status of research in this area, and we have benefited greatly from the existence of these works \cite{SadovskiiReview, IvanovskiiReview, IzyumovReview, IshidaReview, JohnstonReview, PaglioneReview, LumsdenReview, WenReview, StewartReview}.

We intend in this smaller scope review to focus on one particular question among the several fascinating issues surrounding the Fe-based superconductors, namely the symmetry and structure of the superconducting gap.  In the study of cuprate superconductors, the $d_{x^2-y^2}$ symmetry of the gap, with $\cos k_x-\cos k_y$ structure, was empirically established soon after high quality samples were prepared, by penetration depth, ARPES, NMR and phase sensitive Josephson tunneling experiments.  After three years of intensive research on the Fe-based superconductors,  no similar consensus on any universal gap structure has been reached, and there is strong evidence that small differences in electronic structure can lead to strong diversity in superconducting gap structures, including gaps with nodes in some and full gaps in other materials.  The actual symmetry class of most of the materials may be the same, of generalized $A_{1g}$ ($s$-wave) type, probably involving a sign change of the order parameter between Fermi surface sheets in most materials.  In addition, there have been recent suggestions that some related materials, furthest from the nearly compensated semimetal band structure of the originally discovered compounds, may have $d$-wave symmetry.  Understanding both the symmetry character of the superconducting ground states and the detailed structure should provide clues to the microscopic pairing mechanism in the pnictides and thereby a deeper understanding of the phenomenon of high-temperature superconductivity.

A complete review of the literature of even the more focussed problem of gap structure we consider here is beyond the scope of our paper.  We have attempted to select those works we find most relevant (or at least sufficiently representative) to the questions we believe to be important:
\begin{itemize}
\item What do experiments tell us about the pairing symmetry and gap structure of the
Fe-based materials, and what systematic trends can one identify?
\item How do changes in electronic structure drive changes in gap symmetry and structure?
\item What physical effect drives pairing in Fe-based materials? Is more than one important, at least in some materials?
\item What role does disorder play, and how is it related to changes in carrier densities,
electronic structure, and pairing interactions?
\end{itemize}
{If any works in this category have been omitted or slighted, we beg the authors will attribute
it to ignorance or haste rather than malice!}


\subsection{Fe-based superconductors.}

\subsubsection{Comparison with cuprates.}

High-$T_c$ cuprates are known for their high critical temperature, unconventional superconducting state, and unusual normal state properties. The  Fe-based superconductors (FeBS), with $T_c$ up to 55K in SmFeAsO$_{1-x}$F$_{x}$ stand in second place after cuprates and 15K above MgB$_2$. When superconductivity in  the FeBS was discovered, the question immediately arose: how similar are they to cuprates? Let us compare some of their properties.

Both cuprates and FeBS have 2-dimensional lattices of $3d$ transition metal ions as the building blocks. In both cases  orthorhombic distortions can be present at small doping. The main structural difference in these planes is that the $2p$-ligands lie very nearly in the plane with the Cu in cuprates, while in FeBS As, P, Se, or Te lie in nearly tetrahedral positions above and below the Fe plane. The in-plane subset of Cu $d$-orbitals $e_g$ are both present near the Fermi level, and of these, the planar $d_{x^2-y^2}$ is quite dominant (see, however, \cite{Sakakibara}), allowing in principle the reduction of  the multiband electronic structure to a low-energy effective one-band model. In FeBS, on the other hand,  out-of-plane As hybridize well with the $t_{2g}$ Fe $d$-orbitals and all three of them have weight at the Fermi surface.  In addition (as opposed to the cuprates), there is substantial overlap between the $d$-orbitals.  The minimal model is then essentially multiband and that makes FeBS in this respect more similar,  e.g. to ruthenates than to cuprates.

At first glance, the phase diagrams of cuprates and many FeBS are similar. In both cases the undoped compounds exhibit antiferromagnetism, which vanishes  with doping; superconductivity appears at some nonzero doping and then disappears, such that $T_c$ forms a ``dome''. While in cuprates the long range ordered N\'eel phase vanishes before superconductivity appears, in FeBS the competition between these orders can take several forms. In LaFeAsO, for example, there appears to be a first order transition between the magnetic and superconducting states at a critical doping value, whereas in the 122 systems the superconducting phase coexists with magnetism over a finite range and then persists to higher doping. It is tempting to  conclude that the two classes of superconducting materials show generally very similar behavior, but there are profound differences as well. The first striking difference is that the undoped cuprates are Mott insulators, but FeBS are metals. This suggests that the Mott-Hubbard physics of a half-filled Hubbard model  is not a good starting point for pnictides, although some authors have pursued strong-coupling approaches. It does not of course exclude effects of correlations in FeBS, but they may be moderate or small. In any case, DFT-based approaches describe the observed Fermi surface and band structure reasonably well for the whole phase diagram, contrary to the situation in cuprates, especially in undoped and underdoped regimes.

The second important difference pertains to normal state properties. Underdoped cuprates manifest pseudogap behavior in both one-particle and two-particle charge and spin observables, as well as a variety of competing orders. At least for hole-doped cuprates, a strange metal phase near optimal doping is characterized by linear-$T$ resistivity over a wide range of temperatures. In FeBS, different temperature power laws for the resistivity, including linear $T$-dependence of the resistivity for some materials, have been  observed near optimal doping and interpreted as being due to multiband physics and interband scattering \cite{GolubovResistivity}. The FeBS do not  manifest a robust pseudogap behavior in a wide variety of observable properties.

The mechanism of doping deserves additional discussion. Doping in cuprates is accomplished by replacing one of the spacer ions with another one with different valence or adding extra out-of-plane oxygen, e.g. La$_{2-x}$Sr$_x$CuO$_2$, Nd$_{2-x}$Ce$_x$CuO$_2$, and YBa$_2$Cu$_3$O$_{6+\delta}$. The additional electron or hole is then assumed to dope the plane in an itinerant state. In FeBS, the nature of doping is not completely understood: similar phase diagrams are obtained by replacing the spacer ion as in \LOFFA~and \SrKFeAs, or by in-plane substitution of Fe with Co or Ni as in \BFCA~and \BFNA, or by replacing Ba with K, \BFKA. Whether these heterovalent substitutions dope the FeAs or FeP plane as in the cuprates was not initially clear \cite{Sawatzky_09}, but now it is well established that they affect the Fermi surface consistent with the formal electron count doping \cite{Nakamura,Brouet}. Another mechanism to vary electronic and magnetic properties is via the possibility of isovalent doping with phosphorous in \BFAP~or ruthenium in \BFAR. `Dopants' can act as potential scatterers and change the electronic structure because of differences in ionic sizes or simply by diluting the magnetic ions with nonmagnetic ones. But crudely the phase diagrams of all FeBS are quite similar, challenging workers in the field to seek a systematic structural observable which correlates with the variation of $T_c$. Among several proposals, the height of the pnictogen or chalcogenide above the Fe plane has frequently been noted as playing some role in the overall doping dependence \cite{Kuroki_pnictogen_ht_2009, Mizuguhci, Kuchinskii}.
%

It is well established that the superconducting state in the cuprates is universally $d$-wave. By contrast, we review evidence below that the gap symmetry and/or structure of the FeBS can be quite different from material to material. Nevertheless, it seems quite possible that the ultimate source of the pairing interaction in both systems is fundamentally similar, although essential details such as pairing symmetry and the gap structure in the FeBS depend on the FS geometry, orbital character, and degree of correlations.

\subsubsection{Comparison with MgB$_{2}$.}
\label{subsubsec:MgB2}



MgB$_{2}$ was the first example of multigap superconductivity (or at least,
the first one recognized as such). There is little doubt that this property
is shared by FeBS, as discussed below; therefore it is instructive to see
which multiband features have been discovered in MgB$_{2}$ and what
similarities can be found in FeBS.

The thermodynamic properties of MgB$_{2}$ show very characteristic behavior which can easily be understood within multiband BCS theory (see Section~\ref{subsec:multiband}) assuming a weak coupling between bands.
At the critical temperature the larger gap is clearly visible in
thermodynamics, and at a lower temperature, roughly corresponding to the
critical temperature of the weaker band alone, the second gap becomes
manifest in thermodynamic properties. The second gap, while formally appearing at
the same temperature as the first gap, remains very small until a much lower
temperature. Such considerations of course lead one to examine also the opposite situation where the bands are strongly coupled (as  in essentially all theories of superconductivity in FeBS).
 In this case both gaps gradually diminish as T is raised, but one or
both may show non-BCS behavior, and the thermodynamic properties cannot
be accurately described by one gap; the sum of two gaps, on the other
hand, can provide a realistic description.
This is indeed the case in many FeBS, as probed by specific heat, penetration depth, NMR relaxation rate etc. (see Section~\ref{sec:expt_structure}). On the other hand, the picture is additionally clouded, as compared to MgB$_{2}$, because of presumably larger gap anisotropy and pair-breaking effects of impurity scattering, as discussed below.

Another manifestation of  multiband superconductivity is found in the
thermal conductivity. The reduced thermal conductivity $\kappa / T$ is,
generally speaking, zero at $T=0$, if the Fermi surface is fully gapped in
the superconducting state (although pair breaking effects  due to magnetic impurities may, in principle, create mobile quasiparticle states with zero energy \cite{Koshelev}). In MgB$_{2}$, as well as in many FeBS, this is the case. Upon applying magnetic field, Abrikosov vortices form in the system. As soon as these
vortices begin to overlap, the thermal conductivity starts growing. This
happens at field on the order of $H_{c2}/3$. Now, if there are two gaps in
the system, one substantially smaller than the other, one may think that the
vortex overlap will start at much smaller fields. Indeed, the distance
between the vortices is proportional to $H^{-1/2}$, while their size is defined by
the coherence length and thus inversely proportional to $\Delta$. So, the
critical field where the ``weaker'' band will be smaller than that for the ``stronger'' band by a factor of, roughly. $(\Delta_{1}/\Delta _{2})^{2}$, which is, for MgB$_{2}$, about 10. So, the argument goes \cite{Tanatar}, one cannot observe the flat low-field part of $\kappa(H)/T$, and experimentally the dependence looks linear at the smallest accessible $H$. Of course, one must see flattening at $H\lesssim H_{c2}/30$, but so far nobody has observed this.  We only emphasize here that many FeBS studied by this technique show a linear increase of $\kappa / T|_{T\rightarrow 0}$ with $H$ at small $H$, which, in a traditional multiband interpretation, suggest a considerable disparity between the largest and the smallest gaps, or possibly strong gap anisotropy.

Another interesting lesson that one can derive from the MgB$_{2}$ studies is
negative. One of the reasons why a number of theorists were initially
reluctant to accept the two-gap scenario for this material is the fact that
nonmagnetic impurities, in the Abrikosov-Gor'kov theory, should
suppress $T_{c}$ linearly with fairly large slope until the gaps are averaged. 
This effect has not been observed in MgB$_{2}$. While there have been reasonable explanations of why particular impurities may have little effect on $T_{c}$ \cite{Mazin_Andersen}, in retrospect it is clear that the impurity effect is weaker than that expected from the theory in many different cases.

Finally, it is worth looking back at the normal state of MgB$_{2}$. Detailed
quantum oscillation studies \cite{CarringtonKortus} prove unambiguously
that the two band systems ($\pi$ and $\sigma$) are shifted with respect to
each other by up to 100 meV. Similarly, in FeBS quantum oscillations
shows that the hole bands and the electron bands are shifted with respect to
each other by up to 70 meV, so that the hole and the electron Fermi surface
become smaller relative to DFT predictions. This holds both in magnetic \cite{Terashima_dHvA} and nonmagnetic \cite{Coldea_Carrington_dHvA} cases. It has been ascribed to correlation effects \cite{Ortenzi}, but the comparison with MgB$_{2}$
demonstrates that these effects beyond LDA are, if anything, less severe
than in MgB$_{2},$ which is not generally considered to be a correlated
metal.

\subsubsection{Conceptual importance.}


While the FeBS may not signify a particular advance in terms of practical
applications---their $T_{c}$ is only 15K higher than that of MgB$_{2}$,
and, just as the cuprates, they are expensive to make and
difficult to work on---their conceptual value is hard to overestimate. Indeed, fullerides and MgB$_{2}$ clearly belong to a different class than the cuprates, being in certain respects exotic, but still phonon-driven superconductors. Not surprisingly, there had been a growing feeling among physicists that phonon superconductivity will
probably never grow past 50-60K, while true high-temperature superconductivity
is a strong-correlation phenomenon limited to the unique family of layered
cuprates. It had been justly pointed out that the CuO$_{2}$ layers have many
unique properties, largely coming from the fact that Cu is the last $3d$
transition metal and as such is by far the most strongly correlated of all, yet
its simple one-orbital electronic structure provided for a simple and large
Fermi surface when doped. One can point to many aspects in which cuprates are unique, and many people did.

What the discovery of the FeBS brought onto the table was the understanding that
however unique cuprates may be, these features are not  prerequisites for  non-phonon, high-temperature superconductivity. And, if that is true, there are likely many other crystallochemical families to be discovered, some of which may have higher critical temperature or be better suited for applications than cuprates
and FeBS.

In a twisted way, we are lucky that FeBS and cuprate are so different in so
many aspects. This makes it more reasonable to look for those few commonalities which exist and to assume, even without profound theoretical insight, that these
commonalities are important for high $T_{c}$. Some of these obviously include proximity to magnetism and quantum criticality, or substantial anisotropy of the Fermi surface (quasi-2D) and it is has already been argued by many that one should
look for a combination of these factors to search for novel superconductors \cite{IIMazin_Nature}.

\subsubsection{Gap symmetry and structure.}

\begin{figure}
\begin{indented}
\item[]
\includegraphics[width=0.5\columnwidth]{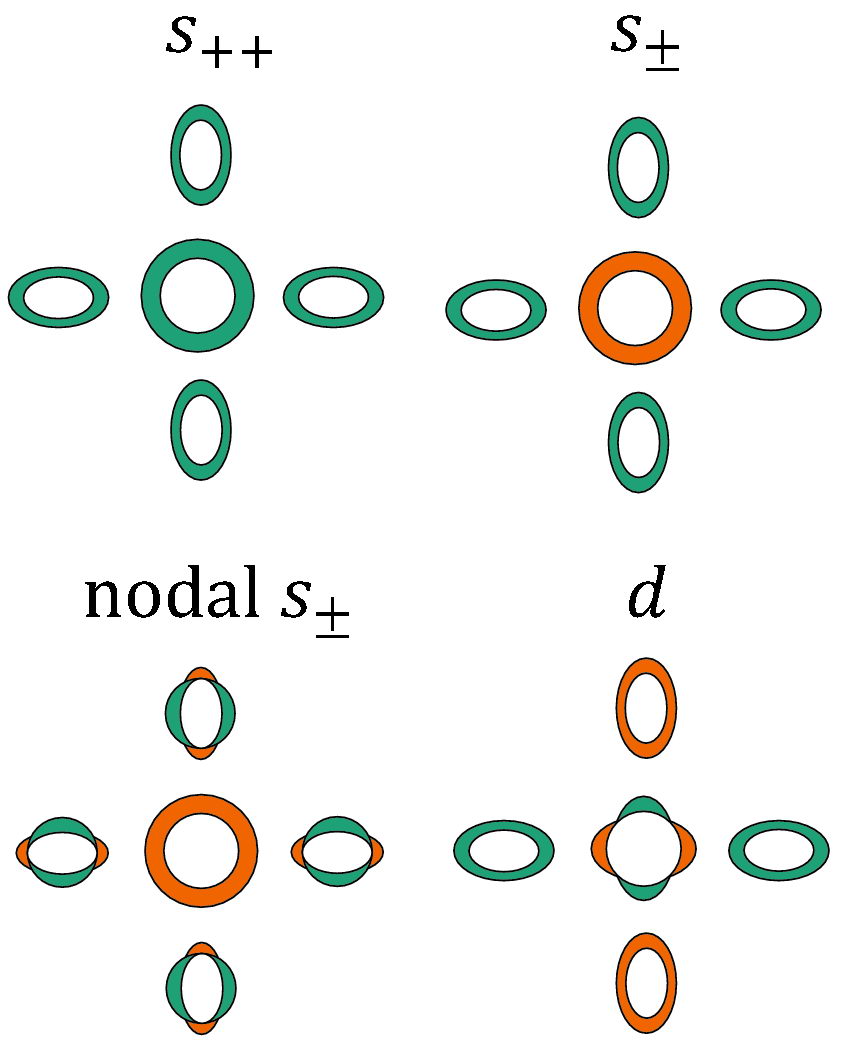}
\caption{Cartoon of order parameters under discussion in the Fe-pnictide superconductors represented in the 2-dimensional, 1-Fe Brillouin zone (see Section 2). Different colors stands for different signs of the gap.}
\label{fig:op_cartoon}
\end{indented}
\end{figure}

The group theoretical  classification of gap structures in unconventional superconductors is somewhat arcane and has been amply reviewed  elsewhere \cite{SigristUeda}.  Here we present the simplest notions relevant to the discussion of symmetry and structure of the order parameters under discussion in the Fe-based superconductors at present.
In the absence of spin-orbit coupling, the total spin of the Cooper pair is well-defined and can be either $S=1$ or $S=0$. Experimental data appear to rule out spin triplet states (see Section~\ref{sec:expt_symmetry}), so we focus on
the spin singlet case.  We focus first on simple tetragonal point group symmetry. In a 3D tetragonal system, group theory allows only for five one-dimensional irreducible representations: $A_{1g}$ (``$s$-wave''), $B_{1g}$ (``$d$-wave'' [$x^2-y^2$]), $B_{2g}$ (``$d$-wave'' [$xy$]), $A_{2g}$ (``$g$-wave'' [$xy(x^2-y^2)$]), and $E_g$ (``$d$-wave'' [$xz,yz$])  according to how the order parameter transforms under rotations by 90$^\circ$ and other operations of the tetragonal group. In figure~\ref{fig:op_cartoon} we have illustrated two of these symmetries, namely $s$-wave and $d_{x^2-y^2}$-wave. Note that the $s_{++}$ state and $s_{\pm}$ states represented all have the same {\it symmetry}, i.e. neither changes sign if the crystal axes are rotated by 90$^\circ$.  By contrast, the $d$-wave state changes sign under a 90$^\circ$ rotation.  Note further that the mere existence of
the single hole and single electron pocket shown lead to new ambiguities in the sign structure
of the various states.  In addition to a global change of sign, which is equivalent to a gauge transformation, one can have individual rotations on single pockets and still preserve symmetry; for example, if one rotates the gap on the hole pocket for the $d$-wave
case in figure~\ref{fig:op_cartoon} by 90$^\circ$ but keeps the electron pocket signs
fixed, it still represents a $B_{1g}$
state.  Note also that $B_{2g}$ states, while not shown in the figure, are also possible by symmetry and would have nodes on the electron pockets.  Further, more complicated, gap functions with differing relative phases on the different pockets become possible when more pockets are present,  and/or when 3D effects are included (See Section~\ref{subsec:dimensionality}).

These {\it symmetry} properties are distinct from gap {\it structure}, a term we use to designate the $k$-dependent variation of an order parameter within a given symmetry class. Gaps with the same symmetry may have very different structures, as also illustrated in the figure, where three different types of $s$-wave states are shown.  The isotropic, fully gapped $s_{++}$ and $s_\pm$ states differ only by a relative phase of $\pi$ in the latter case between the hole and electron pockets.  On the other hand, in the nodal $s$ case, the gap is shown vanishing at certain
points on the electron pockets.  This particular case shows a case we will sometimes refer to as ``nodal $s_\pm$'', in that the sign on the hole pockets is still opposite the average sign on the electron pockets.  Nodes of this type are sometimes described as ``accidental'', since their existence is not dictated by symmetry, but rather by the details of the pair interaction. As such, they can be removed continuously, resulting in either an $s_{++}$ {\it or} an $s_\pm$ state.

\section{Electronic structure.}
\label{sec:bandstructure}

\subsection{First principles.}

\subsubsection{General properties of FeBS electronic structure.}


The basic crystallographic element of the FeBS is the FeAs (where instead of
As one can also have P, Se or Te) plane with an ${a} \times {a}$
square plane of Fe ions, and two $\tilde a \times \tilde a$ square planes of As above and
below (where $\tilde a = {a}\sqrt{2}).$ The minimal unit cell of the entire FeAs
plane is, therefore, also $\tilde a \times \tilde a$ and includes two formula units. In some,
but not all cases the low-energy part of the electronic structure can be
``unfolded'' into a  Brillouin zone (BZ) which is twice as large, corresponding to the ${a}\times {a}$ unit cell, so that the
real band structure can be recovered by folding the 2D Brillouin zone in such
a way that the ``unfolded''
$\mathrm{X}=(\pi/a,0)$ and $\mathrm{Y}=(0,\pi/a)$
points fold on top of each other, forming the
$\tilde{\mathrm{M}}=(\pi/\tilde{a},\pi/\tilde{a})$
point in the  small
Brillouin zone. Here and throughout the article we always use the untilded
notations in the one-Fe unit cell and the ``unfolded'' Brillouin zone, and the tilded notations in the crystallographic unit cells and the corresponding Brillouin zone.

Despite the large variety of crystal structure and chemical compositions, all
FeBS share the same gross features of the electronic structure. These can be
listed as follows:

1. In the nonmagnetic state, the band structure is of semimetallic nature,
with two or more hole band crossing the Fermi level near the $\Gamma$ point
and two electron bands crossing the Fermi level near the $\tilde{M}$ point.

2. Two hole bands, universally present in all superconducting compositions, are
formed by the $xz$ and $yz$ derived Fe band, which are degenerate (without the
spin-orbit effects) at the $\Gamma$ point, but split apart (and acquire some
$z^{2}$ character) in a relatively uniform manner away from it. The two
electron bands take their origin from the downfolding effect described above,
and are formed mostly by the ${x}{z}$ and ${y}{z}$
orbitals, respectively, plus the ${x}{y}$ orbital.

3. As a result, there are always at least two hole Fermi surfaces and two
electron Fermi surfaces, which are well separated in the reciprocal space.
Moreover, their respective centers are removed from each other exactly by
$\Q = \widetilde {(\pi/\tilde a,\pi/\tilde a)}=(\pi/a,0)$. In general, the Fermi surfaces have sufficiently different shapes so that one cannot speak of a good nesting here, only of a quasi-nesting.

On the other hand, many aspects of the electronic structure vary from
material to material. For instance, in some materials another hole band
appears, which may be either of $z^{2}$ character, in which case it is
substantially 3D, or ${x}{y},$ which is even more 2D than the
$xz/yz$ bands.

Different FeBS may have different degree of charge doping, of either sign,
ranging from 0.5 h to 0.15 e per Fe. It appears though that in all  this
compositional range, the general structure of the FS almost always survives. That is, while either the electron or the hole FS shrinks, they never entirely disappear in the superconducting range of dopings, even though the nesting conditions may have drastically deteriorated. It worth remembering that in strongly anisotropic systems the size of a FS has little correlation with its density of states at the Fermi level. In addition, recently two FeBS systems have been discussed that are superconducting but which may lie outside this ``typical'' range; in \KFA~the Fermi surface pockets near the $\mathrm{X}$ point nearly disappear, while the hole pockets around $\Gamma$ are greatly expanded, and \KFS~may (or may not; see Section~\ref{subsubsec:KFS}) be completely lacking the hole pockets.

{By contrast,  the $k_{z}$ dispersion can vary substantially from material to material}. This depends mostly on two factors: the thickness of the ``filler'' layer between the FeAs or FeSe layer, and on crystallographic symmetry. Obviously, materials with the 1111 structure are more anisotropic, in fact nearly 2D, than those with the same $P4/nmm$ symmetry but no filler species, that is, with the 11 structure. Less obviously,
materials with a body center symmetry, such as 122 ($I4/mmm$), have an
additional reason for a 3D character: the downfolding procedure in that case
projects the electronic states near the $\left({\pi/a}, {0}, {0}\right)$ point onto the states near $\left(0,{\pi/a}, {\pi/c}\right)$. Crossings occur at general $k$-point, and
therefore hybridization between these states is not forbidden, but depends on
$k_{z}$. As a result these materials tend to be even more 3D than the 11
compounds, despite having a filling monolayer in the former.  {We explore the
consequences of $k_z$ dispersion for superconductivity in Section~\ref{subsec:dimensionality}.}

\subsubsection{Limitations of DFT calculations.}

First principles calculations were very {important} in the beginning of the
FeBS era, and they have informed the emerging understanding of the physics of the FeBS much more than in the case of cuprate superconductors, although due to somewhat stronger correlations and complexity of materials they have not yet proven as definitively useful as in MgB$_2$. In FeBS, they successfully predicted the right topology of the FS \cite{SinghDu}, as well as the correct magnetic ordering in the normal state \cite{Mazin_etal_splusminus}. The most successful
proposal regarding the pairing symmetry so far has been made based on  band
structure calculations \cite{Mazin_etal_splusminus}. These facts have been instrumental boosting the reputation of the band theory in the superconducting context, but one should remember that while  DFT  is not a snake oil, it is not a panacea either, and nor are any of its generalizations such as DFT+DMFT etc. Let us list below the most important shortcomings and limitations of the DFT calculations as regards FeBS.

1. The DFT is, by construction, a mean-field theory (but not a low-energy
theory, as is sometimes incorrectly asserted). It is a more sophisticated mean
field theory than many, for it includes in the energy functional (and thus in
the mean field potential), all correlation effects and integrates in all fluctuations. On the other hand, the actual implementations
of the DFT, such as the local density approximation or the generalized gradient
approximation, by construction only include those correlations and those fluctuations that
are present in the reference system, the  uniform electron gas, at densities comparable
with the electron densities in real solids. Remember that  the  uniform electron gas
at such densities is very far from magnetism, and even farther from the electron localization
(Wigner crystallization). The corresponding physics is, therefore, largely missing
when DFT is used in a ``local'' approximation. This belongs to two major classes: (i) on-site
Coulomb fluctuations, also called Hubbard correlations, which are included in a
very limited way on the level of the Stoner magnetic interaction (reflecting
the first Hund's rule), and (ii) quantum critical fluctuations; examples of
such are long range ferromagnetic fluctuations in nearly ferromagnetic metals \cite{QCP}.
The hallmark of the former is underestimation of the tendency to magnetism in a
DFT calculation, of the latter - overestimation.

In cuprates, the DFT calculations suffer from the former problem, in FeBS
mostly from the latter. From the density functional point of view, it is
rather curious that despite the fact that the calculated magnetic moments are
large, as opposed to such known cases of near-quantum-critical
materials ZrZn$_{2}$ or Fe$_{3}$Al, yet the effect of such long-range
fluctuations appears to be strong. The explanation is that magnetic moments in
this system are quite soft (in the calculations
they can change from nearly zero to more than a Bohr magneton depending on the
magnetic pattern), and on top of the transverse spin fluctuations typical for strong
antiferromagnets, there are longitudinal
fluctuations characteristic of itinerant magnets and quite efficient in
reducing the ordered magnetic moment.
Moreover, there is a possibility that  other,
so-called ``nematic'' fluctuations, may play an additional role in reducing the ordered moment.

This does not mean that the first DFT problem, underestimation of  on-site
Coulomb correlations, is nonexistent in these materials. It is relatively
mild, and secondary compared to the other deficiency, yet it exists and it
manifests itself, for instance, in the bandwidth renormalization.
From the point of view of this Section, the important corollary of the above
is that superconductivity in FeBS develops not on the background of a
non-magnetic state, but of a paramagnetic state which still has fluctuating
local magnetization of Fe ions. Therefore, the calculated bands and the Fermi
surfaces are true only as long as averaging over these fluctuating quasi-local
magnetic moments is equivalent to dropping the spin-dependent part of the
crystal potential entirely. Experimental evidence
so far has been inconclusive. De Haas - van Alphen experiments generally agree
well with the DFT calculation, up to some uniform shift of different bands
with respect to each other and overall mass renormalization. ARPES derived
Fermi surfaces, while conforming with the general topology, predicted by DFT,
differ in details substantially. It is fair to say that that the DFT bands are
a reasonable, but not exceedingly good approximation of the actual band
structure, even after accounting for the bandwidth renormalization and the
band shifts.
They appear to be renormalized by a factor up to three, and may be even larger
for some systems, and the renormalization appears to be stronger as the system approaches the AFM quantum critical point \cite{Coldea_Carrington_dHvA}. Also, additional repulsive interaction between the holes and the electrons seems to be operative, pushing the (mostly occupied) hole bands down, and (mostly empty) electron bands up. The last effect is responsible for shrinking all Fermi surfaces compared to DFT, but this effect is weak (but is, again, stronger near the quantum critical point) \cite{Ortenzi}.

The band renormalization comes from both the on-site and long-range fluctuations. Existing DMFT calculations, while
qualitatively agreeing among themselves, disagree on the exact share of the total mass renormalization provided by
the on-site vs. long range fluctuations. Indeed, all groups find that the effect comes predominantly from Hund's $J$, and not Hubbard's $U$, and that the 11 family is substantially more correlated than other families. At the same time, the Rutgers group has obtained mass renormalizations closely matching the experiment, leaving basically no room for the long-range fluctuations, while the other DMFT groups' results suggest that both effects provide comparable contributions 
to the total renormalization in pnictides \cite{Anisimov,Georges,Haule}.

\subsection{Minimal band models.}

On the basis of the DFT band structure one can make a simplified model which then can be studied by sophisticated theoretical methods like a Green's function formalism. There is always a trade off between complexity of a model and physical effects captured by it.  Here we discuss several popular models with increasing levels of complexity.

\begin{figure}[ht]
\begin{indented}
\item[]
\includegraphics[width=0.85\columnwidth,angle=0]{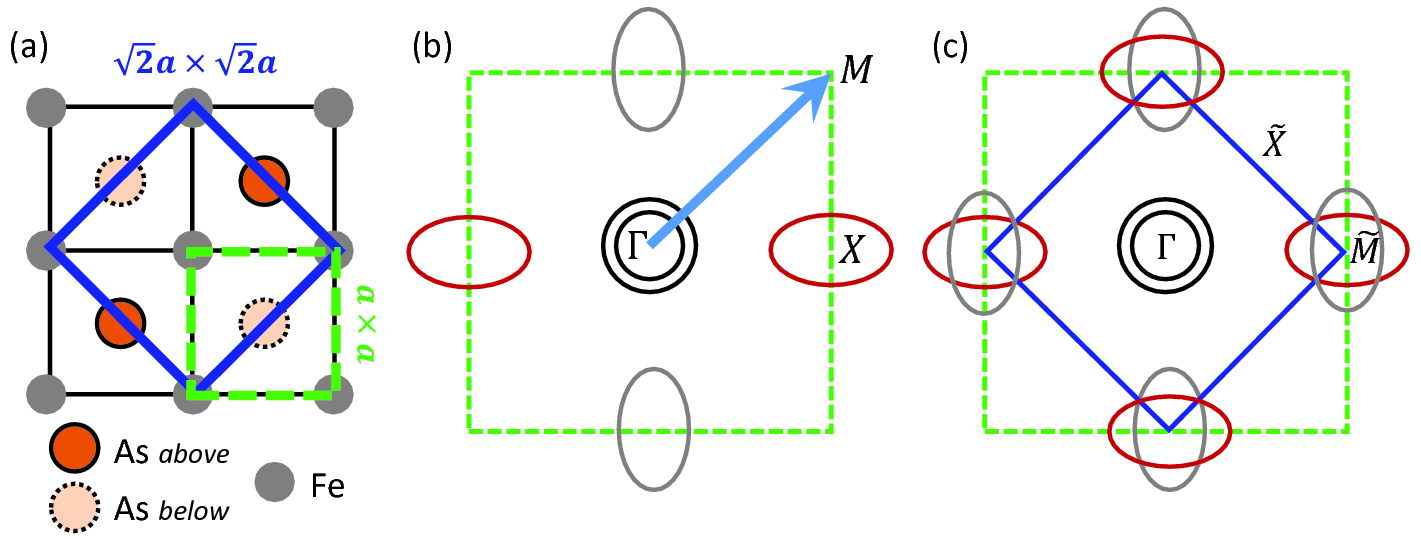}
\caption{(a) FeAs lattice indicating As above and below the Fe plane. Dashed green and solid blue squares indicate 1- and 2-Fe unit cells, respectively. (b) Schematic 2-dimensional Fermi surface in the 1-Fe Brillouin zone whose boundaries are indicated by a green dashed square. Arrow indicates folding wave vector $\Q_F$. (c) Fermi sheets in the folded Brillouin zone whose boundaries are now shown by a solid blue square.
}
\label{fig:folding}
\end{indented}
\end{figure}

According to DFT, FeBS are essentially multiband systems and a minimal model must include both hole and electron bands. The first complication comes from the As, which forms square lattice planes between the lattice sites of, but also above and below,  the square lattice of Fe. This alternating pattern of As makes the correct real space unit cell  twice the one-Fe unit cell. The corresponding 2-Fe BZ is twice as small as the 1-Fe one and called the ``folded BZ'', see figure~\ref{fig:folding}. For the simplest case of single-layer FeBS the folding wave vector is 2-dimensional and equal to $\Q_F = (\pi,\pi)$. Most experimental results and DFT band structures are reported in the folded BZ since crystallographically it is the correct one. However, some experiments sensitive to the Fe positions, like neutron scattering on Fe moments, may have more meaning in the 1-Fe (``unfolded'') zone. Theoretically, the virtue of using this zone is its simplicity, since the number of bands is smaller by a factor of two. From the point of view of symmetry, one might think of the As height as a perturbation: if $z_{As} = 0$, the electronic structure is correctly reproduced in the unfolded BZ, and when $z_{As}$ increases, the procedure of unfolding becomes less and less justified from the structural point of view. Despite the fact that the As displacements from the Fe plane are {\it not} small, use of the Fe-only band structure in the 1-Fe zone is
frequently a good approximation (see, however, Section~\ref{subsubsec:KFS}) since DFT calculations predict the band structure near the Fermi level to be mostly due to Fe $d$-bands, while the As $p$-bands are about 2eV below \cite{SinghDu}.

The simplest model accounting for distinct electron and hole Fermi surfaces  would be a model in the 1-Fe zone with parabolic dispersions \cite{ChubukovReview, Eremin2010},
%
\begin{equation}
 \label{eq:Hparabolic}
 H = \sum_{\k, \sigma, i=\alpha_1,\alpha_2,\beta_1,\beta_2} \varepsilon^{i}_{\k} c_{i \k \sigma}^\dag c_{i \k \sigma}.
\end{equation}
Here, $c_{i \k \sigma}$ is the annihilation operator for an electron with momentum $\k$, spin $\sigma$, and band index $i$, $\varepsilon^{\alpha_{1,2}}_{\k} = - \frac{\k^2}{2 m_{1,2}} +\mu $, $\varepsilon^{\beta_1}_{\k}= \frac{(k_x-\pi/a)^2}{2 m_x} + \frac{k_y^2}{2 m_y} - \mu$, and $\varepsilon^{\beta_2}_{\k}= \frac{k_x^2}{2 m_y} + \frac{(k_y-\pi/a)^2}{2 m_x} -\mu $ are the dispersions of hole $\alpha_i$ and electron $\beta_i$ bands.

The electron pockets have to be distinct since they are located in the different points of the Brillouin zone, but since the two hole pockets around $\Gamma$ point are close in size, some physics can be captured by approximating them as one band. After the folding procedure, this model produces a Fermi surface topology which agrees with the results of DFT calculations, see figure~\ref{fig:folding}. This is useful for qualitative analysis of the physics near the Fermi level, like magnetic susceptibility and formation of the SDW state. The simplest extensions are the tight-binding models which can be formulated in either the folded \cite{Korshunov2008} or the unfolded BZ \cite{Knolle2010}. These can reproduce correctly the Fermi surface and Fermi velocities, but neglect the orbital character of the bands. 
The orbital character of electrons in different bands is important e.g. for a correct analysis of scattering in particle-particle and particle-hole channels. Furthermore, local Coulomb interactions like Hubbard $U$ and Hund's exchange $J$ are momentum-independent only in the orbital representation.

According to DFT analysis, the band structure near the Fermi level consists mainly of Fe $d$-orbitals, since the out-of-plane As $p$ orbitals hybridize most effectively with Fe orbitals with both out-of-plane and in-plane components. These conditions are satisfied for $d_{xz,yz}$ orbitals, so their contribution at the Fermi level is dominant.
The second largest contribution comes from the $d_{xy}$ orbitals.
The other two $d$ orbitals, $d_{x^2-y^2}$ and $d_{3z^2-r^2}$, also contribute at low energies, but their weight at the Fermi surface  is minimal except in some materials near the top of the Brillouin zone.

A possible minimal model for the FeBS is then  one which includes two orbitals, $d_{xz}$ and $d_{yz}$ \cite{Raghu}. This has the virtue of being simple while having mostly correct orbital character along the Fermi surface -- $(0,\pi)$ and $(\pi,0)$ pockets have dominant $d_{xz}$ and $d_{yz}$ contributions, respectively, and hole pockets with a mixture of these two orbitals.  This model  has several significant disadvantages \cite{PALee, s_graser_08}, however. The first one is that the Fermi velocities are incorrect, leading to  incorrect  tendencies towards  superconductivity and SDW formation. Second, it is missing small patches of $d_{xy}$ character at the tips of the electron pockets, which can be important for node formation \cite{Maier_anisotropy} and transport \cite{KemperSelfEnergy}. Finally, a serious flaw  is the position of the larger hole pocket which is located at $(\pi,\pi)$ point of the 1-Fe zone. After the folding, this Fermi surface sheet is centered at the $\Gamma$ point and resembles DFT results for the Fermi surface. But the fact that the two bands forming hole sheets
are not degenerate at the $(0,0)$ point of the unfolded zone contradicts the symmetry of the DFT wave functions.
This problem can be adjusted by adding a $d_{xy}$ orbital to the model \cite{Daghofer}, but this three-orbital model has other pathologies and  fails to reproduce the peak at the nesting wave vector $\Q$ \cite{Brydon}, which is established both experimentally and theoretically. Although the origin of the problem is not obvious, it is related to the matrix elements of transformation from the orbital to the band basis.

\begin{figure}[ht]
\begin{indented}
\item[]
\includegraphics[width=0.4\columnwidth,angle=0]{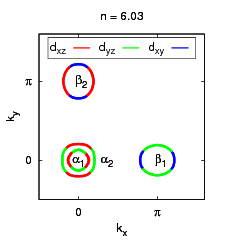}
\includegraphics[width=0.4\columnwidth,angle=0]{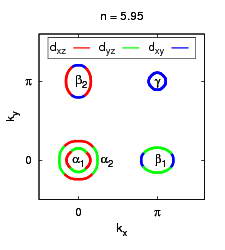}
\caption{Fermi sheets of the five-band model in the unfolded BZ for $n=6.03$ (top) and $n=5.95$ (bottom) with colors indicating majority orbital character (red=$d_{xz}$, green=$d_{yz}$, blue=$d_{xy}$). Note the $\gamma$ Fermi surface sheet is a hole pocket which appears for $\sim 1\%$ hole doping \cite{Kemper_sensitivity}.}
\label{fig:fermisurface}
\end{indented}
\end{figure}

The next step in the direction of more realistic models is to include four or all five Fe $t_{2g}$ orbitals. Models of this type \cite{s_graser_08, k_kuroki_08} work well in reproducing the DFT Fermi surface and band structure, and they are free from the disadvantages described above. The kinetic energy in \cite{s_graser_08} is then given by the Hamiltonian
\begin{equation}
 H_0 = \sum_{\k \sigma} \sum_{\ell \ell'} \left( \xi_{\ell \ell'}(\k) + \epsilon_{\ell} \delta_{\ell \ell'} \right) d_{\ell \k \sigma}^\dagger d_{\ell' \k \sigma},
 \label{eq:H0}
\end{equation}
where $d_{\ell \k \sigma}^\dagger$ creates a particle with momentum $\k$ and spin $\sigma$ in the orbital $\ell$, $\xi_{\ell \ell'}(\k)$ are the hoppings, and $\epsilon_{\ell}$ are the single-site level energies. This model with the parameters obtained by Wannier fits to $k_z=0$ cuts of the 1111 band structure of Cao \textit{et al} \cite{Cao} gives rise to the Fermi surfaces shown in figure~\ref{fig:fermisurface} displayed in the unfolded BZ \cite{s_graser_08, Kemper_sensitivity}. The undoped material has completely filled $d^6$ orbitals corresponding to  electron number $n = 6$. Note for the hole doped case $n = 5.95$ shown in figure~\ref{fig:fermisurface} there is an extra hole FS $\gamma$ around the $(\pi,\pi)$ point.
An important role is played by the orbital matrix elements $a^\ell_\nu(\k)= \langle\ell|\nu \k\rangle$ which relate the orbital and band states. The dominant ($>50\%$) orbital weights $|a^\ell_\nu(\k)|^2$ on the Fermi surfaces are illustrated in figure~\ref{fig:fermisurface} by the colors indicated.
The orbital matrix elements and the small patches of the $d_{xy}$ contribution to the electron sheets play an important role in the formation of nodes in the superconducting order parameter as will be discussed in Section~\ref{subsec:sf}.
Natural generalizations of the multiorbital models are 1) to include dispersion along the $k_z$ direction; and 2) to include the proper effects of the 122 body centered cubic symmetry. These effects  appear to have important consequences for the pairing in \BFKA, where effects of three-dimensionality are more important than in 1111 systems \cite{Graser3D,Kuroki3D}.


\section{Theoretical background.}
\label{sec:theorybackground}

\subsection{Spin fluctuation pairing.}
\label{subsec:sf}

It has become commonplace when a new class of superconductors is
discovered to discuss electronic pairing mechanisms as
soon as there is some evidence that the electron phonon
mechanism is not strong enough to produce observed critical temperatures;
this was the case in both the cuprates and Fe-based superconductors.
Among many candidates for electronic pairing, Berk-Schrieffer \cite{BerkSchrieffer} type spin fluctuation theories are popular because they are relatively simple to express and give some qualitatively correct results in the well-known cases of $^3$He and the cuprates.   The interesting history of the development of this theory in the one-band case has been reviewed by Scalapino \cite{ScalapinoSFhistory}. It is important to keep in mind that this type of description  cannot be regarded as the complete answer even in superfluid $^3$He, where the true pairing interaction contains a significant density fluctuation component, while in the cuprates it is controversial whether the full pairing interaction can be  described by a simple boson exchange theory at all.  Nevertheless spin fluctuation theories can explain  the symmetry of the order parameter in both systems quite well, in part because other interaction channels are projected out in the ground state.  For example, in the cuprates, the $d$-wave nature of the pair wave function follows from the strongly peaked spin susceptibility at $(\pi,\pi)$, characteristic of repulsive local interactions between electrons hopping on a square lattice.  In the Fe-based superconductors, the early realization that the Fermi surface consisted of small, nearly nested electron and hole pockets led to the analogous anticipation of a strongly peaked susceptibility near $\Q=(\pi,0)$, and a corresponding pairing instability with sign change between electron and hole sheets. Below we illustrate the basic equations leading to the canonical $d$-wave case within the one-band Berk-Schrieffer approach, and sketch the generalization to the multiorbital/multiband case. Many authors have obtained similar results with a variety of related methods (see Section~\ref{para:similar_approach}, ``similar approaches'' below).

It is important to emphasize that while spin fluctuation theories come in many varieties and flavors, they share more commonalities than differences. Indeed, as will be discussed later in this Section, in the singlet channel spin fluctuations exchange always leads to a repulsive interaction, and therefore can only realize sign-changing superconducting states. If this interaction is sufficiently strong at some particular momentum it will necessarily result in superconductivity. In case of a single Fermi surface this superconductivity will necessarily be nodal, usually of a $d$-wave symmetry. Examples of this situation are high-$T_{c}$ cuprates and, possibly, overdoped KFe$_{2}$As$_{2}$ FeBS. On the other hand, in a multiband system there may be a possibility to avoid nodes, while still preserving a sign-changing structure. Examples of this are: $d$-wave superconductivity  that can develop in a cubic system with Fermi surface pockets around the $\mathrm{X}$ points and in a hexagonal system with pockets around the $\mathrm{M}$ points \cite{ABG}, $d$-wave superconductivity in a tetragonal system with FS pockets near $\mathrm{X}$/$\mathrm{Y}$ points \cite{k_kuroki_08}, $s_\pm$ superconductivity proposed for bilayer cuprates (where the bonding and the antibonding band have opposite signs of the order parameter) \cite{BulutScalapino,Mazin_bilayer}, and the electron-hole $s_\pm$ superconductivity that can develop in semimetals \cite{AS}.

While all these options had been discussed theoretically many
years ago, all of them have been revisited in connection to FeBS. The last option is now the leading contender for the majority of pnictides and selenides, while $d$-wave superconductivity has been proposed for KFe$_{2}$As$_{2}$ that is on the verge of losing its semimetallic character (see Section \ref{subsec:materials_summary}), a version of the
nodeless $d$-wave state first discussed in \cite{k_kuroki_08}
%
has been proposed for Se-based 122 materials (see Section \ref{subsubsec:KFS}), and the ``bonding-antibonding'' nodeless $s_\pm$ state, analogous to that discussed in \cite{BulutScalapino,Mazin_bilayer} has been also proposed for these selenides. What is important, however, is that as long as the spin fluctuations are strong and nonuniform, some superconducting state will unavoidably form, and the details of the electronic structure and of the pairing interaction will decide which particular symmetry will form, often in close competition with other symmetries.

\begin{figure}[ht]
\begin{indented}
\item[]
\includegraphics[width=0.6\columnwidth]{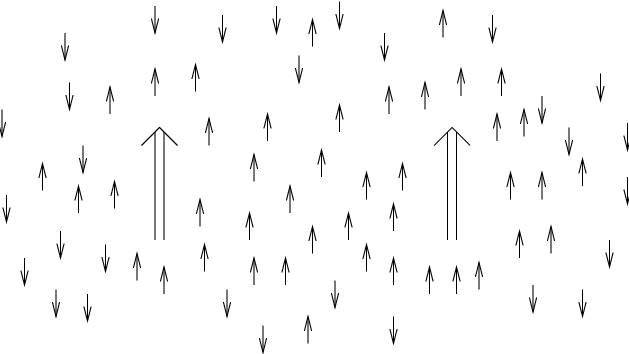}
\caption{Heuristic snapshot of pairing of two test spins by ferromagnetic spin fluctuations.}
\label{fig:heuristic_sf}
\end{indented}
\end{figure}

\paragraph{Historical: ferromagnetic spin fluctuations.} The original
proposal of superconducting pairing arising from magnetic interactions was
put forward by Emery \cite{Emery} and by Berk and Schrieffer \cite{BerkSchrieffer}, who were
interested primarily in transition metal elements and nearly ferromagnetic metals.  
Such systems are considered to be close to a ferromagnetic ordering transition in the Stoner sense, i.e.
their susceptibility may be approximated by $\chi=\chi_0/(1-U\chi_0)$,
where $U$ is  a local Hubbard-like Coulomb matrix element assumed to be
large since $U\chi_0\simeq 1$ ($\chi_0$ is the susceptibility in the absence of interactions).  Physically this means a spin up electron
traveling through the medium polarizes the spins around it
ferromagnetically, lowering the system's energy as illustrated in
figure~\ref{fig:heuristic_sf}.

\begin{figure}
\begin{indented}
\item[]
\includegraphics[width=0.8\columnwidth]{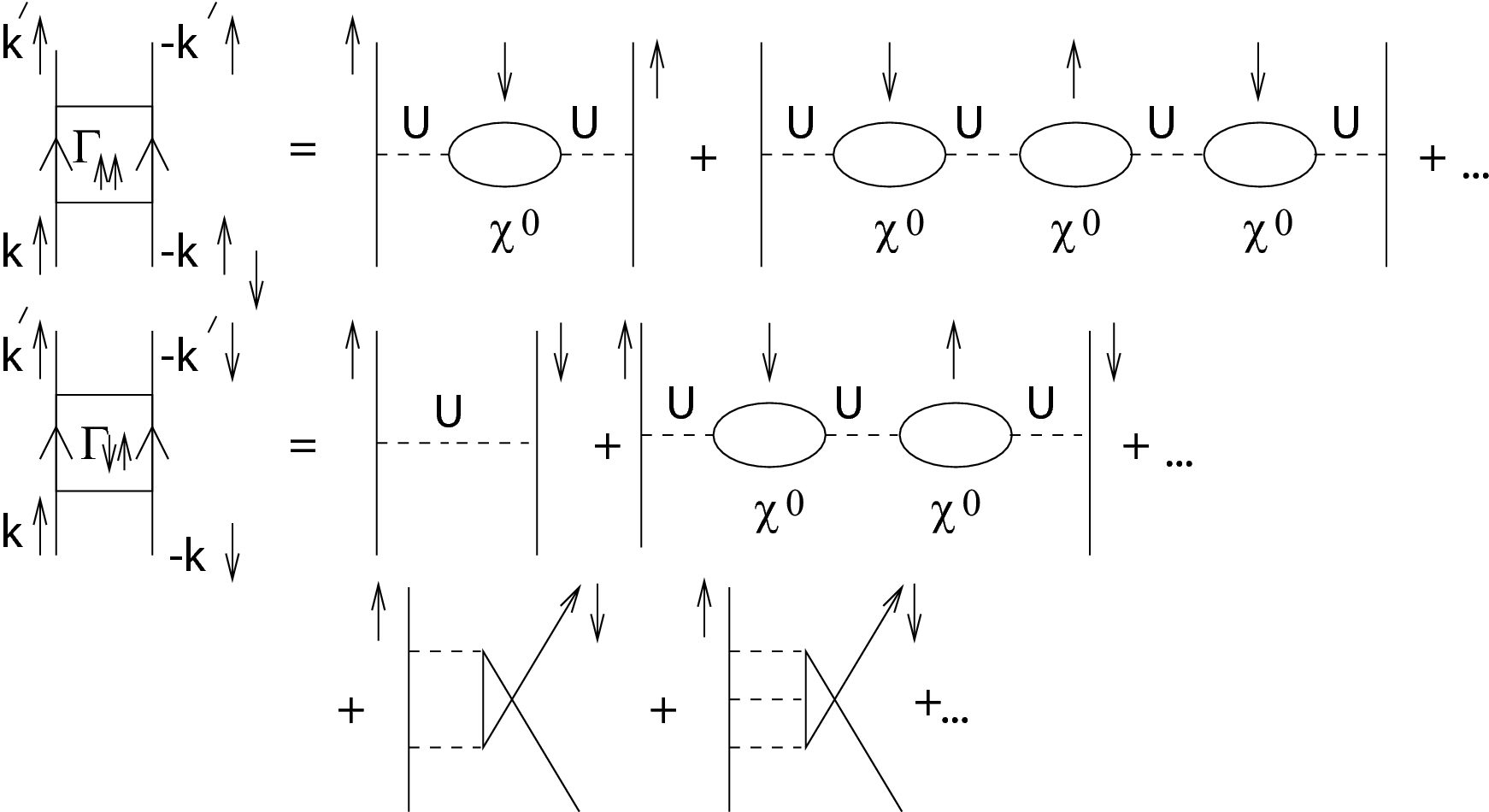}
\caption{Effective pairing interaction between (a) equal spins and (b)
opposite spins.  Solid lines are electron $G$'s, dashed lines Hubbard $U$'s,
i.e. interactions between electrons of opposite spin only.}\label{fig:sf_diagrams}
\end{indented}
\end{figure}

The excitations being ``exchanged'' in such a picture are not well-
defined collective modes of the system such a phonons or magnons,
but rather "paramagnons", defined by the existence of a peak-like
structure in the the imaginary (absorptive) part of the small q susceptibility.
Diagramatically, only the
ring-type diagrams shown in figure~\ref{fig:sf_diagrams} contribute to
the equal-spin channel $\Gamma_{\uparrow \uparrow}$, whereas
both ring and ladder-type diagrams contribute to the effective pairing
vertex in the opposite-spin channel $\Gamma_{\uparrow \downarrow}$.
These series may be summed in the usual way to give:
\begin{eqnarray}
\Gamma_{\uparrow \uparrow} &=& {-U^2\chi_0(\k^\prime-\k) \over
1-U^2{\chi_0}^2 (k^\prime-k)},\label{eq:Gamma1}\\
\Gamma_{\uparrow \downarrow} &=& {U\over 1-U^2{\chi_0}^2 (\k^\prime-\k)}
+ {U^2\chi_0(\k^\prime+\k)\over 1-U^2{\chi_0}^2 (\k^\prime+\k)}\label{eq:Gamma2}\\
&=& U^2\left({3\over 2}\chi^s -{1\over 2}\chi^c\right)+U,\label{eq:Gamma3}
\end{eqnarray}
where we have defined $\chi^s\equiv \chi_0/(1-U\chi_0)$ and $\chi^c=\chi_0/(1+U\chi_0)$, and in the last step we have changed $-\k$ to $\k$ in the second term of $\Gamma_{\uparrow\downarrow}$ because
we assume we work in the even parity (singlet pairing) channel.
The total pairing vertex in the triplet (singlet) channel is $\Gamma_t={1\over 2}\Gamma_{\uparrow\uparrow}$ ($\Gamma_s={1\over 2} (2 \Gamma_{\uparrow\downarrow} -\Gamma_{\uparrow\uparrow} )$).
In the original paramagnon theory, $\chi_0(\q)$ is the  noninteracting
susceptibility of the (continuum) Fermi gas, i.e. the Lindhard function.
This function at small frequency  has a maximum at q=0,
meaning correlations are indeed ferromagnetic.  Thus due to the negative
sign in the equation for $\Gamma_{\uparrow \uparrow}$ (note $\chi_0>0$ and $U\chi_0<1$ to prevent
a magnetic instability), pairing is attractive
in the triplet channel and singlet superconductivity is suppressed, which explains why Pd, for example, does not superconduct \cite{BerkSchrieffer}.

\paragraph{Antiferromagnetic spin fluctuations.} In the context of heavy fermion systems it was realized \cite{ScalapinoHF,VarmaHF} that strong antiferromagnetic spin fluctuations in either the weak or strong coupling limit lead naturally to spin singlet,  $d$-wave pairing. The weak coupling argument has been elegantly reviewed by Scalapino \cite{ScalapinoPhysRep}. Suppose the susceptibility is strongly peaked near some wave vector $\Q$.  The form of the singlet interaction is
\begin{eqnarray}
\Gamma_s(\k,\k') = \frac{3}{2} U^2 \frac{\chi_0(\q)}{1-U\chi_0(\q)} \label{eq:singlet_approx}
\end{eqnarray}
{\it if} we neglect terms which are small near the RPA instability $U\chi_0(\q)\rightarrow 0$ \cite{Nakajima}.
This now implies that $\Gamma_s(\q)$ is also peaked at this wavevector, but is
always repulsive.  Nevertheless, if one examines the BCS gap equation for this interaction
\begin{equation}
\Delta_\k= -{\sum_{\k^\prime}}^\prime \Gamma_s(\k,\k^\prime) {\Delta_\k^\prime \over 2E_\k^\prime } {\rm tanh} {E_\k^\prime \over 2T},
\label{eq:gapeqn}
\end{equation}
one sees immediately that an isotropic state cannot be a solution, but that
if the state changes sign, \begin{eqnarray}\Delta_\k= - \Delta_{\k+\Q},\label{eq:changesign}\end{eqnarray} a solution will be allowed.

In the cuprates, $\chi$ is peaked at $\Q\simeq (\pi,\pi)$, and the two possible states of this type which involve pairing on nearest neighbor bonds only are
\begin{eqnarray}
\Delta_\k^{d,s} & = &\Delta_0 (\cos k_xa \mp \cos k_y a ) .
\end{eqnarray}
Which state will be stabilized then depends on the Fermi surface in question.
So we need to use the fact that the states close to the Fermi surface are the most important in Equation~(\ref{eq:gapeqn}), and examine the pairing kernel for these momenta.  For example, for a $(\pi/a,0)\rightarrow (0,\pi/a)$ scattering, $\Delta_\k^s$ satisfies
Equation~(\ref{eq:changesign}) by being zero, whereas $\Delta_\k^d$ is nonzero and changes sign, contributing to the condensation energy.
It should therefore not be surprising that the end result of a complete
numerical evaluation of Equation~(\ref{eq:gapeqn}) over a cuprate Fermi surface  gives $d$-wave pairing.

An alternative way to approach the question of how a purely repulsive interaction allows for pair formation is to examine the interaction Fourier transformed back to real space, where it shows regions (in the cuprates with $\Q=(\pi,\pi)$, on nearest neighbor sites), where the pair potential becomes attractive \cite{ScalapinoPhysRep} if the interaction is sufficiently nonuniform in  momentum space.

\paragraph{Effect of 2 bands.}
The discussion in Section~\ref{sec:bandstructure} described the unusual, fully compensated Fermi surface of the parent Fe-pnictide materials.  In the Fe-based superconductors, the proximity of the Fe $d$-states to the Fermi level has led many authors to consider a Hamiltonian which takes a kinetic energy $H_0$ consisting of the bands derived from the 5 Fe $d$-orbitals found in DFT, approximated within some tight-binding or other scheme. The 2D Fermi surface in the 1-Fe zone thus obtained is shown in figure~\ref{fig:fermisurface}. Like DFT, this model Fermi surface is characterized by small concentric hole pockets around the $\Gamma$ point and slightly elliptical electron pockets around the $\mathrm{M}$ points.  Mazin and collaborators \cite{Mazin_etal_splusminus}
pointed out that modeling these pockets in the simplest possible way, allowing for one hole and one electron pocket, led to  a very simple and elegant generalization of the ``standard'' argument for $d$-wave pairing in the cuprates (a similar result was reached in a strong-coupling approach by Seo \textit{et al} \cite{Seoetal}).  In a weak-coupling approach, the near-nesting of the hole and electron pockets suggested the existence of a peak in the spin susceptibility at $\Q=(\pi,0)$ in the 1-Fe zone.    The gap equation~(\ref{eq:gapeqn}) then admits a solution with the property~(\ref{eq:changesign}) if there is a sign change of $\Delta_k$ between electron and hole pockets.
In the simplest version of this theory, the anisotropy on each electron sheet is neglected, with the argument that the pockets are small.  This leads to the so-called isotropic ``$s_{\pm}$" state (figure~\ref{fig:op_cartoon}) which has become the leading candidate for the discussion of many of the superconducting properties of these materials.  Note that such a state has the full symmetry of the crystal lattice and is therefore formally an $A_{1g}$ or ``$s$-wave'' state, but
with fundamentally different {\it gap structure} which leads to many nontrivial superconducting properties.

\paragraph{Spin fluctuation pairing in multiorbital systems.}

More realistic analyses of pairing in these systems by electronic interactions soon followed.  Many
authors began with a Hamiltonian consisting of a kinetic energy $H_0$ for the effective Fe bands as described above, plus an interaction $H_{int}$ containing
 all possible two-body on-site
interactions between electrons in Fe orbitals as a good starting
point for a microscopic description of this system,
\begin{eqnarray}
H & = & H_{0}+\bar{U}\sum_{i,\ell}n_{i\ell\uparrow}n_{i\ell\downarrow}+\bar{U}'\sum_{i,\ell'<\ell}n_{i\ell}n_{i\ell'}\\
 &  & +\bar{J}\sum_{i,\ell'<\ell}\sum_{\sigma,\sigma'}c_{i\ell\sigma}^{\dagger}c_{i\ell'\sigma'}^{\dagger}c_{i\ell\sigma'}c_{i\ell'\sigma} \nonumber \\
 &  & +\bar{J}'\sum_{i,\ell'\neq\ell}c_{i\ell\uparrow}^{\dagger}c_{i\ell\downarrow}^{\dagger}c_{i\ell'\downarrow}c_{i\ell'\uparrow} \nonumber
\label{eq:H}
\end{eqnarray}
where $n_{i\ell} =  n_{i,\ell\uparrow} + n_{i\ell\downarrow}$. The Coulomb parameters ${\bar U}$, ${\bar U}'$, ${\bar J}$, and ${\bar J}'$ are in the notation of Kuroki \textit{et al} \cite{k_kuroki_08}, and are related to those used by Graser \textit{et al} \cite{s_graser_08} by $\bar U=U$, $\bar J=J/2$, $\bar U'=V+J/4$, and $\bar J'=J'$. The noninteracting  $H_0$ is given by a tight-binding model spanned by the 5 Fe $d$-orbitals, Equation~(\ref{eq:H0}), which give rise to the Fermi surfaces shown in figure~\ref{fig:fermisurface}.
%
%

In Equation~(\ref{eq:H}), we have distinguished the intra- and inter-orbital
Coulomb repulsion, as well as the Hund's rule exchange $\bar J$ and
``pair hopping'' term $\bar J'$ for generality, but if they
are generated from a single two-body term with spin rotational
invariance they are related by $\bar U'=\bar U - 2 \bar J$ and
$\bar J'=\bar J$.  In a real crystal, such a local symmetry will not always hold.

\begin{figure}[ht]
\begin{indented}
\item[]
\includegraphics[width=0.7\columnwidth,angle=0]{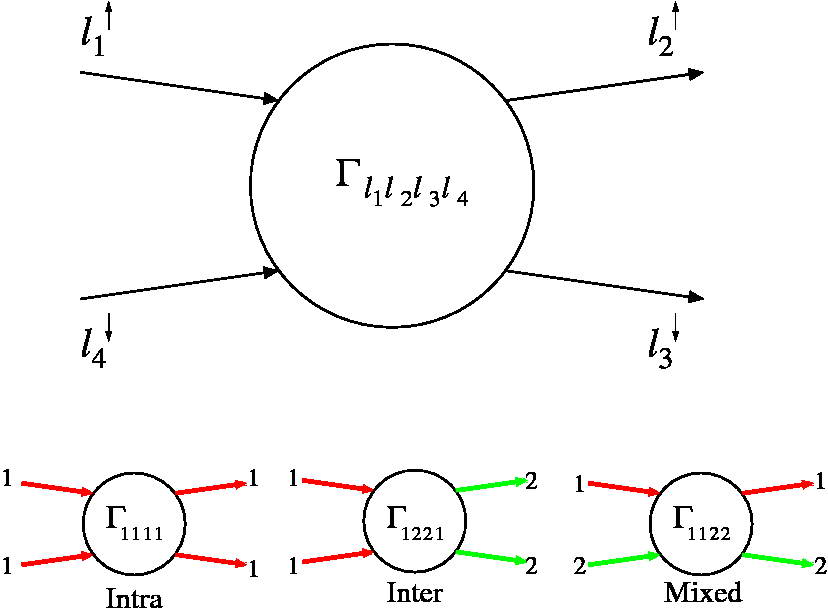}
\caption{Top: pairing vertex $\Gamma_{\ell_1,\ell_2, \ell_3,\ell_4}$ defined in terms of orbital states $\ell_i$ of incoming and outgoing electrons.  Bottom: representative examples of classes of orbital vertices referred to in the text: intra-, inter- and mixed orbital vertices. From \cite{Kemper_sensitivity}.}
\label{fig:vertex}
\end{indented}
\end{figure}

The generalization of the simple 1-band Berk-Schrieffer spin fluctuation theory to the multiorbital case was discussed by many authors \cite{Ueda_etal, Kubo}.
The effective pair
scattering vertex $\Gamma(\k,\k')$ between bands $i$ and $j$ in the singlet channel is
\begin{eqnarray}\label{eq:Gam_ij}
{\Gamma}_{ij} (\k,\k') & = & \mathrm{Re}\left[\sum_{\ell_1\ell_2\ell_3\ell_4} a_{\nu_i}^{\ell_2,*}(\k) a_{\nu_i}^{\ell_3,*}(-\k) \right. \\
&&\left. \times {\Gamma}_{\ell_1\ell_2\ell_3\ell_4} (\k,\k',\omega=0) a_{\nu_j}^{\ell_1}(\k')  a_{\nu_j}^{\ell_4}(-\k') \right] \nonumber
\end{eqnarray}
where the momenta $\k$ and $\k'$ are confined to the various
Fermi surface sheets with $\k \in C_i$  and $\k' \in C_j$.    The
orbital vertex functions $\Gamma_{\ell_1\ell_2\ell_3\ell_4}$ describe the
particle-particle scattering of electrons in orbitals $\ell_1,\ell_4$ into $\ell_2,\ell_3$
(see figure~\ref{fig:vertex}) and in the fluctuation exchange
formulation \cite{n_bickers_89, Kubo} are given by
\begin{eqnarray}\label{eq:fullGamma}
&&{\Gamma}_{\ell_1\ell_2\ell_3\ell_4} (\k,\k',\omega) = \left[\frac{3}{2} \bar U^s
\chi_1^{\rm RPA}  (\k-\k',\omega) \bar U^s + \nonumber \right.\,~~~~~~\,\\
&&\,~~~~~\left.
 \frac{1}{2} \bar  U^s
 - \frac{1}{2}\bar U^c  \chi_0^{\rm RPA}  (\k-\k',\omega)
\bar U^c + \frac{1}{2} \bar U^c \right]_{\ell_1\ell_2\ell_3\ell_4},
\end{eqnarray}
where each of the quantities $\bar U^s$, $\bar U^c$, $\chi_1$, etc
represent matrices in orbital space which depend on the interaction parameters.  This is the
multiorbital generalization of Eq. (\ref{eq:Gamma3}).
Here  $\chi_1^{\rm RPA}$  describes the spin-fluctuation
contribution and  $\chi_0^{\rm RPA}$  the orbital
(charge)-fluctuation contribution, determined by Dyson-type
equations as
\begin{equation}
(\chi_{0}^{RPA})_{st}^{pq} = \chi_{st}^{pq} - (\chi_{0}^{RPA})_{uv}^{pq} (U^c)_{wz}^{uv} \chi_{st}^{wz}
\end{equation}
and
\begin{equation}
(\chi_{1}^{RPA})_{st}^{pq} = \chi_{st}^{pq} +
(\chi_{1}^{RPA})_{uv}^{pq} (U^s)_{wz}^{uv} \chi_{st}^{wz},
\end{equation}
where repeated indices are summed over.  Here $\chi_{st}^{pq}$ is a generalized multiorbital susceptibility (see \cite{Kemper_sensitivity}).

\paragraph{Results of microscopic theory.}

The simplest goal of the microscopic approach is to calculate the critical
temperature $T_c$ via the linearized gap equation and determine the symmetry of the pairing instability there.
If one writes the superconducting order parameter
$\Delta(\k)$ as $\Delta g(\k)$, with $g(\k)$ a dimensionless function
describing the momentum dependence on the Fermi
surface, then $g(\k)$ is given as the stationary solution of
the dimensionless pairing strength functional \cite{s_graser_08}
\begin{equation}
\lambda [g(\k)] = - \frac{\sum_{ij} \oint_{C_i} \frac{d k_\parallel}{v_F(\k)} \oint_{C_j}
\frac{d k_\parallel'}{v_F(\k')} g(\k) {\Gamma}_{ij} (\k,\k')
g(\k')}{ (2\pi)^2 \sum_i \oint_{C_i} \frac{d k_\parallel}{v_F(\k)} [g(\k)]^2 }
\label{eq:pairingstrength}
\end{equation}
with the largest eigenvalue $\lambda$, which provides a dimensionless measure of the pairing strength.
 Here  $\k$
and $\k'$ are restricted to the various Fermi surfaces $\k \in C_i$
and $\k' \in C_j$ and $v_{F,\nu}(\k) = |\nabla_\k E_\nu(\k)|$ is the
Fermi velocity on a given Fermi sheet.

\begin{figure}[ht]
\begin{indented}
\item[]
\includegraphics[width=0.38\columnwidth]{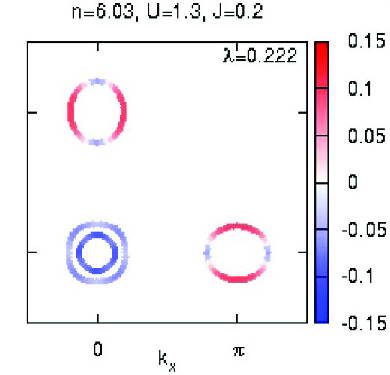}
\includegraphics[width=0.38\columnwidth]{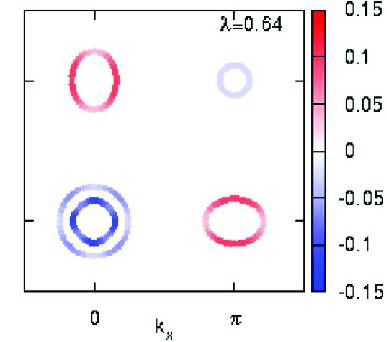}
\vskip 1.5cm
\includegraphics[width=0.6\columnwidth]{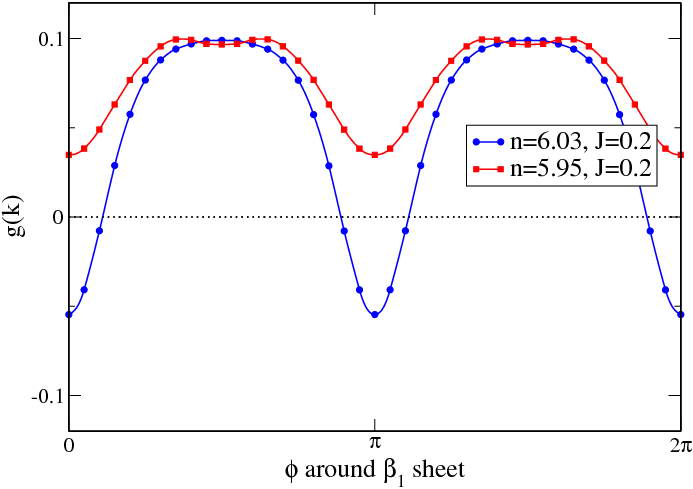}
\caption{Top: false color plots of dimensionless gap function $g(k)$  on various Fermi surface sheets for electron doped $n=6.03$ (left)
and hole doped $n=5.95$ (right).  Bottom: detail of $g(k)$ on  $\beta_1$ pocket for $U=1.3$ and
$n=5.95, \bar J=0.2$ (red squares) and
$n=6.01, \bar J=0.2$ (blue circles).
Here the angle $\phi$ is measured from the $k_x$-axis. From Kemper \textit{et al} \cite{Kemper_sensitivity}.}
\label{fig:evector_vs_angle}
\end{indented}
\end{figure}

In figure~\ref{fig:evector_vs_angle}, we plot the leading dimensionless gap function $g(k)$ derived from the RPA theory
around the electron $\beta_1$ sheet for two different values of the doping, for spin-rotationally invariant parameters $U=1.3$ and $J=0.2$.  The gap on the hole sheets is seen to be essentially isotropic, while on the electron sheets the average of the gap is of opposite sign compared to the hole sheets, and is highly anisotropic, with nodes in the case of electron doping.   One would like to understand the origin of the anisotropy and its doping dependence.

\paragraph{Physical origins of anisotropy of pair state and node formation.}

An unusual element which emerges from the spin fluctuation pairing analysis  based on (\ref{eq:H}) is that  the orbital structure of the Fermi surface can have a significant impact on the anisotropy of the pair state. The intra-orbital scattering of $d_{xz}$ and $d_{yz}$ pairs between the $\alpha$ and $\beta$ Fermi surfaces by $(\pi,0)$ and $(0,\pi)$ spin fluctuations (figure~\ref{fig:vertex}) leads naturally to a gap which changes sign between the $d_{xz}/d_{yz}$ parts of the $\alpha$ Fermi surfaces and the $d_{yz}$ and $d_{xz}$ parts of the $\beta_1$ and $\beta_2$ electron pockets, just as in the early proposal of Mazin \textit{et al} \cite{Mazin_etal_splusminus}, see figure~\ref{fig:fermisurface}.   However, as discussed by Maier \textit{et al} \cite{Maier_anisotropy}, Kuroki \textit{et al} \cite{Kuroki_pnictogen_ht_2009}, and Kemper \textit{et al} \cite{Kemper_sensitivity},
scattering between the  $\beta_1$ and $\beta_2$ Fermi surfaces frustrates the
formation of an isotropic $s_\pm$ gap there.
Furthermore, this anisotropy can also be driven by  the effect of the intraband Coulomb interaction. Finally, inter-orbital pair scattering can also occur, depending upon $\bar U'$ and $\bar J'$. The contributions to the total pairing
interaction arising from mixed and interorbital processes may be
shown explicitly to be subdominant to the intraorbital processes
but important for nodal formation \cite{Kemper_sensitivity}.

On the other hand, the $\beta_1-\beta_2$ $d_{xy}$ orbital frustration
is weaker or does not occur if an additional hole pocket $\gamma$ of $d_{xy}$ character is present (see figure~\ref{fig:fermisurface}); these scattering processes are at ($\pi,0$) in the unfolded zone and therefore support isotropic $s_\pm$ pairing (see figure~\ref{fig:evector_vs_angle}). Kuroki \textit{et al} \cite{Kuroki_pnictogen_ht_2009} took the important step of relating the crystal structure, electronic structure, and  pairing, by noting that the As height above the Fe plane in the 1111 family was a crucial variable controlling the appearance of the $\gamma$ pocket and thus driving the isotropy of the $s_\pm$ state.
It is important to note that
the transition between nodal and nodeless $A_{1g}$ gap structures, investigated by a number of authors \cite{Kuroki_pnictogen_ht_2009, Chubukov_nodal-gapped, Kemper_sensitivity, Thomale_nodal-gapped, Thomale_nodal-gapped2, FaWangLAFP}, does not involve any symmetry
change, and relies only on the details of the pairing interaction.

\paragraph{Similar approaches.}

\begin{figure}[ht]
\begin{indented}
\item[]
\includegraphics[width=0.5\columnwidth,angle=0]{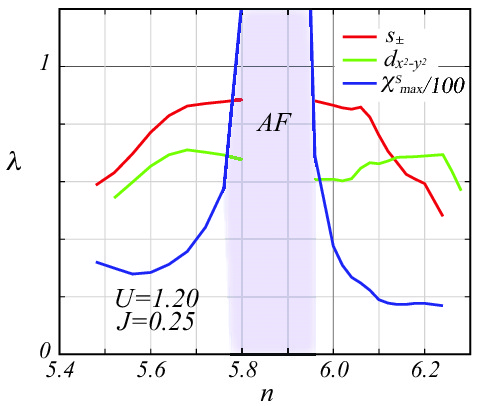}
\caption{Phase diagram in FLEX treatment of SDW and SC instabilities (from \cite{Ikeda_2010A}).}
\label{fig:Ikeda}
\end{indented}
\end{figure}

\label{para:similar_approach} In this discussion we have presented ``spin fluctuation theory" in
terms of an RPA approximation to the pair scattering vertices $\Gamma$, which also includes subleading
charge/orbital contributions via Equation~(\ref{eq:fullGamma}).  It is important to note that other
approaches have obtained very similar results for the Hamiltonian~(\ref{eq:H}).   The most closely related
technique is the FLEX (fluctuation exchange) approximation \cite{n_bickers_89}, which
is a conserving approximation to the Luttinger-Ward functional and the self-energy.
Several authors have applied this approach to the  pairing problem \cite{Ikeda_2008,Ikeda_2009,FernandesFLEX,Ikeda_2010A,Ikeda_2010B}, employing a five-band model for FeBS based
on Wannier fits to DFT results.   Qualitatively, results are similar for the leading pairing instabilities, including the doping dependence \cite{Ikeda_2010A} (see figure~\ref{fig:Ikeda}), although node formation or strong angular anisotropy
have not been observed in FLEX so far.  In addition, FLEX has certain well-known peculiarities which need to be handled carefully.  In particular, the real part of the self-energy, which shifts the band structure in a momentum-dependent way, includes some of the correlations already included in DFT; thus the use of a kinetic energy $H_0$ fit to DFT plus the self-consistent FLEX self energy tends to overcount these correlations.  This problem is traditionally handled by
subtracting the real part of the static self-energy \cite{Ikeda_2010B}.

A further popular approach is the functional renormalization group \cite{Shankar} (fRG), which has
the advantage that it is capable of studying various instabilities of the system
on an equal footing, which is of particular relevance for studying the competition between
SDW and superconductivity in these systems in realistic models. Numerical RG equations
are derived by dividing the Brillouin zone  into
$N$ patches, and then summing at each RG step
over the five one-loop Feynman diagrams
to compute the renormalized 4-point vertex function.
This technique was pioneered in connection with FeBS by Fang \textit{et al} \cite{DHLee_2008}, generalizing numerical work on 1-band systems \cite{Honerkamp}, and has been extended and applied to various FeBS \cite{Thomale_nodal-gapped, Thomale_nodal-gapped2, FaWangLAFP, DHLee_2009, Thomale_K122}.  These works are the intellectual offspring of earlier analytical (logarithmic) RG calculations in two-band models \cite{Chubukovetal, Tesanovic_2008} which were later generalized to include low-order angular harmonics \cite{Chubukov_nodal-gapped} to describe gap anisotropy.  Recently this approach has been extended and compared directly to the RPA results, such that the doping evolution of the fundamental band interactions could be obtained \cite{Maiti}.

\begin{figure}[ht]
\begin{indented}
\item[]
\includegraphics[width=0.4\columnwidth,angle=0]{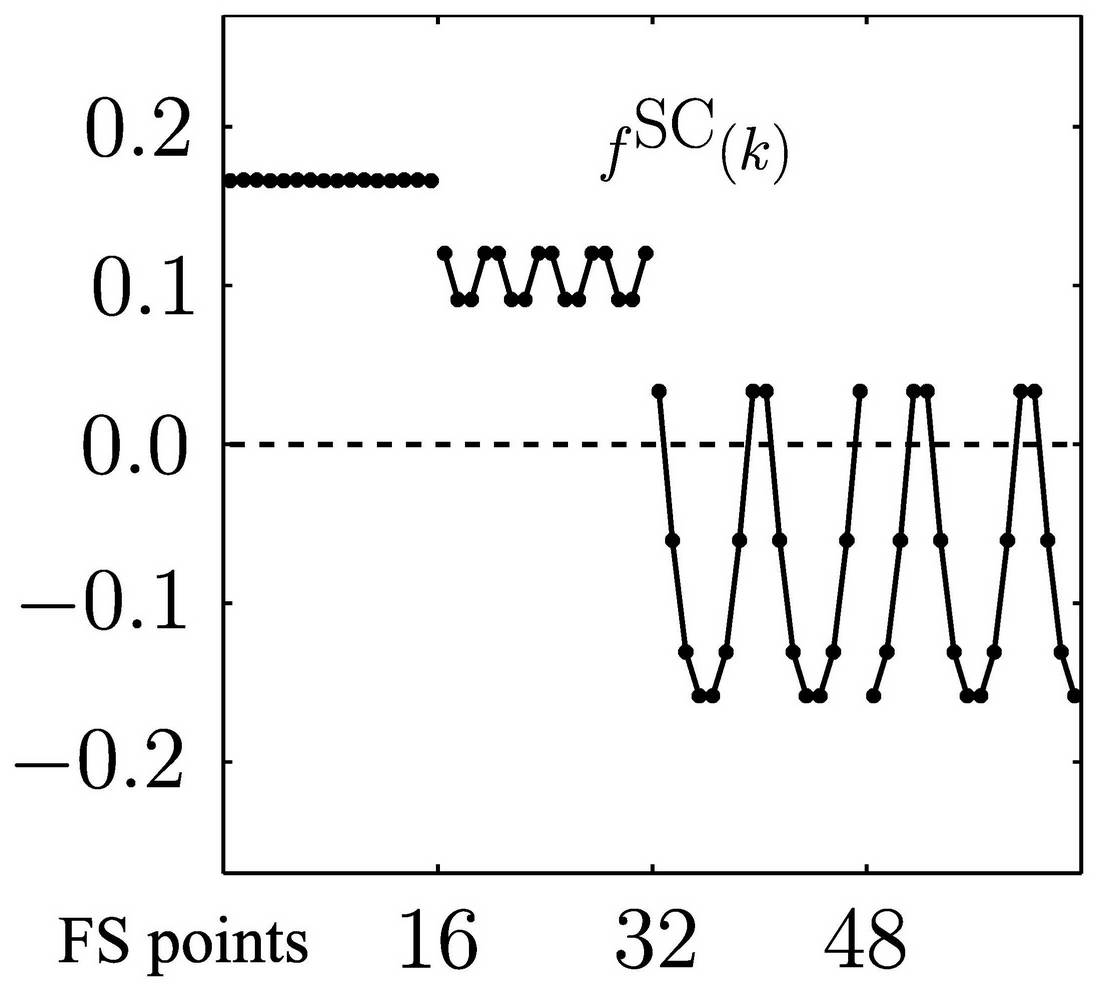}
\includegraphics[width=0.45\columnwidth,angle=0]{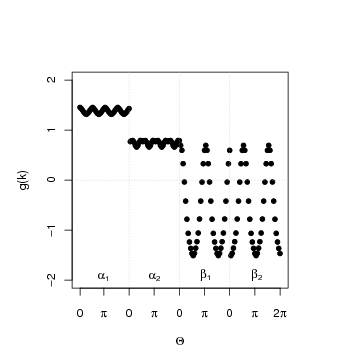}
\caption{Comparison of fRG and RPA results for the gap function $g(\theta)$ vs. $\theta$, with $\theta$ parametrizing each Fermi pocket, for 4-Fermi pocket model of \LFPO / \LOFA.  Left: FRG results from \cite{Thomale_nodal-gapped2} with $\bar U=3.5$, $\bar U'=2.0$, $\bar J=\bar J'=0.7$.  Right: RPA results from \cite{Maier_anisotropy} with  $\bar U=1.67$, $\bar U'=1.46$, $\bar J=\bar J'=0.10$.  Bands $\alpha_1$, $\alpha_2$,
$\beta_1$ and $\beta_2$ correspond to inner and outer hole pockets, and two electron pockets, respectively. Energy units for parameters are in eV, units on vertical axes of both figures are arbitrary.}
\label{fig:fRG-RPA_compare}
\end{indented}
\end{figure}

It is remarkable that rather dissimilar techniques, none of which are controlled in the usual perturbative sense, give such similar results.  To illustrate this, in figure~\ref{fig:fRG-RPA_compare}  we plot results for a FeBS Fermi surface for \LFPO ~or \LOFA~including only the two inner hole pockets, obtained by fRG and RPA techniques.  Although the scale of the interactions are quite different, the ratio $J/U$ is $\ll 1$ in both cases.
Note that the order parameter on both hole pockets is quite isotropic, but that on the electron sheets nodes appear, but consistent with the overall average $s_\pm$ character of the state.

Finally, we comment on so-called ``strong-coupling'' approaches to pairing, based on the $J_1-J_2$ multiple competing exchange model of the magnetism in these systems \cite{SiAbrahams, CFang, Sachdev_orborder}. Note that this term frequently appears in reference to the large size of the Hubbard interaction relative to the bandwidth,  which in the 1-band case at half-filling allows for a description in terms of the spin degrees of freedom as the charge degrees of freedom become localized.   The spin dynamics can become more localized in situations other than the canonical Mott-Hubbard case, however,  and are in fact  well described by DFT, which for the FeBS gives a typical energy scale for the magnetic interactions of order 100 meV or larger.  Thus the moments are large and quite localized.  This is not a contradiction; even in genuinely itinerant systems (elemental 3D metals are a good example) magnetic interactions are essentially local, decaying with distance as a power law.

This magnetic model leads to a competition between the Neel and stripe-collinear
orders, also present in the (itinerant) DFT calculation, corresponding
to the same ground state magnetic pattern and to a similar structure of spin fluctuations in the reciprocal state [a maximum near $(\pi,0)$]. In \cite{Seoetal}, a $t-J_1-J_2$ model with two bands was studied, and the exchange terms were decoupled in mean field in the pairing channel. In this procedure, the nearest neighbor exchange $J_1$ induces competing $\cos k_x + \cos k_y$ and $\cos k_x-\cos k_y$ ($s$- and $d_{x^2-y^2}$-wave) pairing harmonics, while the next nearest neighbor exchange leads to $\cos k_x \cos k_y$ and $\sin k_x \sin k_y$ ($s$- and $d_{xy}$-wave). In the region of the general phase diagram with $J_2 \gtrsim J_1$, $\cos k_x \cos k_y$ was the leading instability for the 2-band Fermi surface, leading to a nodeless $s_\pm$ state, and a similar ground state was also found for a 5-orbital model \cite{Goswami}.

Mean field theories of the strong coupling type (see also models where $U$ and $J_{1,2}$-type terms are treated as independent before the mean field step \cite{Daghofer1, Daghofer}) show an intriguing set of circumstantial agreement with the predictions of itinerant weak-coupling models, despite the lack of logical continuity between the two types of models. Yet it is not that surprising. Indeed, the pairing symmetry in any spin-fluctuation model is mainly defined by matching the structure of the spin fluctuations and the FS geometry. Since both are rather similar in the two approaches, not surprisingly, the main results agree. In fact,  Wang \textit{et al} have shown that in the case with 5 Fermi surface pockets, the  low energy spin and charge excitations in the fRG treatment of the 5-orbital model~(\ref{eq:H}) overlap very well with those of the $t-J_1-J_2$ model. At the  Hamiltonian level, however, there is no way to derive these particular strong coupling models (with unrenormalized kinetic term) from the general model with on-site interactions.

An advantage of such strong coupling approaches is that they capture
the local character and large amplitude of Fe magnetic moments, in agreement with the DFT calculations. They also  have some attractive conceptual
simplicity.  On the other hand, they have several uncontrollable shortcomings, which make proper application of this approach tricky.

First and foremost,  such theories artificially separate the itinerant electrons and the local moments, as if the latter were coming from a separate atomic species. Yet, the moments are formed by exactly the same electrons that form the band structure, which also mediate the magnetic interaction mapped onto the $J_1-J_2$ Heisenberg Hamiltonian. In some papers an attempt to account for this fact is made by adding
spin susceptibility of itinerant electrons to the aforementioned Heisenberg Hamiltonian, which essentially amounts to a double counting.

Second, actual band structure calculations \cite{Mazin2008, Yaresko, Antropov} cannot be mapped onto a Heisenberg Hamiltonian of any range. They can be mapped onto a Heisenberg Hamiltonian with biquadratic exchange, or possibly to a more
complicated Hamiltonian (such as ring exchange), but not onto a pure
Heisenberg model.

Third, ``strong-coupling" models essentially fix the shape of the spin-fluctuation
induced interaction; therefore, the resulting solution  fixes the structure of the gap nodes in momentum space, so that the amplitude,
anisotropy and possible nodes on actual FSs depends only on the  proximity of these FSs to the imaginary nodal lines. This result is not corroborated by the weak coupling calculations and is likely unphysical.

The partial agreement between fRG and the $t-J_1-J_2$ results will therefore remain a curiosity until a more concrete understanding why the similarity of the low energy sectors of the two theories is observed.

\paragraph{Pairing in multiorbital systems from DFT perspective.}

RPA calculations, as well as other approaches discussed above, use the same
model Hamiltonian~(\ref{eq:H}), but resort to different approximate methods to solve the problem of superconductivity emerging from this Hamiltonian. Comparing
results of such different approaches one can get an idea of how accurate these
solutions are. Yet this Hamiltonian itself is a rather uncontrollable
approximation, and one can legitimately ask the question, to what extent the
simplifications introduced when constructing this Hamiltonian are justified.
Indeed, some of the qualitative results discussed above, such as anisotropy
(and possible nodes) of the order parameter, are intimately related to the
details of the model that may or may not be sufficiently universal.
Specifically, there are two aspects of the model that appear to be
qualitatively significant. First, the leading spin-dependent term in this
Hamiltonian, $\bar{U}\sum_{i,\ell}n_{i\ell\uparrow}n_{i\ell\downarrow},$ is
orbital diagonal (there are two other spin-dependent terms, the Hund's term,
proportional to $\bar{J},$ and pair hopping term proportional to $\bar
{J}^{\prime},$ but these are smaller). That makes the pairing interaction
rather sensitive to the orbital composition, with the parts of the Fermi
surface whose orbital content is not matched by that of the rest of the Fermi
surface effectively decoupled from the rest. This is true in RPA, FLEX, fRG or
any other method based on the same Hamiltonian. On the other hand, there is
growing  belief
\cite{Hund} that physics of these
materials is controlled by the Hund rule's coupling rather than by the direct
Coulomb repulsion, as implied for instance in the original expressions~(\ref{eq:Gamma1}),~(\ref{eq:fullGamma}). The second aspect of the model is that it does not include the retardation effects, at least on the RPA level. The large
Coulomb repulsion that in real life is logarithmically reduced in the
calculations needs to be avoided, which can be done, for an on-site local
repulsion, by tuning the order parameter in such a way that it integrates to
zero. These two problems, characteristic for this model, suggest that the
tendency towards gap anisotropy is probably somewhat overestimated in this approach.

It is instructive to compare the effect of direct Coulomb repulsion in
conventional superconductors, such as V, to another $3d$ transition metal with the interaction parameters in the charge channel not much different from Fe based superconductors. There the Coulomb repulsion, obviously strong, is
renormalized as
\begin{eqnarray}
\mu &=& U N_{0} \\
\mu^* &=&\mu /[1+\mu \ln \frac{E_C}{E_b}],
\end{eqnarray}
where $N_{0}$ is the density of states at the Fermi level, $E_C$ is a
characteristic electronic energy (it may be the total band width, or the
plasmon energy, or a combination thereof), and $E_{b}$ is the energy of the
pairing bosons. Given that $\mu \ln \frac{E_C}{E_b} \gg 1,$ $\mu^* \approx 1/\ln \frac{E_C}{E_b} \sim 0.1-0.15$ for typical conventional
superconductors. For most Fe superconductors $T_{c}$ is only a factor of two
or three larger than for the best transition metal superconductors (niobum's
$T_{c}$ is 10 K, that of some binary alloys reaches 23 K), and the full
d-band widths are comparable, so if the Coulomb repulsion were uniform one
could apply similar reasoning and conclude that for these systems $\mu^* \sim
0.15-0.2$. If, on the other hand, the unrenormalized Coulomb repulsion is
different in the interband and intraband channels, the renormalization
equation should be written differently, namely
\begin{equation}
\mu_{ij}^*=\mu _{ij}-\sum_{n}\mu _{in}\ln \frac{E_{C}}{E_{b}} \mu_{nj}^*  \label{mu}
\end{equation}%
where $i$, $j$, and $n$ are band indices. The solution in the limit $|\mu
_{i\neq j}-\mu _{ii}|\ln \frac{E_{C}}{E_{b}}\gg 1$ (which may or may not
fold for FeBS) reads:
$\mu_{ii} \approx 1/\ln \frac{E_{C}}{E_{b}}$, $\mu_{i\neq j} \approx const/\ln
^{2}\frac{E_{C}}{E_{b}}$.
Thus the renormalization in a multiband system has two effects: (1) it
strongly reduces the effect of Coulomb repulsion in general and (2)
it suppresses interband repulsion compared with intraband. The
latter effect has important implication -- effect of `Coulomb
avoidance'. Indeed, if the Coulomb repulsion does not depend on the
wave vector, the condition for it to cancel out of the equations on
$T_c$ is that the order parameter, when averaged over the entire FS
and all its sheets, integrates to zero. If the Coulomb repulsion is
only present in the intraband channel, than to avoid its effect on
$T_c$ entirely, the order parameter must integrate to zero in each
band separately.


It should be kept in mind that the preceding argument implicitly relies
upon an analogy with the Eliashberg theory, assuming that all relevant
interactions can be separated into groups: a pairing interaction (boson
exchange) that is restricted to sufficiently low energies (recall that the
spin resonance in FeBS occurs at the energies around 30-40 meV), and is
also subject to the Migdal theorem, and a direct Coulomb repulsion that
exists at all energies, and gets renormalized after the ladder diagrams are
summed. Most existing approaches explicitly or implicitly make this very
assumption, with the notable exception of the renormalization group
techniques. However, in reality for electronic (as opposed to a phonon)
superconductivity all interactions emerge from electrons themselves, and
such separation is not always possible. Nor is it always possible to limit
the approximation to  the ladder diagrams. In more complex cases, formally involving
parquet diagrams, the solution may depend on whether the Coulomb interaction
is first screened and then renormalized or first renormalized and then
screened, etc \cite{Gunnarsson}. On the other hand, in the renormalization
group approach interactions are not separated into these channels, and in
principle all retardation effects are supposed to be included. While it is
very difficult to make a one-to-one correspondence between such approaches
and RPA (and similar) calculations, is quite obvious that they are based on
 substantially different physical assumptions. Why, nevertheless, results
obtained in RPA and in functional renormalization group approaches are
quantitatively similar, is unclear and needs to be understood.

Given these uncertainties involved in RPA calculations described in the previous sections, as well as other similar approaches, it is interesting to look at the problem at hand not from the Hubbard model point of view, but from the opposite, density functional one. Indeed many believe that, as opposed to the high $T_{c}$ cuprates, DFT is a reasonable starting point for these materials, and in some sense it is more consistent to use DFT for the full susceptibility as long as we use the DFT band structure for the noninteracting one. For instance, in classical semiconductors, such as Si, DFT (even the exact DFT) underestimates the fundamental gap, and thus yields an incorrect noninteracting susceptibility, yet the full susceptibility calculated entirely within DFT is by definition exact \cite{MazinCohen}. Note that within DFT there are
only two electronic interactions: the charge interaction is defined as
$\delta^{2}E_{tot}/\delta n(\rr) \delta n(\rr')$, which includes the Coulomb (Hartree) and the exchange-correlation
interactions, and the spin interaction is defined as  $I_{xc}(\rr,\rr')=\delta^{2}E_{tot}/\delta n_{\uparrow}(\rr) \delta n_{\downarrow}(\rr')$. Therefore, RPA is exact in DFT. Obviously,
LDA and GGA approximations to DFT are not exact, yet one may think that the
full spin susceptibility calculated, say, within LDA-DFT is not a bad
approximation. The only caveat is that in a quantum critical material LDA
deviates from the exact DFT in one systematic way: the reduction of the Hund
rule coupling due to long range spin fluctuations is not included. Emprically,
it can be accounted for by scaling $I_{xc}$ down by 15-20\%. After that, one
can use Equations~(\ref{eq:Gamma1}),~(\ref{eq:Gamma2}) using $I_{xc}(\rr,\rr')=I_{xc}(\rr) \delta(\rr-\rr')$ instead of $U.$

One can work in the orbital basis again, just as one does in the Hubbard
model, except that the spin-dependent term now reads $\sum_{i,\ell,\ell
^{\prime}}I_{\ell\ell^{\prime}}n_{i\ell\uparrow}n_{i\ell^{\prime}\downarrow}.$
It is easy to show that $I_{\ell\neq\ell^{\prime}}=I_{\ell\ell}/3,$ so some
orbital dependence remains. Whether it will be enough to provide for nodal
lines, is unclear. The charge channel should also be revisited in this case. The Coulomb repulsion term should be added in form of a matrix in band indices, $\mu_{ij}^{\ast}$, as defined in Equation~(\ref{mu}). The diagonal elements of this matrix should be taken as 0.15-0.20, and nondiagonal between 0 and the
diagonal ones, depending on how different are the matrix elements of $U$ in
the band representation.

Although the course of action outlined above is straightforward, so far it has
not been tried yet. In principle, such calculations could be very useful for
understanding the stability of the qualitative results, such as nodal
structure, with respect to principal approximation, because the Hubbard model
and the DFT in many aspects represent two opposite approximations.

\subsection{Alternative approaches.}


Historically, many other pairing agents have been proposed proposed as mediators for unconventional (non-phonon) superconductivity \cite{Ginzburg_Kirzhnits}. Arguably, historically the first suggestions were those of Little \cite{Little} and Ginzburg \cite{Ginzburg}, who proposed, respectively, quasi-1D metal chains and quasi 2D metal planes embedded in highly polarizable media. This was dubbed ``excitonic
superconductivity'', after the simple physical picture of Cooper pairs living in the metallic subsystem, and the intermediate bosons being excitons localized in the surrounding nonmetalic media.  In the context of FeBS,  this proposition
was recently brought back into limelight by Sawatzky \textit{et al}, who pointed
out that As and Se are large ions and thus have large polarizability \cite{Sawatzky_09, Sawatzky_pairing}. This model is still being discussed; the arguments usually brought up against it include the fact that Fe-As hybridization is not small, as the model requires,  that ion-ion interaction should also be subject to screening by polarizable As ions, but no anomalous phonon softening is
observed, and, finally, that superconductivity seems to be always
adjacent in the phase diagram to antiferromagnetism.

Other proposals rely on long wavelength electron charge fluctuations known as
plasmons, particularly on  acoustic plasmons. Theories of this type were attempted for cuprates, for MgB$_{2}$, and, most recently, for intercalated graphites. So far this model, however, has not had any confirmed success. One shortcoming that plagues the papers advocating
this mechanism is that they hardly ever address the lattice stability, yet any sort of
overscreening (and essentially any attractive interactions in the charge
channel can be considered as a sophisticated overscreening) tends to
overscreen phonons as well and render them unstable. This is of course a
problem for other ``excitonic'' mechanisms as well; one can achieve pairing and keep phonons stable only through invoking vertex corrections that appear in electron-electron but not in electron-ion vertices.

\begin{figure}[ht]
\begin{indented}
\item[]
\includegraphics[width=0.6\columnwidth]{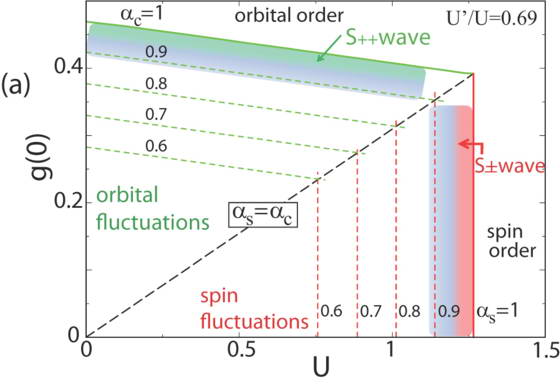}
\includegraphics[width=0.4\columnwidth]{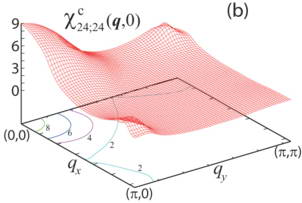}
\includegraphics[width=0.4\columnwidth]{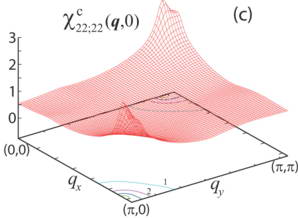}
\caption{ (a) Phase diagram for pairing in Hubbard-Holstein model \cite{Kontani_Hubb_Holst}. $g(0)$ is the bare electron-electron interaction due to electron-phonon coupling. (b) Interorbital susceptibility mixing $xz$ and $xy$ orbitals; (c)  Intraorbital susceptibility for $xz$ orbitals.}
\label{fig:Kontani}
\end{indented}
\end{figure}

Of all alternatives to the spin fluctuation model, the one that has received most
attention in the context of FeBS is the orbital fluctuations model of Kontani and co-workers \cite{Kontani_neutron,Kontani_Hubb_Holst,Kontani_bond_angle}. The possible importance of orbital fluctuations was pointed out by several authors early on due to the possibility of orbital ordering in the Fe $d$ states at the orthorhombic transition \cite{Zaanen_orborder,Sachdev_orborder,WeiKu_orborder,Barzykin_orborder}.
The stripe magnetic order will drive an orbital ordering
to some degree even in itinerant models. {\it Fluctuating}  order of this type, taken in isolation, can in principle lead to an attractive mechanism for pairing, although it is hard to disentangle such orbital
fluctuations from spin fluctuations of the same symmetry.

It is useful to note that orbital fluctuations of this type are present in the standard fluctuation exchange approach~(\ref{eq:fullGamma}), and of course driven by the interorbital Coulomb matrix elements $\bar U'$, $\bar J'$.   It is normally assumed (and verified by ab initio calculations \cite{Imada}) that $\bar U> \bar U'$, but one can ask, taking $\bar U'$ as an independent parameter, what happens to the effective electron pairing vertex?
%
One may show that for sufficiently large $\bar U'$,$\bar J'$ the instability in the charge/orbital channel dominates the spin channel contribution to $\Gamma_{ij}(\k,\k')$, Equation (\ref{eq:Gam_ij}), even for purely electronic interactions. In the former channel the interorbital pair vertex becomes peaked at $(0,0)$ (compare figure~\ref{fig:Kontani}), such that the leading instability occurs in the $A_{1g}$ channel but without sign change, i.e. a $s_{++}$ state.

These strong orbital fluctuations are unphysical in the required limit $\bar U' > \bar U,$ $\bar J' > \bar J$, relying on electronic interactions alone.  However, as pointed out by Kontani and Onari \cite{Kontani_Hubb_Holst}, certain in plane Fe phonons can in principle ``bootstrap'' the interorbital processes such that they dominate the spin fluctuation part of the interaction (a different phonon was considered by Yanagi \cite{Yanagi}). The electron-phonon coupling to these phonons is included in an RPA-type calculation \cite{Kontani_Hubb_Holst}, and it is found indeed that effective interorbital couplings $\tilde U'$, $\tilde J'$, renormalized by the effective electron-electron interaction due to phonons $g(0)$, enhance $s_{++}$ pairing.  The phase diagram of this model for fixed $\tilde U'/\tilde U$ is shown in figure~\ref{fig:Kontani}.  Note that orbitals 2 and 4 in the Kontani-Onari scheme are $xz$ and $xy$.

\begin{figure}[ht]
\begin{indented}
\item[]
(a)~\includegraphics[width=0.24\columnwidth]{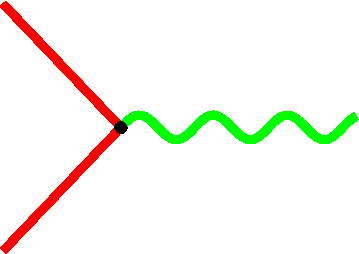} 
(b)~\includegraphics[width=0.22\columnwidth]{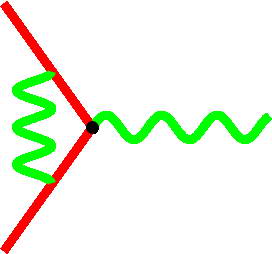}
(c)~\includegraphics[width=0.22\columnwidth]{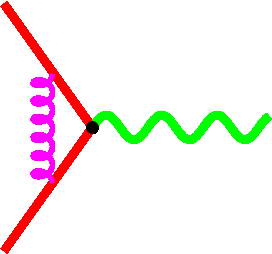}
\caption{{Diagrams for the electron-phonon vertex renormalization. Straight red lines are electron propagators, wavy green line is the phonon propagator, and the curly magenta line is a high-energy electronic excitation (orbital fluctuation).} }
\label{fig:KontaniDiagrams}
\end{indented}
\end{figure}

While the concept of orbital fluctuation pairing is quite clear, there
are several open questions concerning this idea. Probably the most
intriguing one is that how model calculations of Kontani and Onari should
be reconciled with the density functional calculations.
Indeed, electron-phonon coupling in the linear harmonic adiabatic
approximation is included in the standard linear response calculations
of the phonon frequencies and coupling strength. These calculations find
only very moderate coupling strength for representative FeBS \cite{Matteo}, and include diagrams of the type  shown in figure~\ref{fig:KontaniDiagrams}(a),
where the vertex is computed adiabatically, as a derivative of the electronic
Green's function with respect to ionic displacement. An enhancement
found by Kotani and Onari can only come from vertex corrections.
However, vertex corrections of the type shown in figure~\ref{fig:KontaniDiagrams}(b) are excluded by virtue of the
Migdal theorem, and  vertex corrections of the type
shown in figure~\ref{fig:KontaniDiagrams}(c), where the
springy line is a high-energy electronic excitation (like an orbital fluctuation) are
included in the adiabatic density functional calculations (any orbital
repopulation due to a static ionic displacement is accounted for). Thus,
the slow (compared to the electronic scale) orbital fluctuations are
excluded by the Migdal theorem, and fast ones (compared to the phonon
scale) are already included in the DFT calculations. This appears to leave rather
little room for strong renormalization of the electron-phonon coupling.

Similarly, one can ask how such a huge enhancement will affect the
corresponding phonon self-energy. Indeed pretty much the same vertex
corrections enter the equation on the phonon frequency and one would
expect that if the density functional theory fails so miserably in
calculating the electron-phonon coupling for a particular mode, it
should also drastically overestimate the frequency of this mode compared
to the experiment. Yet this is definitely not the case \cite{Reznik}.

Thus, the question to which extent this interesting orbital-fluctuation
model is operative in real compounds remains open.
It is possible that, due to varying interactions strengths, orbital fluctuations dominate in some of the FeBS and spin fluctuation in the others.  The authors of these works believe that orbital fluctuations, and thus $s_{++}$ pairing, are
dominant in most of these systems, pointing to weak impurity scattering \cite{Onari}, the broad neutron scattering resonance observed in most FeBS \cite{Kontani_neutron}, and the natural explanation \cite{Kontani_bond_angle} of the Lee plot (maximum of $T_c$ within family at tetrahedral Fe-As angle \cite{Leeplot}) as evidence in favor of this scenario. We postpone the question of whether or not $s_{++}$ pairing has in fact been observed to Section~\ref{subsec:signchange}.

\subsection{Multiband BCS theory.}
\label{subsec:multiband}

The famous BCS formula is derived in the assumption that the pairing amplitude
(superconducting gap, order parameter) is the same at all points on the Fermi
surface. The variational character of the BCS theory makes one think that
giving the system an additional variational freedom of varying the order
parameter over the Fermi surface should always lead to a higher transition
temperature. For a case of two bands with uniform order parameters in each of
them this problem was solved first in 1959 by Matthias, Suhl, and
Walker \cite{Suhl} and by Moskalenko \cite{Moskal}. It can be easily generalized
onto a general $\k$-dependent order parameter. In the weak coupling
limit it reads
\begin{equation}
\Delta(\k)=\int\Lambda(\k,\k')\Delta(\k')F[\Delta(\k'),T] d\k',
\label{Deltak}
\end{equation}
where summation over $\k$ implies also summation over all bands crossing
the Fermi level. A strong coupling generalization in the spirit of Eliashberg
theory is straightforward. Here the matrix $\Lambda$ characterizes the pairing
interaction, and $F=\int_{0}^{\omega_{B}}dE\tanh(\frac{\sqrt{E^{2}+\Delta^{2}%
}}{2T})/\sqrt{E^{2}+\Delta^{2}}$ . The intermediate boson frequency sets the
cut-off frequency. Assuming that the order parameter $\Delta$ varies little
within each sheet of the Fermi surface, while differing between the different
sheets, Equation~(\ref{Deltak}) is reduced to the original expression of \cite{Suhl,Moskal}:
\begin{equation}
\Delta_{i}=\sum_{j}\Lambda_{ij}\Delta_{j}F(\Delta_{j},T),
\label{Deltai}
\end{equation}
where $i,j$ are the band indices and $\Lambda$ is an \textit{asymmetric}
matrix related to the \textit{symmetric }matrix of the pairing interaction $V$,
$\Lambda_{ij}=V_{ij}N_{j}$, where $N_{i}$ is the contribution of the $i$-th
band to the total DOS. It can be shown that in the BCS weak coupling limit the
critical temperature is given by the standard BCS relation, $kT_{c}%
=\hbar\omega_{D}\exp(-1/\lambda_{eff}),$ where $\lambda_{eff}$ is the largest
eigenvalue of the matrix $\Lambda.$ The ratios of the individual order
parameters are given by the corresponding eigenvector. Note that although the
matrix $\Lambda$ is not symmetric, its eigenvalues (but not eigenvectors!) are
the same as those of the symmetric matrix $\sqrt{N}V\sqrt{N}$.

Furthermore it is evident from Equation~\ref{Deltai} that unless all $V_{ij}$ are the same, the temperature dependence of individual gaps does not follow the canonical BCS behavior. For instance, in a two-band superconductor where the intraband coupling dominates, the smaller gap opens initially at a very small value, and only at a temperature corresponding to its own superconducting transition (not induced by the larger gap) it starts to grow. This effect is gradually suppressed as the interband coupling approaches the geometrical average of the intraband couplings, but as the interband coupling starts to dominate the gaps again show non-BCS temperature dependence. In this limit, however, it is the larger gap that deviates more from the BCS behavior.

It was realized in 1972 \cite{AS} that Equations~(\ref{Deltak}) and~(\ref{Deltai}) may have solutions even when all elements of the interaction matrices $\Lambda$ are
negative, i.e., repulsive. The simplest example is an off-diagonal repulsion:
$V_{11}=V_{22}=0$, $V_{12}=V_{21}=-V<0$. In this case the solution reads:
$\lambda_{eff}=\sqrt{\Lambda_{12}\Lambda_{21}}=|V_{12}|\sqrt{N_{1}N_{2}},$ $\Delta_{1}(T_{c})/\Delta_{2}(T_{c})=-\sqrt{N_{2}/N_{1}}$. At lower
temperatures the gap ratio becomes somewhat closer to 1.

It is important to bear in mind that actual FeBS materials have more than two
bands --- rather four or five. These may all have different gap
magnitudes, and possibly angular dependences, as discussed in Section~\ref{sec:expt_structure}.

\subsection{Disorder in multiband superconductors.}
\label{subsec:disorder}
\begin{figure}[ht]
\begin{indented}
\item[]
\includegraphics[width=0.7\columnwidth]{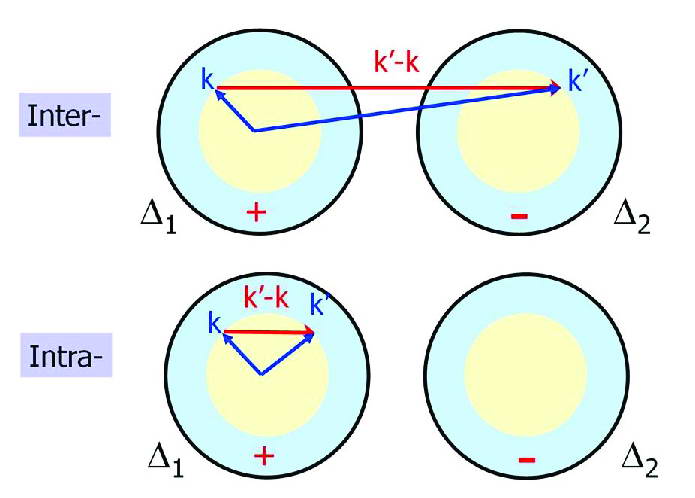}
\caption{ Schematic representation of two Fermi surface pockets
with superconducting gaps indicated with their signs.  Top: interband scattering
by impurities mixes $\Delta_1$ and $\Delta_2$.  Bottom: intraband scattering
mixes  states on each pocket.}
\label{fig:impurities_2band}
\end{indented}
\end{figure}

Experiments on impurity substitution \cite{Kawabata,Sato,SCLee} and
proton irradiation \cite{Tarantini} have given the impression that $T_c$ is suppressed generally more slowly than the maximum rate obtained for
pure interband scattering in the symmetric model, which is identical to the AG
universal $T_c$ suppression curve.  It is worth advising the reader to interpret $T_c$ suppression
experimental results with caution, for several reasons.  First, in some cases not all the nominal concentration
of impurity substitutes in the crystal.  Second, ``slow" and ``fast" $T_c$ suppression cannot be determined by plotting $T_c$ vs. impurity concentration, but only vs. a scattering rate directly comparable to a theoretical scattering rate (see below), which is generally difficult to determine.  The alternative is to plot $T_c$ vs. residual resistivity change $\Delta \rho$, but a) this is only possible if the $\rho(T)$ curve shifts rigidly with disorder, and b) if  comparisons with theory include a proper treatment of the transport rather than the quasiparticle lifetime \cite{GraserTc}.  Finally, the effect of a chemical substitution in a Fe-based superconductor is quite clearly {\it not} describable solely in terms of a
potential scatterer, but the impurity may dope the system or cause other electronic structure changes which influence the pairing interaction.

Given the many uncertainties present in the basic modeling of a single impurity, as well as the multiband nature of the Fe-based materials, it is reasonable to assume that systematic disorder experiments may not play a decisive role in identifying order parameter symmetry as they did, e.g. in the cuprates.  Nevertheless, one can perhaps draw useful qualitative conclusions about the effects of impurities on the various types of superconducting states under discussion by attempting to study models of pairbreaking by impurities which are generalizations of the conventional Abrikosov-Gor'kov (AG) approach \cite{AG}.

\subsubsection{Intra- vs. interband scattering.}

In conventional 2-band superconductors with two different isotropic gaps,
nonmagnetic impurities can either scatter quasiparticles between
bands or within the same band.  Interband processes (see figure~\ref{fig:impurities_2band}) will average
the gaps and can thus lead to some initial $T_c$ suppression, after which $T_c$
will saturate until localization effects become important. Interband scattering
is much more profound in a sign-changing 2-band system \cite{Muzikar, Golubov, kulic},
where nonmagnetic impurities with interband component of the scattering potential
are pairbreaking even if the gaps and densities of states are equal on both bands (symmetric model).
In such a situation, $T_c$ will eventually be suppressed to zero at a finite critical concentration
as in the theory of scattering by magnetic impurities in a 1-band $s$-wave system \cite{AG}. In the context of the FeBS, these general considerations were pointed out early on by
several groups \cite{Mazin_etal_splusminus,Parker2008, Chubukovdisorder, SengaKontani, Bang_disorder}.
A given type of chemical impurity in a given host will be characterized crudely by an effective interband potential $u$ and an intraband potential $v$, and results for various quantities in the superconducting state will depend crucially on the size and relative strengths of these two quantities.

Most calculations are performed in the framework of the $T$-matrix approximation to calculate the average impurity self-energy for pointlike scatterers $\hat{\Sigma}^{imp}(\omega_n)$,
\begin{equation}
 \hat{\Sigma}^{imp} =n_{imp}  \hat{\mathbf{U}} + \hat{\mathbf{U}} \hat{{G}}(\omega_n) \hat{\Sigma}^{imp}(\ii\omega_n), \label{eq:tmatrix}
\end{equation}
where $\hat{\mathbf{U}} = \mathbf{U} \otimes \hat\tau_3$, $n_{imp}$ is the impurity concentration, and the $\tau_i$ are Pauli matrices in particle-hole space. Here ${\mathbf{U}}$ is a matrix in band space, frequently taken for simplicity to be represented by constant intra- and inter-band potentials $v$ and $u$, respectively, such that $(\mathbf{U})_{\alpha \beta} = (v-u) \delta_{\alpha \beta} + u$. This completes the specification of the equations which determine the  Green's functions

\begin{equation}
\hat{G}(\k, \omega)^{-1} =  \hat{G^0}(\k, \omega)^{-1} - \hat\Sigma^{imp}(\k, \omega),
\label{Greensfctn}
\end{equation}
where $\hat{G}^0$ is the Green's function for the pure system. Note that the self-consistent $T$-matrix approximation includes
only diagrams corresponding to multiple scattering from a single impurity, and is well-known to have some pathologies in two dimensions \cite{JLTPreview}.
In the context of impurities in an $s_\pm$ state, it has been claimed to produce inaccurate results in the  statistics of subgap states \cite{Koshelev}.
 Nevertheless, for qualitative purposes--and we will argue below that one cannot go beyond a qualitative analysis
here anyway--it seems quite adequate.

\subsubsection{Effect on $T_c$.}

\begin{figure}[ht]
\begin{indented}
\item[]
\includegraphics[width=0.8\columnwidth]{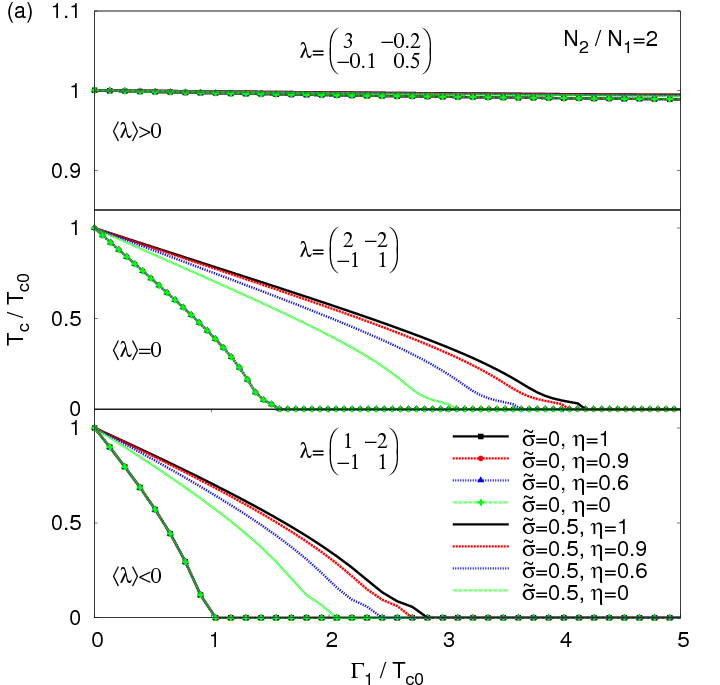}
\includegraphics[width=0.8\columnwidth]{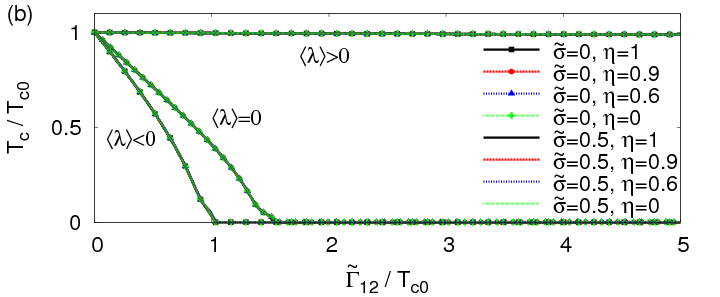}
\caption{ Critical temperature for various $\tilde{\sigma}$ and $\eta$ as a function of (a) the impurity scattering rate $\Gamma_{1}$ and (b) the effective interband scattering rate $\tilde{\Gamma}_{12}$. The parameters are: $N_2/N_1 = 2$, coupling constants are
\\ for $\la \lambda \ra>0 $: $\lambda_{11} = 3$, $\lambda_{12} = -0.2$, $\lambda_{21} = -0.1$, $\lambda_{22} = 0.5$,
\\ for $\la \lambda \ra=0 $:  $\lambda_{11} = 2$, $\lambda_{12} = -2$, $\lambda_{21} = -1$, $\lambda_{22} = 1$,
\\ for $\la \lambda \ra<0 $: $\lambda_{11} = 1$, $\lambda_{12} = -2$, $\lambda_{21} = -1$, $\lambda_{22} = 1$. From \cite{Efremovdisorder}.}
\label{fig:impurities_universal}
\end{indented}
\end{figure}

Properties in the presence of disorder can depend sensitively on the coupling constants $\lambda_{ij}$ of the two-band superconductor (see Section~\ref{subsec:multiband}) as well, since these enter the BCS gap equations including impurities.  In general even the two-band  problem can seem to be quite dependent on many parameters which are  difficult to determine as a practical matter. Recently a simplification was pointed out in \cite{Efremovdisorder} whereby the suppression of $T_c$ can be expressed solely in terms
of a universal pairbreaking parameter
\begin{equation}
 \tilde{\Gamma}_{12} = \Gamma_{1(2)} \frac{(1 - \tilde{\sigma})} {\tilde{\sigma} (1 - \tilde{\sigma}) \eta \frac{(N_1 + N_2)^2}{N_1 N_2} + (\tilde{\sigma} \eta - 1)^2},
 \label{eq:f}
\end{equation}
where $\tilde{\sigma} =(\pi ^{2}N_{1}N_{2}u^{2})/(1+\pi ^{2}N_{1}N_{2}u^{2})$
and $\Gamma_{1(2)} = 
n_{imp}\pi N_{2(1)}u^{2}(1-\tilde{\sigma})$ are  cross-section and normal state scattering rate parameters,
respectively. The parameter $\eta=v/u$ is the ratio of intra-band to inter-band scattering. In the weak scattering (Born) limit, $\tilde{\sigma} \to 0$, while for $\tilde{\sigma} \to 1$ the
unitary limit (strong scattering) is reached. Note the strange result  that $\tilde\Gamma_{12}\rightarrow 0$ in the unitary limit, i.e. nonmagnetic impurities do not affect $T_c$ in an $s_\pm$ state \cite{kulic}.
{While this may be an artifact of the 2-band model used, it is an indication that
a rather robust set of parameters may produce significantly weaker effects of $T_c$ suppression than expected.}

Expressed in terms of~(\ref{eq:f}),  all $T_c$ suppression curves collapse onto one of three ``universal" curves, depending on whether the average pairing strength parameter $\langle\lambda\rangle$ is positive, negative, or zero, as shown in figure \ref {fig:impurities_universal}.  It is clear that in the case where intraband scattering dominates, even in the ``standard" $s_\pm$ scenario with $\langle\lambda\rangle<0$, the $T_c$ suppression will be much slower than the AG result; thus experimental results need not be taken as evidence against the $s_\pm$ state. In the interesting and relatively unexplored $\langle\lambda\rangle>0$ case, a transition below $T_c$ from $s_\pm$ to $s_{++}$ with increasing disorder is possible \cite{Efremovdisorder}.

\subsubsection{Anisotropic states.}
\label{subsubsec:anisotropic}

\begin{figure}[ht]
\begin{indented}
\item[]
\includegraphics[width=0.7\columnwidth]{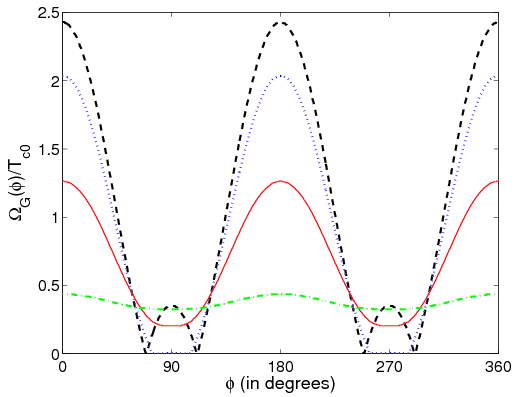}
\caption{(Color online)  Normalized spectral gap
$\Omega_G(\phi)/T_{c0}$ vs. angle $\phi$ on the Fermi surface for
an extended $s$-wave state $\Delta(\phi) = \Delta_0(1+r\cos 2\phi)$, with $\Delta_0/T_{c0}=1$, $r=1.3$ and Born limit scattering rate $\Gamma/T_{c0}=0$
(dashed), 0.3 (dotted), 1.0 (solid), and 3.1 (dashed-dotted). From \cite{Mishra_09}. }
\label{fig:nodelifting}
\end{indented}
\end{figure}

There is substantial evidence from low-energy thermodynamics and transport experiments (Section~\ref{sec:expt_structure}) that low-energy quasiparticle excitations are present in many Fe-based materials, indicating either very small minimum gaps or true nodes of the order parameter on one or more Fermi surface sheets.  The effect of nonmagnetic impurities on $s$-wave states of this type also depends on the character of the scattering, in particular
whether or not inter- or intraband processes dominate.
If intraband scattering processes are considered by themselves, they simply average the angular structure of the order parameter on each Fermi surface sheet, as in the conventional $s$-wave case \cite{Markowitz}.  $T_c$ will fall initially and then saturate.   If gap nodes are present, they will be lifted by the averaging process at a critical value of disorder (figure~\ref{fig:nodelifting}), and  give rise to a crossover at the lowest temperature from power laws in $T$ to exponential behavior \cite{Mishra_09} with increasing disorder.  Thus if intraband scattering dominates, such nodal or near-nodal systems will display, in the clean limit, the low-energy excitations characteristics of nodes, while dirty systems will be gapped with reduced $T_c$.

In the early literature  on disorder in Fe-based system it was frequently assumed, to the contrary,that the order parameter was isotropic $s_\pm$, and that interband scattering dominates.  In this case, under special circumstances (see Section~\ref{subsubsec:singleimp}),  bound states at the Fermi level may be induced and mimic the effect of nodes in
some experiments \cite{Parker2008,Chubukovdisorder}.  In such a situation, the opposite behavior with disorder is to be expected: clean systems will display exponential $T$-dependence and dirty systems the power laws expected from impurity-induced residual densities of states.

\subsubsection{Single impurity problem.}
\label{subsubsec:singleimp}

\begin{figure}[ht]
\begin{indented}
\item[]
\includegraphics[width=0.8\columnwidth]{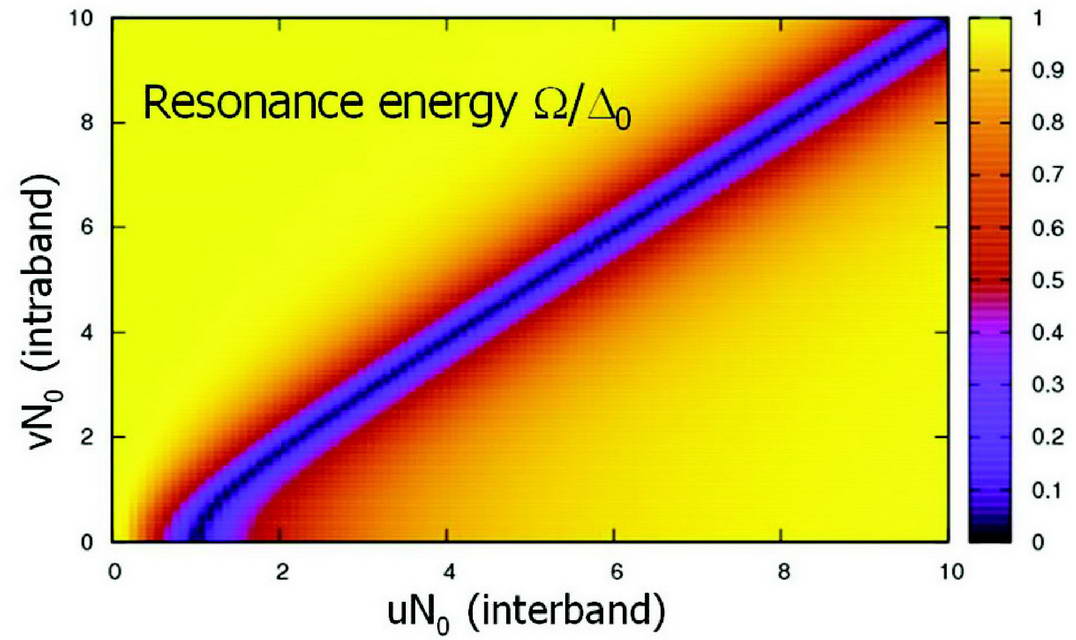}
\caption{Energy of single impurity bound state $\Omega/\Delta_0$ in symmetric $s_\pm$ superconductor with gaps $\pm \Delta_0$ as a function of interband ($u$) and intraband ($v$) scattering (A.F. Kemper, private communication).}
\label{fig:boundstate}
\end{indented}
\end{figure}

In the most general and presumably realistic situation, anisotropic multiband order parameters are present with both intra- and interband scattering. Intraband scattering probably dominates in most situations (see below) and for intermediate to strong impurity potentials, interband scattering effects are largely irrelevant.  To understand why this is the case, we consider the single impurity problem in a symmetric $s_\pm$ state. Equation~(\ref{eq:tmatrix}) for $\hat\Sigma^{imp}$ is essentially identical as the $T$-matrix for a single impurity, whose poles at $\Omega_0$ indicate the existence of impurity bound states \cite{Balatskyreview}.  In general, energies nested near the gap edge correspond to weak pairbreaking, while energies near the Fermi level correspond to strong pairbreaking.  A plot of the single impurity resonance energy position in a symmetric $s_\pm$ state is given in figure~\ref{fig:boundstate}, and shows that in order to influence the states near the Fermi level a very specific fine tuning of interband and intraband scattering is required.  For the symmetric model, this corresponds to $\eta=u/v=1$ in the intermediate to strong potential range, but this criterion will be different in the asymmetric model $N_1\ne N_2$, $\Delta_1\ne \Delta_2$.  It thus seems {\it a priori} unlikely that the impurity band in an isotropic $s_\pm$ state explanation for experiments indicating low-lying quasiparticle states is correct.  For further discussion, see Section~\ref{sec:expt_structure}.

The single impurity problem has been considered in more detail in a 2-band model in \cite{Bang_disorder, Matsumoto2009, Tsai2009, Ng2009, LiWang2009, DZhang2009, Akbari2010, TZhou2009} and in a 5-band model in \cite{Ogata_1imp}.  These calculations show a much richer structure of the local density of states around a single impurity than is found, e.g. in the one band $d$-wave problem \cite{Balatskyreview}, as might be expected, and a complicated dependence on interaction and impurity parameters.  It seems unlikely, given this complexity, that the combination of STM imaging of  impurity states \cite{Balatskyreview} and theories of this type will be able to provide definitive information on order parameter symmetry or structure in these systems.

One approach to reducing the number of parameters and making theories of this type more predictive has been to try to calculate impurity potentials from first principles methods.  For example, Kemper \textit{et al} \cite{Kemper_Co} calculated the nonmagnetic and magnetic impurity potentials for a Co substituting for an Fe in \BFA~within density functional theory.   Nakamura \textit{et al} \cite{IkedaArita_imp} later performed similar calculations for several impurity types.  In principle, such calculations can provide important input into phenomenological treatments of disorder by specifying $u$ and $v$ in band space, but band-resolved results of this type have not yet been given.  Kemper \textit{et al} found in their calculations the nonmagnetic potential was significantly larger than the magnetic one, and that the interband scattering was perhaps a factor of three smaller than intraband.  In orbital space, results from \cite{Kemper_Co} and \cite{IkedaArita_imp} disagree substantively for the Co potential, but it is not clear whether this arises from the fact that the former calculations were performed in the spin-polarized phase, or due to a different treatments of the nonlocal LDA potential \cite{IkedaArita_imp}.

\subsubsection{Magnetic impurities.}

{A qualitative rule of thumb when considering the effect of magnetic and
nonmagnetic impurities on multiband, anisotropic superconductors is
as follows: When a nonmagnetic impurity scatters a pair from one point on
the FS into another point, such that the order parameter does not change
sign, scattering is not pair breaking; if the order parameter flips
its sign, it is pair breaking. For a magnetic impurity, the opposite is
true: scattering with an order parameter sign change is not pairbreaking, otherwise it is. However, quantifying this rule of thumb may be
complicated and results are sometimes counterintuitive.}

In terms of concrete calculations for magnetic impurities in $s_\pm$ states,
Golubov and Mazin \cite{Golubov} showed that for the symmetric model considered above, $T_c$ is suppressed by
magnetic and  nonmagnetic impurities at the same rate in the disorder averaged theory.
The analogous single impurity problem was treated by Akbari \textit{et al} \cite{Akbari_magnetic_imp}
within an Anderson model approach to a rare earth impurity in such a system in the $s_\pm$ state.  In
both situations only quantitative differences in the scattering from magnetic impurities relative to the usual $s$-wave
case were found. In special symmetric situations, interband magnetic scattering in an $s_\pm$ state can give rise to arbitrarily weak pairbreaking, whereas intraband scattering is strongly pairbreaking as expected from AG theory \cite{LiWang2009}.


\subsubsection{Orbital effects.}

In Section~\ref{subsubsec:singleimp} above it was argued that fine tuning of intra-and interband pairing amplitudes was required
in order to create substantial pairbreaking in an $s_\pm$ state, e.g. create bound states near the Fermi level.
In a more realistic approach, however, these parameters ( $u$ and $v$) should not be considered arbitrary,
For example, if one starts from a local atomic picture where an impurity is assumed to modify the orbital occupation energies of the Fe-derived $d$ states, and then transforms to a band basis, it is easy to see that intraband scattering terms of the same order as interband ones are automatically generated by the unitary transformation from orbitals to
bands.  This was the basis of the argument made in \cite{Onari} that interband scattering generically leads to $u\simeq v$ and therefore to large $T_c$ suppression in the symmetric $s_\pm$ model; this was taken as evidence against $s_\pm$ pairing.
However as seen in Section~\ref{subsubsec:singleimp},   a fine tuning  is required to produce significant pairbreaking, so we consider it generically unlikely that for a given set of four or five Fermi surface pockets, with differing density of states and order parameter magnitudes, that the condition for maximal (Abrikosov-Gor'kov like) $T_c$ suppression will be achieved accidentally.  This point of view is also borne out by the argument in \cite{Efremovdisorder} and exhibited in figure~\ref{fig:impurities_universal}.

\subsection{Dimensionality.}
\label{subsec:dimensionality}

In this section we will concentrate on the qualitative effects due to 3D
dispersion of the electronic bands. It should be kept in mind that
dimensionality also plays a role in magnetism, in particular, it may affect the
extent of the magnetic part of the phase diagram and magnetic-orthorhombic
splitting \cite{MazinSchmalian}, and also it can manifest itself through
anisotropy of elastic properties, phonon dispersion and electron-phonon
coupling (which does not seem to be the case in FeBS). Again, we will not
discuss these effects, but only the effects of dimensionality on the band
structure, and, via the latter, on superconductivity.

Such effects can be, roughly speaking, divided into three groups. First, there
is a generic issue of the correspondence between the number of carrier and
their density of states. Let us compare conventional superconductors,
MgB$_{2}$ and B-doped diamond. Both  exhibit hole-doped covalent bonds, and
electron-phonon matrix elements are practically identical. Note that it is
rather difficult to dope such bonds, so the number of carriers is small in
both cases. Yet in the quasi-2D $\sigma$-bands of MgB$_{2}$ this small amount
creates a sizeable DOS (recall that in ideal 2D parabolic bands DOS does not
depend on the carrier concentration at all), and a critical temperature of 40
K, while in {Boron-doped diamond} the DOS remains small, according to the small number of
holes, and so does $T_{c}$.

The second effect is the geometry of the Fermi surface. As we know, in the
spin-fluctuation model the structure of the order parameter is defined by the
interplay between the $q$-dependence of the spin fluctuations and the shape of
the Fermi surface. This is an interesting possibility that has been explored
mostly by that part of the community which traces the spin fluctuations to  nearest- and second
nearest neighbor superexchange (see Section~\ref{para:similar_approach}). In that case the nodal lines are fixed in reciprocal space at $k_{x}=\pm\pi/2$, and at $k_{y}=\pm\pi/2$. If at some particular $k_{z}$ a Fermi surface expands as to cross these lines, actual gap nodes develop.

\begin{figure}[ht]
\begin{indented}
\item[]
\includegraphics[width=0.52\columnwidth]{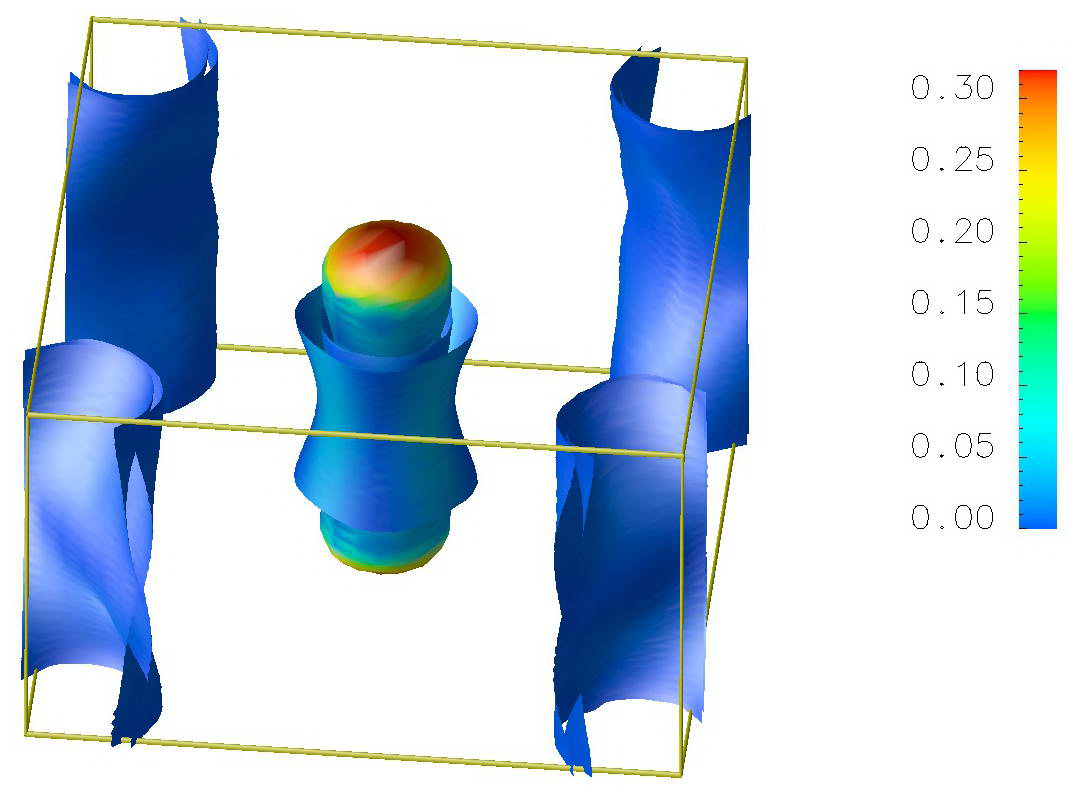}
\caption{$d_{z^2}$ orbital character on the Fermi surface of 5\% electron doped Ba-122 according to DFT. The two outer hole Fermi surfaces are clipped at $\pm \pi/2c$ to show the parts carrying most of the $d_{z^2}$ character.}
\label{fig:orb_character}
\end{indented}
\end{figure}

Finally, the last group of effects is related not so much to possible changes with $k_{z}$ of the Fermi surface \textit{shape} but to the orbital composition of the states forming the Fermi surface. These may come from two sources. First, near the $\Gamma$ point, besides the ubiquitous $xz/yz$ band, occasionally other bands may cross the Fermi level, including the $z^{2}$ band that is very dispersive in the direction. This band hybridizes with the $xz/yz$ band everywhere except the high-symmetry planes, leading to Fermi surface pockets that rapidly change their character as $k_{z}$ changes (see figure~\ref{fig:orb_character}).

Another, more subtle source of 3D effect related to the orbital character of the bands comes from the fact that the electron bands, as discussed in Section~\ref{sec:bandstructure}, are never pure $xz/yz$, but always have an admixture of the $xy$ symmetry, in the outer barrel. This fact was first noted and explained by  Lee and Wen \cite{PALee}, and elaborated in a review paper by Andersen and Boeri \cite{Ole}. The relevant physics also controls the warping and the twisting of the electron FSs.

\begin{figure}[ht]
\begin{indented}
\item[]
\includegraphics[width=0.49\columnwidth]{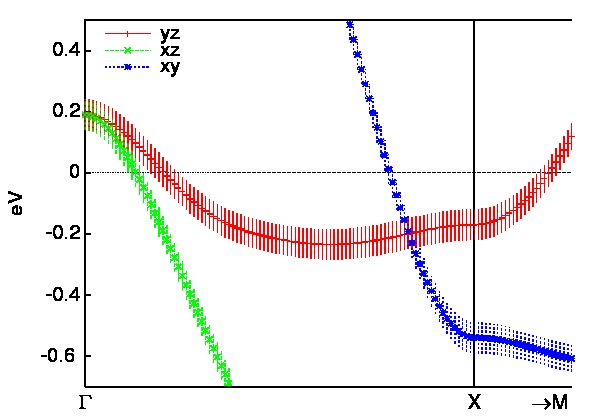}
\caption{Band dispersion in the unfolded BZ corresponding to a 1-Fe unit cell. Major orbital contributions are labeled.}
\label{fig:figEk}
\end{indented}
\end{figure}

To understand this, we will start as usual with the unfolded band structure (figure~\ref{fig:figEk}), corresponding to a single Fe unit cell (figure~\ref{fig:orb_character}(b)). The unfolded Fermi surface geometry of the electron pockets is, essentially, defined by their ellipticity and its variation with $k_{z}$. The ellipticity at a given $k_{z}$ in the unfolded zone is determined by the relative position of the $xy$ and $xz/yz$ levels of Fe, and the relative dispersion of the bands derived from them. Indeed,
the point on the Fermi surface located between $\Gamma$ and $\mathrm{X}$ has a purely $xy$ character, while that between $\mathrm{X}$ and $\mathrm{M}$ a pure $yz$ character. At the $\mathrm{X}$ point the $xy$ state is slightly below the $yz$ state, but has a stronger dispersion, therefore depending on the system parameters and the Fermi level the corresponding point of the Fermi surface may be more removed from
$\mathrm{X}$, or less. In the 1111 compounds, for instance, the dispersion
of the $xy$ band is not high enough to reverse the natural trend, so the Fermi
surface remains elongated in the $\Gamma \mathrm{M}$ (1,0) direction.

Both $xy$ and $xz/yz$ orbitals point away from the Fe-Fe bond  (as opposed to
the $x^{2}-y^{2}$) orbital, therefore their hopping mainly proceeds via As
(Se) $p-$orbitals. The $xy$ states near the $\mathrm{X}$ point mainly hop through the $p_{z}$ orbital (see \cite{Ole} for more detailed discussions), and $xz$ ($yz$) via $p_{y}$ ($p_{x}$) orbitals (not because of the orbitals' shape, but because of their phases at $\mathrm{Y}$[$\mathrm{X}$]). If there is a considerable interlayer
hopping between the $p$ orbitals, whether direct (11 family) or assisted (122
family), the ellipticity becomes $k_{z}$-dependent. For instance, in FeSe
there is noticeable overlap between the Se $p_{z}$ orbitals, so that they form
a dispersive band with the maximum at $k_{z}=0$ and the minimum at $k_{z}=\pi/c$. Obviously, hybridization is stronger when the $p_{z}$ states are
higher, therefore the Fermi surface ellipticity is practically absent in the
$k_{z}$=0 plane, while it is rather strong in the $k_{z}=\pi/c$ plane, which
leads to formation of the characteristic ``bellies''
in the Fermi surface of FeSe. On the other hand,
$p_{x,y}$ orbitals in FeSe hardly overlap in the neighboring layers, so the
$xz$ and $yz$ bands have very little $k_{z}$ dispersion, so that the inner
barrels of the electronic pockets in this compound are practically 2D.

In 122, the interlayer hopping proceeds mainly via the Ba (K) sites, and thus
the $k_{z}$ dispersion is comparable, but opposite in sign (for instance, at
the $\Gamma$ point the hopping amplitudes from Ba $s$ to As $p_{z}$ orbitals
above and below have opposite signs, while those for the As $p_{x,y}$ orbitals
have the same signs) for the $xy$ and $xz/yz$ bands. {Note
this  difference between the 122 and 1111 materials is related simply to the structural difference
 between ``in-phase" and ``out of phase" FeAs layers in the unit cell}. As a result, when going
from the $k_{z}=0$ plane to the $k_{z}=\pi/c$ plane the longer axis of the
Fermi pocket shrinks, and the shorter expands. In BaFe$_{2}$As$_{2}$ the
average ellipticity is very small, while the As-Ba overlap is large, so that
the actual ellipticity changes sign when going from $k_{z}=0$ to $k_{z}=\pi/c$.
On the other hand, in Se based 122 systems the Se-K hopping is quite small,
so ellipticity is small for all $k_{z}$.

Importantly, the symmetry operation that folds down the single-Fe Brillouin
zone when the unit cell is doubled according to the As (Se) site symmetry is
different in the 11 and 1111 structures, as compared to the 122 structure. In
the former case, the operation in question is the translation by $({\pi},{\pi},0)$,without any shift in the $k_{z}$ direction, in the latter by $({\pi}, {\pi}, {\pi})$. Thus the folded Fermi surface in 11 and in
1111 has full fourfold symmetry, while that in the 122 has such symmetry only
for one particular $k_{z}$, namely $k_{z}=\pi/2c$. Furthermore, in 122 the
folded bands are not degenerate along  $\tilde{M}\tilde{X}$
as they were in 11/1111.
Finally, there is a considerable (at least on the scale of the superconducting
gap) hybridization when the folded bands cross (except for $k_{z}=0$). As a
result, although the band structure calculations for 122 materials actually
produce two detached (except for one plane) cylinders for the electronic FSs,
one cannot ``unfold'' these two cylinders as
if one of them was folded down into the other. The actual folded FSs
intersect, yet we do not observe these intersections because of hybridization
induced by the As (Se) potential.

All these effects of 3-dimensionality of the electronic structure manifest themselves in the
3D gap structure, tending to complicate the simple 2D theoretical pictures.
We comment below on some of the most significant ways in which this occurs.

%

%

\section{Gap symmetry.}
\label{sec:expt_symmetry}

The first question to be asked regarding the pairing state in a novel
superconductor is, what is the symmetry of the order parameter? Of course, the
symmetry itself does not fully describe the structure of the gap. For
instance, even a full-symmetry ($s$-wave, or, synonymously, $A_{1g},$
symmetry) order parameter may have ``accidental'', that is, not required by
symmetry, nodes, as long as these nodes transform into each other by all
point group operations. Yet, establishing the right symmetry is arguably
the most important step towards uncovering the full gap structure.

\subsection{Triplet or singlet?}
\label{subsec:tripletorsinglet}
In materials with an inversion center, so-called centrosymmetric, any Cooper
pair can be characterized by its parity, in the sense that the spatial part of
its wave function (the order parameter) can either change sign or remain the
same under the inversion operation. Electrons being fermions, the full wave
function is always antisymmetric. Since inversion corresponds to swapping the
pair components, if they have total spin $S=1$ (triplet), the spatial part of the wave
function should be odd, and if they have  $S=0$ (singlet) it should be even.

A triplet Cooper pair 
with $S_z=\pm 1$ can screen
an external magnetic field, just as individual electrons can. The spin-orbit
interaction can prevent this for some directions, but not for  others. On
the other hand, a singlet pair has no net spin and does not contribute to
magnetic susceptibility  as $T\rightarrow 0$. Thus for singlet superconductivity one can expect the
uniform spin susceptibility to diminish below $T_{c}$. The easiest and the
most accurate way to probe the latter is {via} the Knight shift. This
experiment has been performed on several FeBS including \BFCA~\cite{Ning}, \LOFFA~\cite{Grafe}, PrFeAsO$_{0.89}$F$_{0.11}$~\cite{Matano}, \BFKA~\cite{MatanoBKFA, Yashima}, \LFA~\cite{Jeglic111, Li111}, and BaFe$_2$(As$_{0.67}$P$_{0.33}$)$_2$~\cite{NakaiPdoped},
and it was found that the Knight shift decreases in all crystallographic directions. This effectively excluded triplet symmetries such as $p$-wave or $f$-wave.

\subsection{Chiral or not?}

Another way to classify the superconducting state is according to whether it
breaks  time-reversal symmetry  or not, and according to whether it is chiral (finite expectation value of the magnetic moment operator in the ground state) or not. A singlet chiral superconducting state is allowed in tetragonal symmetry \cite{SigristUeda}, and is transformed under the symmetry operations as $xz + \ii yz$ (of course the complex conjugate state, $xz - \ii yz$, is also allowed). There are several ways to detect $T$-breaking and chirality, but the simplest is probably the $\mu SR$ spectroscopy. This technique is sensitive to small local magnetic fields, as long as they are static. In a chiral superconductor, as long as the balance between the two conjugate states is broken, which should be happening near crystallographic defects, spontaneous orbital current appear, and should be visible by the $\mu$SR technique. Such an effect was observed, for instance, in Sr$_{2}$RuO$_{4}$ \cite{Mackenzie},  but not in FeBS \cite{Amato2009}.
 In principle,
a chiral state can also be generated by mixing two 1D representations, e.g. a so-called
$d+id'$ (note a $s+id$ state, also possible in principle, is $T$-breaking but is not chiral). However, this requires separate transition temperatures corresponding to
the two representations; since this has not been observed, we do not discuss this possibility
further.


\subsection{$d$ or $s$?}

Having excluded triplet ($p$ and $f$) symmetries, as well as the chiral state, we are left with the following singlet pairing possibilities (in a 3D system with tetragonal symmetry):
$s$-wave [$A_{1g}$]; $d(xy)$  [$B_{2g}]$; $d(x^{2}-y^{2})$ [$B_{1g}$]; $g$-wave $(xy(x^2-y^2))$ [$A_{2g}]$;
and
$d(xz\pm yz)$ [$E_{g}$]. It is important to note that, while $d$-wave does not necessarily imply
the existence of gap nodes, in combination with a quasi-2D Fermi surface centered
around the $\Gamma Z$ line such nodes are unavoidable, either vertical for the
$A$ and $B$ symmetries, or horizontal, for the $E$ symmetry. As will be
discussed later in this review, the surface probes, such as ARPES and tunneling show full gaps with no nodes, and, at least for
some compounds bulk probes show exponential low-temperature behavior.

There are also direct experiments that provide evidence against $d$-wave. The
$d$-wave representations, $B_{1g}$ and $B_{2g}$, should not exhibit any Josephson
current {when weakly coupled to a known $s$-wave superconductor}, by symmetry, if the current flow precisely along the $z$ axis.
However, such current was observed in the 122 single crystals \cite{Greene_tunneling}. This
observation is difficult to explain away by deviation from the correct
geometry, because the observed current was strong and showed a well-defined
Fraunhofer diffraction.

Another piece of evidence comes from the so-called anomalous Meissner, or
Wohlleben effect. This effect was predicted in the beginning of the cuprates
era \cite{Geshkenbein_Meissner} and since then has been routinely observed in $d$-wave superconductors. In a nutshell, this effect appears in polycrystalline samples with random orientation of grains. For any $d$-wave superconductor one expects roughly 50\% of weak links to have a zero phase shift, and 50\% a $\pi$ phase shift. One can show that in this case the response to a weak external magnetic field is paramagnetic, i.e., opposite to the standard diamagnetic superconducting
response. This effect has been searched for in FeBS \cite{KAM}, but not found.

These separate pieces of evidence strongly suggest that the pairing symmetry is
$s$, and not $d$. However, we want to stress that direct testing similar to
that performed in cuprates, namely a single-crystal experiment with a 90$^\circ$
Josephson junction forming a closed loop, is still missing, and it is highly
desirable for experimentalists to perform this ultimate test. {In addition, it should
be borne in mind that no law of nature forbids different FeBS materials from having
different order parameter symmetries, although our previous experience with other
novel superconductors tends to argue against this.  Indeed, there are several proposals
that, while most  FeBS are $s$-wave, those  with unusual Fermi surfaces with only
one type of pocket can be $d$ wave, see Section~\ref{subsubsec:KFS}.}

\section{Gap structure.}
\label{sec:expt_structure}

\subsection{Does the gap in FeBS change sign?}
\label{subsec:signchange}

Even though there is convincing evidence that the point symmetry of the order
parameter is for most, if not for all
compounds, $s$-wave, it does not tell us much about the actual structure of
the order parameter and the excitation gap. As opposed to the $d$-wave case, where
 nodes are  mandated on the hole pockets by symmetry, in an extended $s$-wave scheme they
may appear on either type of pocket if higher harmonics in the angular expansion of the order parameter
are sufficiently large. As discussed in Section~\ref{subsec:sf},
there are microscopic reasons  why this may be the case.
Moreover, since nodeless $s_{\pm}$, nodeless $s_{++}$, and an extended $s$ with accidental nodes all belong to the same symmetry class, the difference between them is only quantitative (but important).
In this regard, several experiments appear relevant.

\subsubsection{Spin-resonance peak.}
\label{subsec:resonance}
One obvious effect that was mentioned even in the very first paper proposing
the $s_{\pm}$ scenario \cite{Mazin_etal_splusminus} and later elaborated in detail \cite{Korshunov2008,Maier,Maier2}, is the neutron spin resonance. Neutron scattering is a powerful tool to measure the dynamical spin susceptibility $\chi_s(\q,\omega)$.
%
For the local interactions (Hubbard and Hund's exchange, see Equation~(\ref{eq:H})), $\chi_s$ can be obtained in the RPA from the bare electron-hole bubble $\chi_0(\q,\omega)$ by summing up a series of  ladder diagrams to give
%
\begin{eqnarray}
\chi_s(\q,\omega) = \left[I - U_s \chi_0(\q,\omega)\right]^{-1} \chi_0(\q,\omega),
\label{eq:chi_s_sol}
\end{eqnarray}
where $I$ is a unit matrix in orbital space and all other quantities are matrices  as well.

The fact that $\chi_0(\q,\omega)$ describes particle-hole excitations
has  interesting consequences in the case of an unconventional superconducting state.
Excitations are gapped below approximately $2\Delta_0$; (at $T=0$) only above this
threshold does  $\mathrm{Im}\chi_0(\q,\omega)$ become non-zero. The term arising from the anomalous Green functions is proportional to
\begin{equation}
\sum_\k \left[1 - \frac{\Delta_\k \Delta_{\k+\q}}{E_{\k} E_{\k+\q}}\right]...
\label{eq:cohfac}
\end{equation}
where $...$ represents the kernel of the BCS susceptibility (see e.g. \cite{schrieffer}).
At the Fermi level, $E_{\k} \equiv \sqrt{\varepsilon_\k^2 + \Delta_\k^2} = |\Delta_\k|$. If $\Delta_\k$ and $\Delta_{\k+\q}$ have the same sign, the coherence factor in square brackets in~(\ref{eq:cohfac}) vanishes,    leading to
a smooth increase of the magnetic response with frequency above the $T=0$ threshold of $\Omega_c = \min \left(|\Delta_\k| + |\Delta_{\k+\q}| \right)$. In case of unconventional superconductors \cite{MonthouxScalapino}, when for a given $\q$, $\sgn \Delta_\k \neq \sgn \Delta_{\k+\q}$, the coherence factor is non-zero and the imaginary part of $\chi_0$ possesses a discontinuous jump at $\Omega_c$. Due to the Kramers-Kronig relations, the real part exhibits a logarithmic singularity. For a range of interaction values entering the matrix $U_s$, $\mathrm{Im}\chi_0 = 0$ and non-zero $\mathrm{Re}\chi_0$ result in the divergence of $\mathrm{Im}\chi_s(\q,\ii\omega_m)$ according to Equation~(\ref{eq:chi_s_sol}).  Such an enhancement of the spin susceptibility is called a ``spin resonance''. The corresponding peak appears at a frequency below $\Omega_c$ with the exact position $\Omega_{res}$ pushed below  $\Omega_c$ by an amount which scales with the strength  of $U_s$.

\begin{figure}[ht]
\begin{indented}
\item[]
\includegraphics[width=0.85\columnwidth]{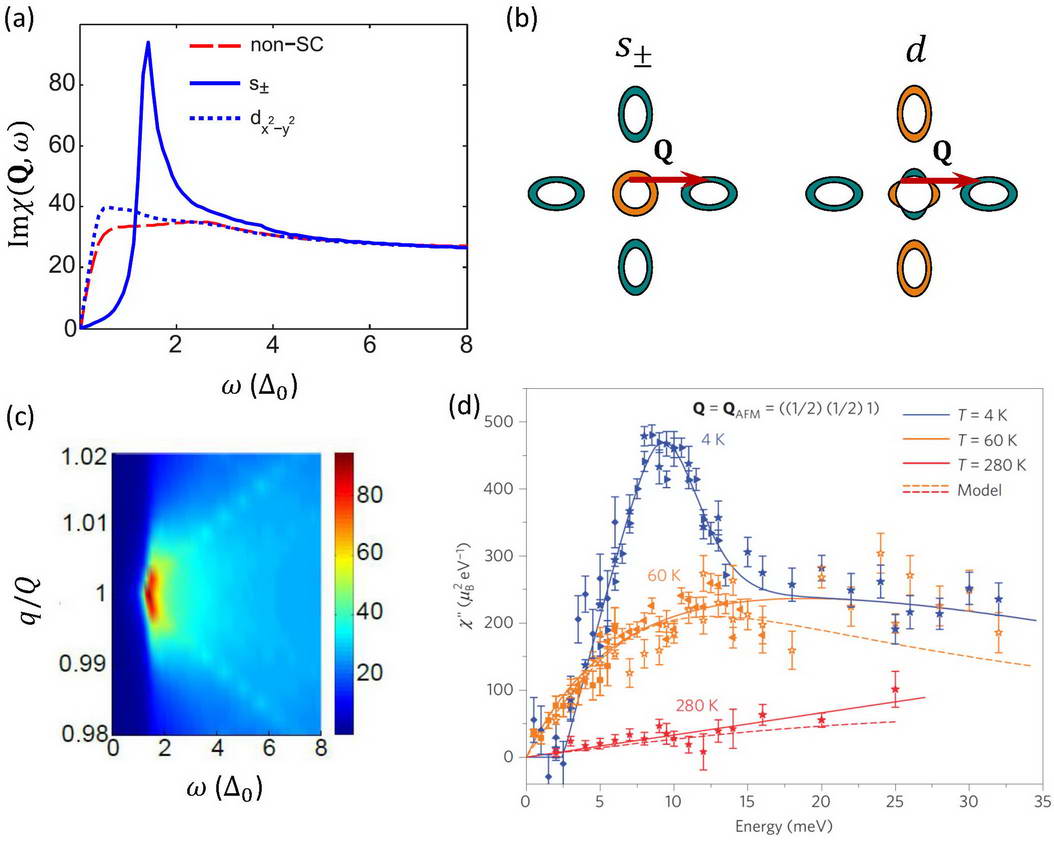}
\caption{(a) Calculated $\mathrm{Im}\chi(\q=\Q,\omega)$ in the normal state and for the $d_{x^2-y^2}$ and $s_\pm$ pairing symmetries \cite{Korshunov2008}. In the latter case, the resonance is clearly seen around $\omega = 2\Delta_0$. (b) Cartoon of the order parameter symmetries and a wave vector $\Q$ connecting different Fermi sheets. (c) $\mathrm{Im}\chi(\q,\omega)$ for the $s_\pm$ state as a function of frequency $\omega$ and momentum along the $(1,1)$ direction \cite{Korshunov2008}. (d) Experimental neutron data showing appearance of the spin resonance in BaFe$_{1.85}$Co$_{0.15}$As$_2$ below $T_c = 25$K \cite{Inosov}.}
\label{fig:resonance}
\end{indented}
\end{figure}

Scattering between nearly nested hole and electron Fermi surfaces in
FeBS produce a peak in the normal state magnetic susceptibility at or
near $\q = \Q=(\pi,0)$. For the uniform $s$-wave gap, $\sgn \Delta_\k =
\sgn \Delta_{\k+\Q}$ and there is no resonance peak.  For the $s_\pm$
order parameter, 
$\Q$ connects Fermi sheets with  mostly different signs of  the gaps,
see figure~\ref{fig:resonance}(b). This fulfils the resonance condition
for the interband susceptibility, and a well defined spin resonance
peak is formed (compare normal and $s_\pm$ superconductor's response in
figure~\ref{fig:resonance}(a)). Moreover, the intraband bare
susceptibilities are small at this wave vector due to the direct gap,
i.e. no states at the Fermi level can be connected by intraband
scattering with wave vector $\Q$. Therefore, a single pole will occur
for all components of the RPA spin susceptibility at $\Omega_{res} \leq
\Omega_c$ and a spin exciton forms \cite{Korshunov2008,Maier}.

 In the case of the $d_{x^2-y^2}$ superconducting gap under discussion in FeBS,
the situation is more complicated. $\Q$ connects states rather close to the nodes of the order parameter on the hole sheets (see figure~\ref{fig:resonance}(b)) and the overall gap in $\mathrm{Im}\chi_0$ determined by $\Omega_c$ is significantly reduced. Still, the resonance condition can be fulfilled due to the fact that for some $\k$'s $\Delta_\k = -\Delta_{\k+\Q}$. However, because of the smallness of
$\Omega_c$ the discontinuous jump in $\mathrm{Im}\chi_0$ is vanishingly small or zero. Thus the total RPA susceptibility  shows a moderate
enhancement with respect to the normal state value, as seen in
figure~\ref{fig:resonance}(a). The same holds for $d_{xy}$- and
$d_{x^2-y^2} + \ii d_{xy}$-wave symmetries \cite{Korshunov2008} and a
triplet $p$-wave \cite{Maier}.  Of course, for states
besides $s_\pm$ a resonance can occur at a {\it different}
wave vector $\Q$ connecting surfaces where the gap changes sign;
in the $d$-wave case a resonance has been predicted for the
wave vector  $\q \approx (\pi,\pi)$ connecting the two electron pockets \cite{Maier2}.

Thus, the resonance peak at $(\pi,0)$ is pronounced only for the
sign-changing $s$-wave order parameter like $s_\pm$. Such a distinct
behavior for the $s$- and $d$-wave gaps can be clearly resolved via the
inelastic neutron scattering experiments and therefore it is a direct
probe for the gap symmetry in FeBS \cite{Korshunov2008,Maier,Maier2}.
This situation is similar to the high-$T_c$ cuprates and heavy fermion
superconductors where a bound state (spin resonance) with a high
intensity also forms below $T_c$ \cite{rossat, sato, broholm}.


The existence of the spin resonance in FeBS was first calculated
theoretically \cite{Korshunov2008, Maier} and subsequently discovered
experimentally. No pronounced features were observed in the earliest work \cite{Qiu}, presumably due to sample quality issues. This was followed,
however, by many reports of well-defined spin resonances near $(\pi,0)$
in 1111, 122, and 11 families of FeBS \cite{Inosov, ChristiansonBKFA, Lumsden, ChristiansonBFCA, Park, Argyriou, QiuFeSeTe, Babkevich}. Although the ratios of $2\Delta_0/T_c$ vary from material to material \cite{InosovDeltaTc}, the gross features of the spin excitations are similar: they are gapped below $\Omega_c$ at $T<T_c$, and there is an enhancement at $\Omega_{res}$, which
vanishes at temperatures above $T_c$. Results for BaFe$_{1.85}$Co$_{0.15}$As$_2$ shown in figure~\ref{fig:resonance}(d) are representative in many respects. For this material, $\Omega_{res}/2\Delta_0 = (0.79 \pm 0.15)$ \cite{Inosov} which is close to 0.64, which has been claimed to be a universal value for cuprates, heavy-fermion superconductors, and FeBS \cite{InosovDeltaTc,Yu} (note there is no compelling theoretical reason for this to be the case). While the resonance in FeBS and cuprates are similar in many aspects, there are some differences. For instance, the temperature evolution of $\Omega_{res}$ in \BFCA~is BCS-like without a signature of the pseudogap \cite{Inosov}. Also, because in cuprates the AFM wave vector $\Q_{AFM}$ connects different parts of the same Fermi surface, for $\q < \Q_{AFM}$ the gap becomes smaller than at $\Q_{AFM}$ since we are closer to the $d$-wave nodes;  the resonance therefore shows downward dispersion. On the other hand, in FeBS with the $s_\pm$ gap symmetry the resonance disperses upwards, see figure~\ref{fig:resonance}(c).

 There are still a few puzzles
connected with the spin resonance. In
figure~\ref{fig:resonance}(c), we show the total RPA spin susceptibility
as a function of both momentum and frequency for  almost perfectly
nested Fermi surfaces. Note that the $s_\pm$ gap changes only slightly
on the hole and electron Fermi sheets and can be considered nearly as a
constant. Therefore, one always finds $\Delta_\k = -\Delta_{\k+\q}$ as
long as the wave vector $\q < \Q$ connects the states on the distant
Fermi surfaces. The nesting condition is very sensitive to the variation
of $\q$ away from $\Q$ and already at $\q \approx 0.995 \Q$, the
$\mathrm{Re}\chi_0(\q,\Omega_r)$ is much smaller than its value at $\Q$.
As a result the resonance peak is confined to the nesting wave vector
and does not disperse very much as occurs in high-T$_c$ cuprates.
Therefore, one expects that when a system doped away from the perfectly
nested case, the spin resonance should become incommensurate with $\q
\neq \Q$. This is however not the case in the 1111 and Co-doped 122 families
where it stays at $\Q$ independently of doping (within experimental
accuracy) \cite{Lumsden, ChristiansonBFCA, Inosov, Park}. On the other hand, incommensurability was found in the Fe(Se,Te) \cite{Argyriou}, and recently in the K-doped 122 system as well \cite{Osborn}.
Note that in the latter case the K-doping was far from the optimal doping studied in \cite{ChristiansonBKFA}, which may explain why the incommensurability was easier to detect. However, there is currently no clear understanding of why the resonance appears to be commensurate in some cases and not others.

Another puzzle is connected with the anisotropy in the spin space observed with polarized neutrons in the non-magnetic phase of \BFNA~\cite{Lipscombe}. It was found that $\mathrm{Im}\chi_{+-}$ and  $2\mathrm{Im}\chi_{zz}$ are different, displaying different resonance frequencies and intensities. This contradicts spin-rotational invariance (SRI) condition $\left<S_+ S_-\right> = 2\left<S_z S_z\right>$ which must be obeyed in the paramagnetic system. The relation $\mathrm{Im}\chi_{+-} > 2\mathrm{Im}\chi_{zz}$ was confirmed by measurements of the NMR spin-lattice relaxation rate in the perpendicular magnetic fields \cite{MatanoBKFA,LiBKFA}. One possible solution to the puzzle could be the presence of the spin-orbit interaction, which can break the SRI as it does in Sr$_2$RuO$_4$ \cite{Eremin2002}.

Recently, it was suggested that the theoretically predicted peak for the isotropic
$s_\pm$ state is is too sharp and too strong compared to the maximum observed in the experiment. Onari \textit{et al} \cite{Kontani_neutron} proposed an alternative explanation for the spin resonance that does not involve a sign change of the order parameter.
They noted
that if there is a collapse of the scattering rate below the pairbreaking edge, the redistribution of the spectral weight upon entering the superconducting state can lead to the enhancement of the spin response
below $T_c$ as compared to the normal state. This effect does not
represent a true spin resonance in the sense that there is no divergence
in $\mathrm{Im}\chi$, but depending on the parameters one can gain
significant enhancement, and the observed resonance is indeed generally
broader than predicted in theories of the neutron response in a clean
$s_\pm$ state.

On the other hand, the similar spin resonance in cuprates, albeit
somewhat sharper than that in FeBS, is also rather broad, and in this case there there is little doubt that the scattering involves
a sign-changing gap. Broadening of a spin excitation can of course arise from many  sources, the most obvious one in this case being significant anisotropy of the $s_\pm$ gap. Another problem with the explanation of \cite{Kontani_neutron} is that it may require a special form of scattering in the normal state,
$\mathrm{Im}\Sigma(\q,\omega) = A (\pi T + \omega)$, and in addition one
needs to fine-tune the parameter $A$.  The exact effect of various assumptions regarding the scattering has in fact been the subject of some debate \cite{Kuroki_recent_neutron,Kontani_recent_neutron}. The fact that extremely similar features of the spin excitations are observed in all families of FeBS would seem to argue
against the possibility of an isotropic $s_{++}$-wave gap.


\subsubsection{Josephson junctions.}
\label{subsec:josephson}

While in cuprates the existence of the neutron resonance mode was a strong
argument in favor of  $d$-wave pairing, really instrumental in establishing
that beyond reasonable doubt were Josephson-based experiments. A direct probe
of the symmetry of the order parameter was performed by creating a current
loop that included two Josephson contacts, one at the $a$ face of a crystal,
and the other at the $b$ face. It is easy to show that if the phase difference
across both contacts is $0$, the allowed values of magnetic flux through the
loop, expressed in flux quanta, are integer, while if one of the two contacts
has a phase shift of $\pi$, they are half-integer. The two cases are easily
distinguishable in the experiment.

The problem with applying this technique to FeBS is that in the $s_{\pm}$
case there is no direction where symmetry would be dictating the phase of the
order parameter. The $x$ and the $y$ directions are indistinguishable by
symmetry. On may think about a Josephson loop with the contacts in the $ab$
plane, but along inequivalent directions (for instance, $[10]$ and $[11]$), in the
hope that the numbers that define the Josephson current (orbital composition
of the wave function, relative gap sizes, etc) will conspire in such a way
that the current along one direction will be dominated by holes and the other
by electrons \cite{ParkerPRL,Wu}. Unfortunately, such contacts are not only
difficult to make but also there is no guarantee that the numerics will work
out right; existing theoretical estimates are on the borderline. Some other
designs have been proposed, but they either incorrectly (and too favorably)
estimate the condition of the $\pi$ contact formation, or are even less practical.

\begin{figure}[ht]
\begin{indented}
\item[]
\includegraphics[width=0.3\columnwidth]{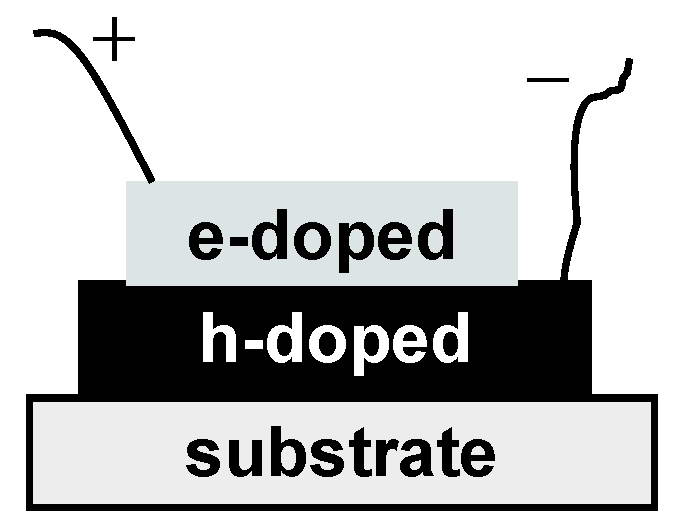}
\caption{Sandwich design suggested in \cite{ParkerPRL}.}
\label{fig:sandwich}
\end{indented}
\end{figure}

More promising is another suggestion \cite{ParkerPRL}. This one utilizes a
so-called sandwich design (figure~\ref{fig:sandwich}), wherein an epitaxial film of a hole-doped FeBS is grown on top of an electron-doped one (or vice versa). If the parallel momentum is conserved at the interface (which is why epitaxial growth is
necessary), the phase coherence is established among the holes in  both
slabs, and correspondingly among electrons. A point contact to the hole-doped
slab will be dominated by the hole current, because of the prevalence of this
type of carriers, and the  one to the electron-doped slab by the electron
current. If these contacts are now connected in a loop, the desired phase
difference is achieved.

So far no such (or similar) experiment has been performed. However, there is
an experiment that presents indirect evidence that Josephson loops with a $\pi$
phase shift can be formed in these materials \cite{Chen}. In this experiment, rather than carefully preparing two Josephson contacts that are dominated by hole and
electron currents, respectively, one measures a very large number of
randomly formed contact pairs, in hope that some of them will accidentally
fulfil the condition that the two contacts have the required phase shift. In
general, a point contact to an electron-doped sample (on which the experiment
\cite{Chen} was performed) will be dominated by electron current. Nevertheless,
there is a possibility that some of the randomly formed contacts will have a
sufficiently thick tunneling barrier, in which case the current may be
dominated by the states near the zone center, i.e. the hole states. Thus, one
expects in such an experiment to see a small, but not negligible fraction of
the formed loops to exhibit the $\pi$ phase difference. This is exactly what
was reported in the experiment \cite{Chen}.

\subsubsection{Quasiparticle interference.}
\label{subsec:qpi}

Another important experiment providing information on gap structure is the so-called
quasiparticle interference scattering (QPI). The idea of this method is
simple: any sort of impurity or defect in a metal is screened by the
conducting electrons. This leads to the well known Friedel oscillations of
the charge and spin density around the imperfection. In real space,
interference among such oscillations stemming from random impurities is  currently
unresolvable in these systems, but the Fourier transform of the measured electron density
will reflect the structure of the charge susceptibility in  reciprocal
space. A natural way to map the electron density near the Fermi level is by
scanning tunneling spectroscopy. The theory of this effect in a $d$-wave superconductor was proposed by Wang and Lee \cite{WangLeeQPI} for a single impurity, and subsequently established for a finite density of impurities \cite{Capriotti,ZhuQPI}.

This technique can be used to gain information on the phases of the
superconducting order parameter. Indeed, let us begin with a simple BCS theory
with a uniform gap. Let us further assume that we have tuned our tunneling
bias to a voltage slightly above the gap value. The quasiparticle density of
states is enhanced at this voltage, showing a coherence peak as the voltage
approaches the gap value. As with any generalized susceptibility, there are
coherence factors involved \cite{NunnerQPI,HanaguriColeman}. It  turns out that for scattering from magnetic impurities, {or from order parameter suppressions, including vortices}, the coherence factors are proportional to $(u_{k}u_{k'}+v_{k}v_{k'})$, that is, they are constructive when $\Delta_{k}\Delta_{k'}>0$, and destructive otherwise. Just as with  impurity pair breaking effects, discussed in Section~\ref{subsec:disorder}, the situation is reversed when the impurity is nonmagnetic, and the coherence factor is proportional to $(u_{k}u_{k'}-v_{k}v_{k'})$. Thus, magnetic impurities and vortices emphasize processes that scatter pairs without flipping the sign of the order parameter, and the nonmagnetic defects emphasize the sign-changing processes. We do not have a tool to change dynamically the impurity concentration, but we can introduce vortices by
applying an external magnetic field, and then the QPI features associated with
the same-sign scattering will be enhanced in comparison with those due to the
sign-flip scattering \cite{PeregBarneaFranz}.

In the Fe-based superconductors, theoretical predictions for the dispersion of the QPI $\q$-peaks have been made for models with electron and hole pockets \cite{early_QPI} in the presence of $s_\pm$ superconducting order \cite{DHLee_2009,FaWangLAFP}, in the SDW state with no superconducting order \cite{Knolle} and in the coexistence phase \cite{Akbari}. All give differing signatures depending on the evolution of the contours of constant quasiparticle energy in the various reconstructed Fermi surfaces. The QPI signatures depend strongly on the sign change of the gap, but also on details of the Fermi surface. A  problem that prevents QPI from being as useful a tool as it was in cuprates, is that in a $d$-wave superconductor the tunneling current at low biases is dominated by ``hot spots'' on the underlying Fermi surface where  the superconducting gap is exactly equal to the bias voltage \cite{HanaguriColeman}. In an isotropic nodeless superconductor there are no ``hot spots'' and the entire theoretical picture is therefore blurred compared to QPI in cuprates.

Experimentally, QPI measurements on  Ca-122 lightly doped with Co \cite{Davis_Ca122} in the magnetic phase revealed strong breaking of  tetragonal symmetry, qualitatively consistent with the observed SDW \cite{Knolle}
and with DFT calculations with the observed stripelike magnetic order \cite{MazinArgyriou}. It was also pointed out that scattering in this system may be affected by anisotropic impurity states around
the Co sites imaged in the experiment, enhancing or modifying the background anisotropy caused by the stripelike magnetism.

In the superconducting state of   an Fe(Se,Te) superconductor near
optimal doping, a QPI experiment in a varying magnetic field
was performed by Hanaguri \textit{et al} \cite{Hanaguri2}.
They found three features, one associated with the hole-electron scattering (the smallest momentum), and two associated with two different electron-electron scattering options.
The last two features grow with respect to the first one, which led Hanaguri \textit{et al} to conclude that the holes and electrons have opposite signs of the order parameter, as dictated by the $s_{\pm}$ model.

It should be noted \cite{Mazin_Singh} that the last two wave vectors coincide with
the two smallest reciprocal lattice vectors, so QPI features at these vectors
(if any) will  coincide with Bragg peaks. Hanaguri \textit{et al} \cite{HanaguriReply} argued that the
corresponding features can be decomposed into sharp peaks reflecting the Bragg
reflection, and broader paddings than must arise from QPI. The problem
however remains that both the sharp peaks and the paddings show similar
dependence on magnetic field (although the Bragg peaks should be insensitive
to the vortex concentration), which makes one suspect the real effect of
magnetic peaks is suppression of the small-moment feature, rather than
enhancing the other two. Thus the issue of whether Hanaguri \textit{et al} have really observed QPI features and can make conclusions regarding the order parameter is open.

Finally, there is another intriguing aspect of the QPI spectroscopy.
If the order parameter has nodes or deep minima, and there is a good
experimental reason to believe that this is the optimally doped Fe(Se,Te) \cite{Zeng}, the same mechanism that creates hot spots, dominating the QPI
picture in cuprates, will kick in as soon as the bias voltage is larger than
the gap minimum. After that, the spectrum should be dominated by these hot
spots, creating, similarly to the cuprates, a complicated patter of multiple
very sharp spots, dispersing with the bias. No trace of this effect as
been observed.

\subsubsection{Coexistence of magnetism and superconductivity.}
\label{subsec:coexistence}

This discussion will not be complete without mentioning an experiment that
Mother Nature has performed for us, namely that in the phase
diagram of the Co-doped BaFe$_{2}$As$_{2}$ there is a well established range
with microscopic coexistence of weak antiferromagnetism and superconductivity \cite{Bobroff}. Moreover, the ``backbending'' of the SDW instability line in the coexistence phase, observed in Co-doped Ba-122, indicates that the magnetism and superconductivity are carried by the same electrons so that the two instabilities compete for the same carriers. It can be shown \cite{Fernandes,Vorontsov} that in this case an $s_\pm$ superconductivity can easily coexist with an SDW state, but $s_{++}$ can only for a very narrow range of parameter. Thus, $s_\pm$ appears to be a much more natural state given that the coexistence appears in a large part of the phase diagram, and probably exists also in other FeBS, although for the other materials direct microscopic probes (e.g., NMR) are still missing.

This is a quantitative argument. One can also make another, slightly more subtle,  qualitative  argument.
It was noted already some time ago \cite{Bulaevskii} that if conventional ($s_{++}$ in our language) superconductivity develops on the background of a spin density wave, the order parameter develops nodes. These appear everywhere where new band crossings occur due to SDW symmetry lowering (see \cite{Overhauser} for a more detailed discussion).
On the other hand, when we introduce such an SDW in an otherwise nodeless $s_{\pm}$ superconductor, it can be shown that no nodes develop even if in the new, downfolded Brillouin zone, the ``$s_{+}$'' and the ``$s_{-}$'' bands cross \cite{Parker}.

The importance of this theorem can be appreciated if we remember that the
in-plane thermal conductivity in this very part of the phase diagram show the
clear absence of any vertical nodal lines (there are indication of possible
horizontal or so-called ``$c$-axis" nodal lines, but the above mentioned SDW-induced BZ band folding
creates full vertical lines of new band crossings, where the hole and the
electron Fermi surfaces now overlap). Thus, if the order parameter has the same sign
on all FS pockets, in the coexistence region vertical nodal lines must appear,
and they must show up in the in-plane thermal conductivity. If that does not
happen, it leaves only one possibility: an $s_{\pm}$ order parameter.

\subsection{Evidence for low-energy subgap excitations.}

\begin{figure}[ht]
\begin{indented}
\item[]
\includegraphics[width=0.8\columnwidth]{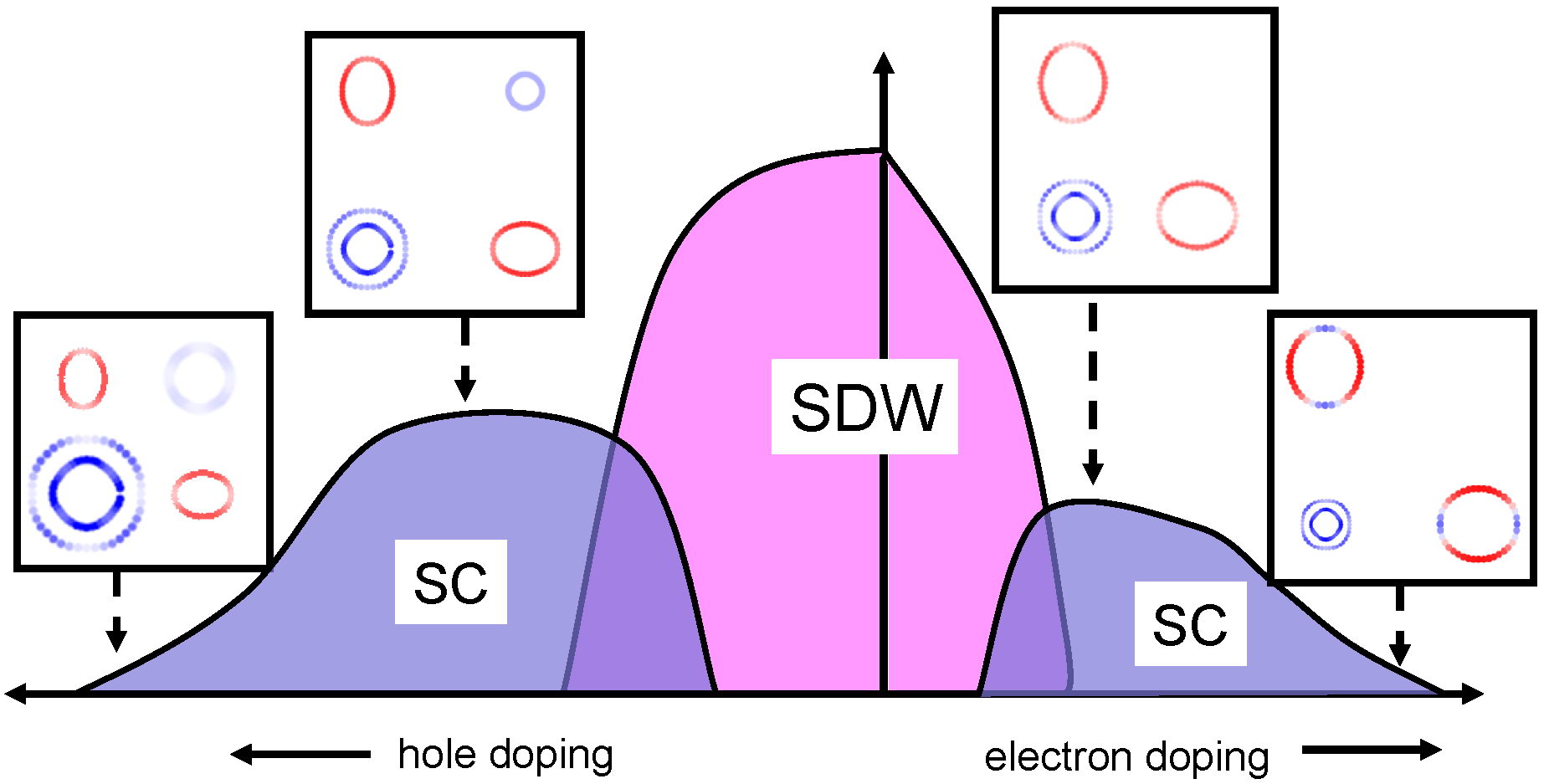}
\caption{Schematic  phase diagram of Fe-based superconductors  vs.
doping, with  order parameter expected from 2D spin fluctuation theory plotted in one quadrant of Brillouin zone as false color on Fermi surface [red=+, blue=-].}
\label{fig:bigpicture}
\end{indented}
\end{figure}

In the few years of experimental studies on Fe-pnictide superconductors,
the hope that one might quickly identify a universal form of the superconducting order parameter was confounded by an unexpectedly wide diversity of experimental results, with some consistent with  fully gapped behavior, and others providing evidence for very low energy excitations consistent with gap nodes.  Early discussion focussed on the possibility that variations could be explained exclusively by the effect of disorder (see Section~\ref{subsec:disorder}), so that some varying results on different samples of the same material could be explained in this way.  It may still be true that in some situations disorder plays a key role and needs to be understood.  However, in the past year or so different experimental probes, particularly penetration depth experiments and thermal conductivity, both bulk probes, are now providing a consistent picture of the  evolution of the low-energy quasiparticle density across the phase diagram of the 122 materials, and to a lesser extent in other families as well.  This suggests a picture in which the gap structure is sensitively related to the details of the Fermi surface as it evolves across the phase diagram from hole- to electron-doped systems (see figure \ref{fig:bigpicture}).

From the point of view of spin fluctuation theory outlined in Section~\ref{sec:theorybackground}, this evolution is relatively
easy to anticipate.
As we move away from the parent compound, the spin pairing interaction weakens. On the other hand, in the ordered phase superconductivity is suppressed entirely by the competition with magnetism for states near the Fermi level, until the SDW amplitude is sufficiently weak, when $T_c$ can begin to grow.
The optimal doping is thus expected to be not far from the antiferromagnetic quantum critical point, as it is indeed in reality. One can also understand on a similarly qualitative level the tendency to node formation in the overdoped regime.
Suppose that local interaction parameters do not vary significantly with doping, as might be expected if they are derived generally from Fe atomic  orbitals (and possibly ligand polarization effects, see \cite{Sawatzky_09}).  Generally speaking, the highest pairing strengths are predicted for  systems that have taken full advantage of the available condensation energy, i.e. optimally doped systems should be maximally isotropic.  Note that this may be a gap close to the idealized  isotropic $s_\pm$ state, or one with considerably more anisotropy, depending on details of the band structure and the interactions themselves.   Optimal doping is then determined by a compromise between ``nesting", which is generally maximal for the undoped parent compound, and the proximity of the ordered SDW. As one overdopes the system, nesting deteriorates, and $T_c$ decreases.  As discussed in Section~\ref{sec:theorybackground}, the dominant orbital interaction is between $d_{xz}$ and $d_{yz}$ orbitals on the electron and hole pockets, while the subdominant one which drives nodal behavior derives from the $d_{xy}$ interactions.  The former interactions are weakened with overdoping while the latter remain constant, leading to a relative enhancement of the frustrating interactions and a tendency towards nodes which grows with overdoping.  Finally, in the common 1111 and 122 systems, there is a further effect which drives an overall asymmetry of the $T$ vs. doping phase diagram.  This is the existence of the additional $d_{xy}$ hole pocket which appears with sufficient hole doping.  As explained in Section~\ref{sec:theorybackground}, this implies that hole doped systems should be generally more isotropic than electron-doped ones. The above ``standard scenario" should now be tested against experiment.

\subsubsection{Penetration depth.}
\begin{figure}[ht]
\begin{indented}
\item[]
\includegraphics[width=0.85\columnwidth]{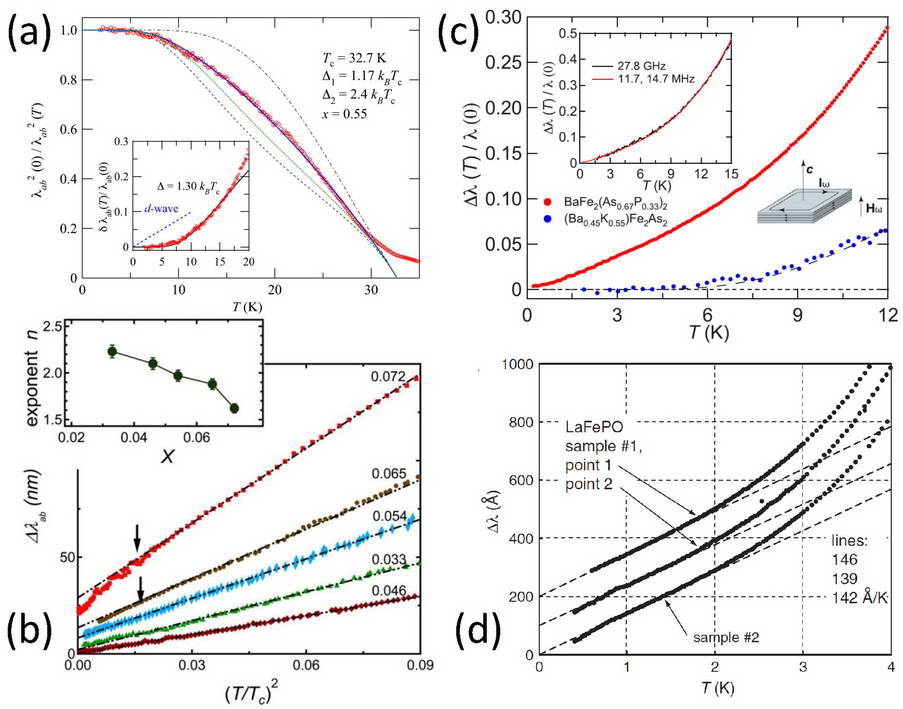}
\caption{Superfluid density temperature variation of Fe-pnictide superconductors.  (a) $(\lambda(T_{min})^2/\lambda(T)^2$ vs. $T$(K) for optimally doped \BFKA~\cite{Hashimoto2009BKFA}; (b) $T-$dependent change in penetration depth $\Delta\lambda(T)$(nm) vs. $(T/T_c)^2$ for \BFNA~\cite{CMartin_2010}; (c) $\Delta\lambda(T)/\lambda(0)$ for optimally doped \BFAP~\cite{k_hashimoto_09}; (d) $\Delta\lambda(T)$(\AA) for \LFPO~\cite{hicks09}.}
\label{fig:pendepth}
\end{indented}
\end{figure}

Magnetic penetration depth measurements, summarized in figure~\ref{fig:pendepth}, are bulk probes of  quasiparticle excitations which can provide evidence for nodal structures or small gaps.
In general, fits to theory over the entire temperature range are difficult
particularly for multiband systems, and depend sensitively on details, so
information obtained at low temperatures is simpler to relate directly to
gap structure.  The relation of different gap nodal structures to power laws
in temperature  $\Delta \lambda\sim T^n$ was pointed out by Gross \textit{et al} \cite{Grossetal}. In a fully gapped system, at low temperatures relative to the smallest full gap in the system an exponentially activated behavior is expected; if this gap is very small, however, the penetration depth can typically be fit to a power law in $T$ over some intermediate temperature range.  Another factor complicating the interpretation is disorder; at the lowest temperatures, impurity scattering can lead to a $T^2$ dependence \cite{Grossetal} {\it if} a residual density of states at the Fermi level is induced (see Section~\ref{subsec:disorder}).  Thus fits to low-temperature power laws at low but not asymptotically low temperatures may not be completely straightforward to interpret, but do provide evidence for low-lying quasiparticle excitations.  Only in the case that a true linear power law $\Delta\lambda\sim T$ is observed may one make definitive statements about the existence of (line) nodes.

In both the LaFePO system \cite{j_fletcher_09,hicks09} and in
BaFe$_2$As$_{1-x}$P$_x$ \cite{k_hashimoto_09}, a linear-$T$ dependence
of the low-$T$ penetration depth $\Delta \lambda(T)$ was reported.  By contrast,  in  \BFCA~and \BFNA, $\Delta\lambda$ was  initially reported to vary close to $T^2$ over most of the phase diagram \cite{r_gordon_08,CMartin_2010}; these power laws are in contrast to the activated temperature dependencies expected for an isotropic gap. While $T^2$ is the power law one naively expects for a (dirty) line nodal state, one may
also show that in an isotropic ($s^\pm$) superconductor,
disorder can create subgap states \cite{Golubov} under certain conditions, depending on the ratio of inter- to intraband impurity scattering \cite{Mishra_09} (see Section \ref{subsec:disorder}).
. If these states are at the Fermi level,  a fully gapped $s^\pm$ state will also lead to
$\Delta \lambda \sim T^2$. Fits of the same or very similar data on these systems \BFCA~and \BFNA~near optimal doping are also possible for an isotropic multigap model \cite{BFCA_full_gap_pendepth}, and at optimal doping the $T^2$ fit is rather poor, suggesting a small true gap. In this context, it is worth noting that a number of multigap fits---to penetration depth, specific heat, and other observables---in the literature violate BCS theory by taking  arbitrary ratios of the gaps $\Delta_i/T_c$ as fit parameters. Kogan \textit{et al} \cite{KoganMartinProzorov2009} have warned that unphysical results can be obtained by this procedure even if the gaps are truly isotropic, since the various gaps are coupled through the BCS gap equation.
At the moment, substantial experience has been accumulated by researchers from various groups that indicates that full solution of a multiband Eliashberg equations in realistic cases always yields at least one gap that is larger than the isotropic gap with the same $T_c$,
and one smaller \cite{KoganMartinProzorov2009}.

Finally, there are some systems where a large full gap has been reported.  For example, in optimally doped \BFKA, a minimum gap of $1.3 k_BT_c$ was extracted in \cite{Hashimoto2009BKFA, kfe2as2}, and similar behavior was found for \LFA~\cite{LiFeAspendepth}. Early reports of exponential behavior in 1111
systems with any rare earth except La were probably ``contaminated'' by the magnetic susceptibility of the rare earth ion \cite{SmFeAsOFpendepth}, and \LOFFA~itself has been reported to have a power law $T$ dependence close to $n=2$ \cite{LOFFApendepth}.

Experimental attempts have been made to correlate disorder to the low-$T$ penetration depth to see if conclusions could be drawn regarding the structure of the underlying order parameter. Hashimoto \textit{et al} \cite{Hashimoto2009BKFA} reported not only the sample which fit well to an exponential $T$-dependence, but a second sample, considered dirtier due to its smaller $T_c$, which exhibited a $T^2$ behavior.  This was interpreted as pairbreaking caused by interband scattering in an $s_\pm$ state.  A similar model was employed  to study changes in low-$T$ power laws with disorder, explicitly calculating the low-energy density of states induced by interband impurity scattering and its effect on the penetration depth and $T_c$ simultaneously \cite{Proz_Voront_disorder}.  While these studies are suggestive, they cannot be regarded as conclusive regarding either the structure of the order parameter or the nature of the disorder scattering, due to the uncertainties regarding the difficulty of determining impurity model parameters, see Section~\ref{subsec:disorder}.

\subsubsection{Specific heat.}

Specific heat was pioneered \cite{Moler} as a tool for investigating the gap structure on YBCO. Since the density of states of an unconventional superconductor with line nodes varies as $N(\omega) \sim \omega$, the temperature  dependence of the Sommerfeld coefficient $\gamma=\lim_{T\rightarrow 0} C/T$ in a clean superconductor with lines of nodes varies as $T$ as  $T \rightarrow 0$.  This is a difficult measurement since disorder generally gives rise to a residual density of states $N(0)$ which induces a linear-$T$ term in $\gamma(T)$ below some disorder scale.
It is therefore sometimes more useful to examine the field dependence, which is also quite sensitive to low-energy excitations.

\begin{figure}[ht]
\begin{indented}
\item[]
\includegraphics[width=0.7\columnwidth]{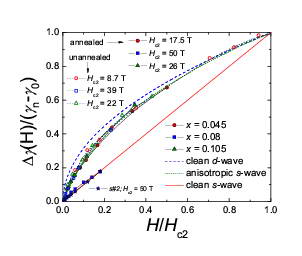}
\caption{Magnetic field dependence of the low-temperature specific heat of \BFCA~(circles, $x = 0.045$; triangles, $x = 0.105$; squares, $x = 0.08$). Empty and full symbols represent unannealed and annealed data, The dotted, solid, and dashed lines described the field dependencies of the low-temperature specific heat according to clean $s$-, anisotropic $s$-, and clean $d$-wave. From \cite{Gofryk2011}.}
\label{fig:Gofryk}
\end{indented}
\end{figure}

In a clean nodal system, the theory of Volovik \cite{Volovik} predicts $\gamma\sim  \sqrt{H}$  in a clean superconductor with lines of nodes (disorder changes this behavior slightly, to  $\gamma \sim H \log H$ \cite{Kuebert} for a disordered superconductor with lines of nodes).  These power laws can be derived very easily  from the Doppler shift of the low-energy nodal quasiparticles in the superflow field of the vortex lattice. For a fully gapped superconductor, $\gamma$ should vary linearly with H at low fields due to the localized Caroli-de Gennes-Matricon states in the vortex cores.  This probe provides a first indication of the variability of thermodynamic properties in the Fe-pnictide materials:  in \BFKA~$\gamma\sim H$ \cite{Mu2009} (implies fully gapped superconductivity), while $\gamma \sim H^{1/2}$  in \LOFFA~\cite{Mu2008} (implies nodal superconductivity).  Recently Gofryk \textit{et al} \cite{Gofryk} performed measurements on \BFCA~across the electron doping range, and reported a nonmonotonic dependence of the density of excitations with doping.  At optimal doping, a very weak field dependence consistent with a small gap or weak nodes was reported, with quasiparticle contributions increasing on either side of optimal doping (figure~\ref{fig:Gofryk}).  Note that the underdoped sample is in the SDW-SC coexistence phase. In the isovalent analog system \BFAP, an early report of linear-$H$ behavior which appeared inconsistent with penetration depth measurements \cite{StewartPdoped1} reporting linear $T$ behavior as discussed above \cite{k_hashimoto_09}, as well as evidence for nodes from NMR \cite{NakaiPdoped} and thermal conductivity measurements \cite{k_hashimoto_09}, was recently superseded by a measurement reporting a small Volovik term at low fields crossing over to linear behavior at higher fields \cite{StewartPdoped2}. This work underlined the importance of determining on which sheets the nodes occur, since sheets with smaller mass and longer relaxation times will dominate transport, while larger mass alone will determine specific heat. Theoretically, the field crossover of $\gamma$ was examined in a semiclassical multigap $s_\pm$ framework \cite{ybang} whose validity is questionable since it neglects the contribution of the core states, which must contribute significantly in a fully gapped superconductor.   Fits to the $H^{1/2}\rightarrow H$ behavior were however obtained within a multiband Eilenberger approach \cite{StewartPdoped2} assuming a highly anisotropic state on one band.

\begin{figure}[ht]
\begin{indented}
\item[]
\includegraphics[width=0.85\columnwidth]{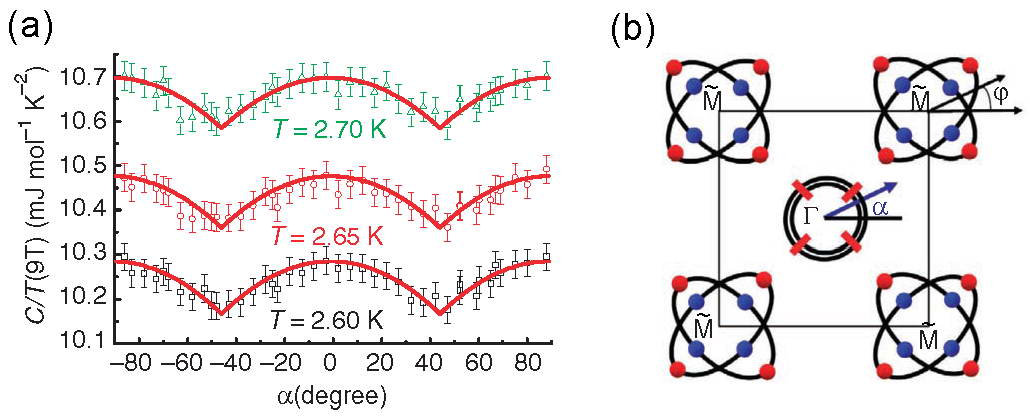}
\caption{(a) Angle dependence of specific heat coefficient $C/T$ for FeSe$_{0.45}$Te$_{0.55}$ at $H=9$T; (b) possible positions of gap minima consistent with measurements from \cite{Zeng}.}
\label{fig:Zeng}
\end{indented}
\end{figure}

If the system has gap nodes or deep minima, the semiclassical theory of specific heat of an unconventional superconductor predicts \cite{Vekhter1999}  that the measured specific heat should oscillate as a function of magnetic field direction relative to the crystal axes.  At the lowest temperature and fields, it is generally expected that the minima in the specific heat will correspond to fields pointing in nodal directions.  It should be noted, however, that this result depends sensitively on the phase space available for quasiparticle scattering, and need not be universal for arbitrary superconducting states.  The roles of minima and maxima in the specific heat can also reverse as a function of temperature or field, as is found even in the well-known $d$-wave case \cite{VekhterVorontsov,Boyd}. Early on in the discussion of the symmetry of the Fe-based superconducting order, it was noted that phase sensitive experiments of the type which led ultimately to the definitive determination of $d$-wave pairing in the cuprates \cite{Tsuei} would be difficult in the new systems, both due to sample preparation difficulties and because of the complexity of interpreting Josephson-based experiments in multiband systems.  As an alternative, it was proposed that specific heat oscillations might provide important information as to the $k$-space structure of the order parameter \cite{Graser_spht}. This experiment was first perfomed on the Fe(Te,Se) system by Zeng \textit{et al} \cite{Zeng} (figure~\ref{fig:Zeng}).
The positions of the specific heat minima along the $\Gamma-\mathrm{M}$ axis are consistent \cite{Graser_spht,Vekhter_FeSe,Chubukov_FeSe} with an anisotropic gap with minima at these angles, as predicted by spin fluctuation theories (see Section~\ref{sec:theorybackground}).

\subsubsection{Thermal conductivity.}

The experimental probe of the bulk
order parameter that has so far been performed at the lowest temperatures
is thermal conductivity. In a system of normal conducting electrons, the thermal conductivity varies linearly with $T$.  While in principle thermal currents are carried by phonons as well, at low $T$ the contribution to the thermal conductivity from phonons $\kappa_{ph}$ typically varies as $T^3$ or nearby power, depending on the phonon mean free path; thus at sufficiently low $T$, any linear-$T$ term in $\kappa$ may be attributed to electronic excitations.  In a superconductor, this term provides information on both the superconducting gap structure and the role of disorder.  The technique has been applied extensively over many years to cuprates and unconventional superconductors \cite{TailleferReview}.  In the presence of a magnetic field which creates a vortex state, quasiparticles are Doppler shifted as in the case of the specific heat,  not only in the density of states but also in the lifetime.  In a clean nodal superconductor, a field dependence of $\Delta\kappa\sim H\log H$ is expected \cite{Kuebert_kappa}.

\begin{figure}[ht]
\begin{indented}
\item[]
\includegraphics[width=0.85\columnwidth]{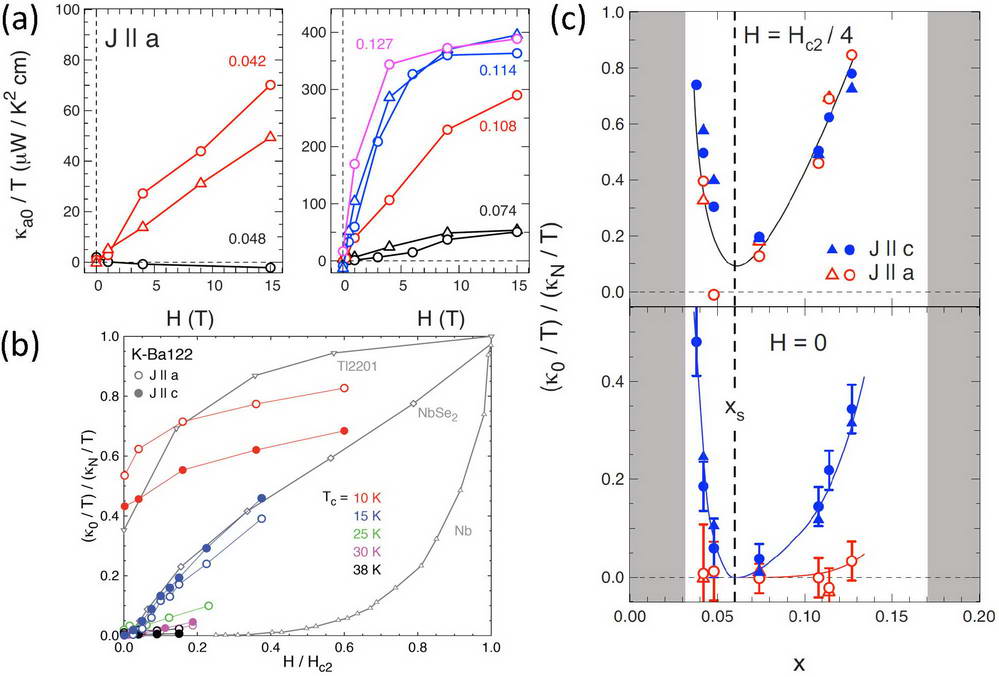}
\caption{(a) Field dependence of in-plane $\kappa_a/T$ in \BFCA~from \cite{Reid}. (b) Same for both normalized $\kappa_a$ and $\kappa_c$ in \BFKA, \cite{Reid2011K-doped}. (c) Normalized $a$ and $c$ thermal conductivities as functions of doping, \cite{Reid}.}
\label{fig:taillefer}
\end{indented}
\end{figure}

In principle, then, thermal conductivity should be one of the best probes of low-energy quasiparticle excitations arising from gap structure.  However, even taken alone the thermal conductivity data on the various Fe-based systems present a complex picture. In the 122 systems, the  $ab$-plane thermal conductivity data for both electron- and hole-doping exhibit zero or extremely small linear-$T$ term in zero magnetic field, reflecting the apparent absence of any nodes in the superconducting gap. The field dependence, however, is significantly stronger than that expected for a large-gap superconductor \cite{x_luo_09, Ding2009, Tanatar2010} (see, however, a discussion in Section~\ref{subsubsec:MgB2}), particularly away from optimal doping (see, e.g. figure~\ref{fig:taillefer}(a)). Mishra \textit{et al} \cite{v_mishra_09} then proposed that such results could most naturally be explained in terms of a gap with A$_{1g}$ symmetry with no nodes but deep minima on the electron sheets. Bang proposed that such strong field dependence could also be explained phenomenologically by an isotropic ``$s$-wave'' state with very small gap on one
Fermi surface sheet \cite{ybang} (on the other hand,  such an explanation appears to be ruled out by $c$-axis thermal conductivity, see below). To provide a scale to interpret statements about the size of the low-$T$ thermal conductivity in the superconducting state, we remind the reader that in a 2D nodal superconductor
\begin{equation}
\frac{\kappa_{ab}}{T}\vert_{T\rightarrow0} \simeq a N_\mathrm{nodes} {k_B^2m^*\over \hbar d }  \left[\frac{v_{F,ab}^{2}}{k_Fv_{\Delta,ab}}\right]_\mathrm{node},
\label{eq:kappalowT}
\end{equation}
where $a$ is a dimensionless constant which depends on the nodal phase space, $d$ is the distance between planes, $N_\mathrm{nodes}$ is the number of distinct nodal surfaces, assumed equivalent, and
$m^*$ is the effective mass for motion of quasiparticles in the $ab$-plane.  The conclusion
of all experiments on near-optimally doped K- doped or Co-doped Ba-122 systems was that the measured linear term was much less than expected from this expression, i.e. $a \ll 1$.

Recently, Reid \textit{et al} \cite{Reid} showed that in the {\it same} samples where the ab-plane  thermal conductivity was very small (consistent with zero within experimental error), a significant $c$-axis linear-$T$ thermal conductivity was reported (note that ``significant" here refers to values normalized to
normal state values determined by the Wiedemann-Franz law; absolute values are still of the same order as quoted error bars for the in-plane conductivity).  This is paradoxical at first glance, because if order parameter nodes exist, they should influence transport properties in both directions, the only difference being a weight factor of $v_{F,i}^2$ for $i=ab,c$, as seen, e.g. in Equation~(\ref{eq:kappalowT}).  The only possible interpretation \cite{Reid} is that the nodes are located on flared portions of the Fermi surface where the $c$-axis velocities are very high. Mishra \textit{et al} \cite{Mishra2011} confirmed this picture and pointed out that that the result $a\ll 1$ above in the ab-plane implies that the phase space for the nodes producing the $c$-axis signal must be very small. Their result is consistent with ``weak nodes", which exist over a small portion of the Fermi surface rather than running the length of the Fermi cylinders.  These could be the ``V-shaped" or ``loop" nodes found in \cite{Graser3D,Kuroki3D} on the hole pockets near the Z point, or small loop nodes on electron pockets as suggested in \cite{MazinDevereaux}.
{In figure~\ref{fig:taillefer}(c) the data of Reid \textit{et al} are plotted as a function of doping at $H=0$ and $H=H_{c2}/4$.  It is seen that the normalized thermal
conductivity, reflecting the existence of low-energy quasiparticles, increases dramatically away from optimal doping for currents along the $c$-axis in
zero field.  In nonzero field, however, the response is quite isotropic.  One is tempted to conclude that only one Fermi surface sheet plays a role in thermal
transport at higher field, such that the thermal current is isotropic when normalized; this is far from obvious, however, given that
the field is aligned in the $c$ direction in both cases, so that the averaging over the inhomogeneous response function is quite different
for the two directions.  These issues are discussed in \cite{Mishra2011}.}

 The existence of nodes on the flared portion of the Fermi surfaces may appear an unlikely accident, but
it is easy to see within the context of spin fluctuation theory that  changes in orbital character on the
Fermi surface tend to produce nodes because of the strong tendency of like orbitals to pair.  As discussed in Section~\ref{subsec:dimensionality}, while the 122 systems look quite similar to the 1111 systems at $k_z=0$
(except for differences in ellipticity), at higher $k_z$ other bands, particularly the $z^2$, mix strongly, such that at the top of the Brillouin zone, where the hole pockets are most flared, there is a strong admixture of several bands whose weight varies around the sheet. This is most likely the origin of possible weak nodes in the hole band \cite{Graser3D,Mishra2011}.

The above discussion has applied exclusively to the Ba-122 system doped with K or Co. The measurements on optimally doped \BFAP~find a strong linear-$T$ term, suggesting a strong nodal component \cite{k_hashimoto_09}, consistent with
other measurements on this material. Measurements in FeSe find a small linear term \cite{FeSe_kappa}, which the authors reported as consistent with nonzero gap, but whose size is of order that found by Reid \textit{et al} for the $ab$ plane values. We are not aware of any thermal conductivity measurements on the 111 or 1111 families.

\begin{figure}[ht]
\begin{indented}
\item[]
\includegraphics[width=0.8\columnwidth]{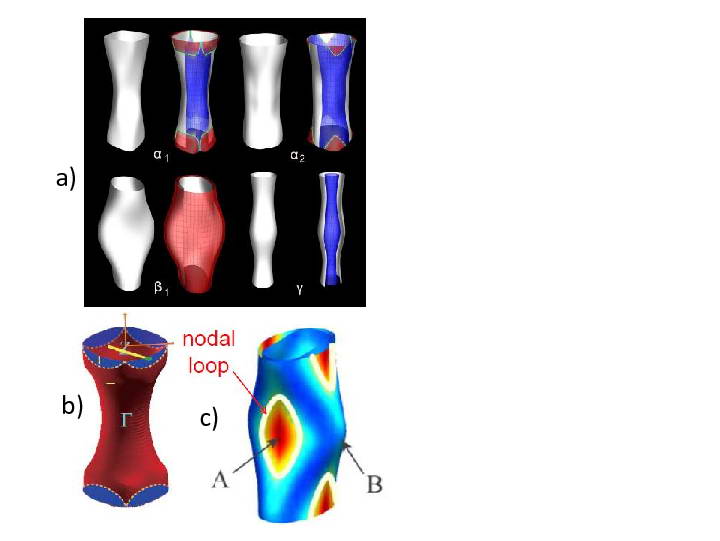}
\caption{(a) Nodal structure on $\alpha$ (hole) and $\beta$ (electron) pockets in 3D spin fluctuation calculation for \BFCA~by Graser \textit{et al} \cite{Graser3D}; (b) Similar result for $\alpha$ sheet of \BFAP~\cite{Kuroki3D}; $\kappa_a$ and $\kappa_c$ in \BFKA, \cite{Reid2011K-doped}; loop-like nodes around point of maximum (A), rather than minimum (B) Fermi velocity  on the outer $\beta$ pocket in \BFAP~deduced from angle-dependent thermal conductivity \cite{Yamashita}, similar to that found in \cite{MazinDevereaux} from analysis of Raman scattering on \BFCA.}
\label{fig:3Dnodes}
\end{indented}
\end{figure}

The oscillations of thermal conductivity in a rotating magnetic field provide similar information to angle-dependent specific heat oscillations.  They tend to be easier to observe and somewhat less straightforward to interpret \cite{VekhterReview}.  At this writing, this experiment has only been performed on P-doped Ba-122, where significant oscillations are observed \cite{Yamashita}, and interpreted in terms of loop nodes on the electron pockets. As mentioned above, both penetration depth and thermal conductivity experiments have provided evidence for
nontrivial 3D nodal structures in the \BFCA~and now the \BFAP~systems.  A variety of such structures have been suggested, by microscopic theory \cite{Graser3D,Kuroki3D}, and phenomenology \cite{MazinDevereaux,Yamashita}, which we summarize in figure \ref{fig:3Dnodes}.

For the most part, we have not discussed experiments in the so-called ``coexistence phases'' of the FeBS phase diagram where superconductivity and magnetic order are simultaneously present. In part this is because the microscopic homogeneity of these phases is not firmly established, and in part because theoretical calculations to predict transport properties have not yet been performed. We note that the spin fluctuation theory calculations for the instability line of $T_c$ vs.
doping in the absence of magnetic order do not show a suppression of superconductivity in the
underdoped regime (see, e.g. figure~\ref{fig:Ikeda}), suggesting that the suppression is due to the
competition with magnetic order. Recent experimental results in
this regime also imply that the magnetic ordering plays an important role in
gap structure in the coexistence regime.  Penetration depth measurements on \BFCA~early on reported a sharp rise
in the coefficient of the $T^2$ term in $\Delta\lambda$ in the underdoped regime \cite{r_gordon_08}.  More recently, thermal transport measurements \cite{Reid} reported a similar sharp increase in the normalized linear-$T$ $\kappa_c$
term (figure~\ref{fig:taillefer}(c)).  While these two results on underdoped \BFCA~have not been understood completely within a single
model (see however \cite{Mishra2011}), they are strong indications that the number of quasiparticles increases in the coexistence phase due to a strengthening of nodal behavior.  On the hole-doped side, the effect is even stronger and appears quite abruptly in $\kappa_{ab}/T$ \cite{Reid2011K-doped}.  This effect will be important to understand theoretically to complete the picture of the 122 materials, but should be approached with caution because issues of inhomogeneity are not settled, particularly on the K-doped side.

\subsubsection{The ARPES ``paradox''.}

Low-energy excitations whose existence is implied by the above measurements should
be visible in angle-resolved photoemission spectroscopy (ARPES).  In fact, ARPES is arguably the most direct probe of the superconducting gap structure.  Yet at this writing no ARPES experiment \cite{l_zhao_08, h_ding_08, t_kondo_08, d_evtushinsky_09, k_nakayama_09, l_wray_08, borisenko_2010, Shin2011} has reported the existence of gap nodes or even significant gap anisotropy, in dramatic contrast to the cuprate case, where ARPES was one of the experiments providing definitive evidence for $d$-wave superconductivity \cite{Damascelli}.
The disagreement  between bulk probes and ARPES on Fe-based superconductors
is an important point which requires resolution if we are to  rely on both types of measurements to study superconductivity, as we have in the past.  There are several possibilities to explain the discrepancy:

\begin{itemize}
\item {\it Surface electronic reconstruction} One possibility is that the electronic structure at the surface is different from that of the bulk.  This was the point of view advocated by Kemper \textit{et al} \cite{Kemper_sensitivity}, who calculated the bulk and surface band structures of \BFA~from density functional theory (DFT), and discovered that the surface bands included an additional pocket of $xy$ orbital character at the Fermi level, and gave arguments to the effect that such a pocket would stabilize an isotropic pair state.
\item{\it Surface depairing.} Given the common assumption that the ground state of many (but probably not all) Fe-based superconductors display anisotropic $A_{1g}$ order, it is easy to see that the anisotropic component of the gap (whether on hole or electron sheet) will be destroyed  by in-plane intraband scattering by the rough surface, since it does not conserve the parallel (to the surface) quasiparticle momentum.  Thus, as one approaches the surface, the superconducting gap associated with a pair of momentum $\k, -\k$ should become isotropic and the nodes lifted.  The phenomenon is similar to that observed for the same type of superconducting order in the presence of intraband scattering by impurities \cite{Mishra_09}.
    Surface scattering and electronic reconstruction should be smaller for non-polar surfaces which occur, e.g. in the LiFeAs material. In this case, however, spin fluctuation theory \cite{theory_LiFeAs}, thermal conductivity \cite{tanatar_2011}, penetration depth \cite{LiFeAspendepth}, ARPES \cite{borisenko_2010} and STM \cite{STMLiFeAs} are all in agreement that the gap is fully developed.

\item{\it Resolution issues.}  Gaps on the hole sheets around the $\Gamma$ point should be imaged with relative ease by ARPES, and it is possible to imagine reconciling the largely isotropic gaps found on these sheets with thermodynamic measurements, since other evidence points primarily to nodes on the electron sheets, as discussed above. It is noteworthy that the nominally highest resolution experiments using laser sources find full gaps but otherwise qualitatively different results than synchrotron-based ARPES \cite{Shin2011}. Most calculations predicting nodal effects are done in the 1-Fe zone; when folding such states one gets gaps from 2 electron pockets centered at the $\mathrm{M}$ points. Given that the resolution around these points (not probed by laser ARPES) is typically several meV, it seems possible that the anisotropy of the gaps on the electron sheets might be missed if the spectral peaks from both sheets were broadened into one with averaged--and hence isotropic--dispersion.

    We note that although ARPES experiments to date are apparently providing unreliable measures of the superconducting gap anisotropy, this does not necessarily mean that they are inconsistent with bulk gap scales, as shown by the comparison of two gaps extracted from ARPES and specific heat on LiFeAs \cite{stockert_11}.
\end{itemize}

\subsubsection{NMR $1/T_1$.}

\label{subsec:1T1}
\begin{figure}[ht]
\begin{indented}
\item[]
\includegraphics[width=0.85\columnwidth]{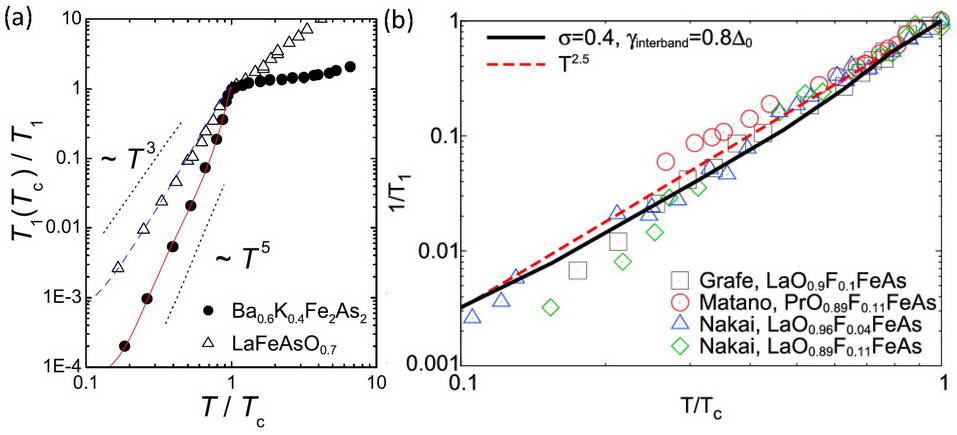}
\caption{Temperature dependence of $1/T_1$ in FeBS.
(a) Experimental results for two classes of materials, 1111 and 122, from \cite{Yashima}.
(b) Log-log plot summarizing experimental data from several groups \cite{Matano,Grafe,Nakai}, theoretical curve for the $s_\pm$ superconductor with intermediate strength of impurity scattering ($0 \leq \sigma \leq 1$) and pairbreaking parameter $\gamma_{interband} = 0.4 \Delta_0$, and $T^{2.5}$ curve to demonstrate a power-law dependence (from \cite{Parker2008}).}
\label{fig:nmr}
\end{indented}
\end{figure}

In addition to the Knight shift, which allows one to distinguish between singlet and triplet pairing (see Section~\ref{subsec:tripletorsinglet}), NMR can probe the spin-lattice relaxation rate $1/T_1$ that corresponds to a the spin susceptibility integrated over the Brillouin zone,
\begin{equation}
\frac{1}{T_1 T} \propto \lim_{\omega \rightarrow 0} \sum_{\q} \frac{\mathrm{Im}\chi(\q,\omega)}{\omega}.
\end{equation}
As in the case with the spin resonance, Section~\ref{subsec:resonance}, $1/T_1$ carries information about the underlying gap symmetry and structure. For example, an isotropic $s$-wave state is characterized by a Hebel-Slichter peak just below $T_c$ and an exponential low-$T$ temperature dependence. It is well-known that  $d$-wave superconductors exhibit weak or absent peak and demonstrate $T_1^{-1}\sim T^3$ behavior for $T\ll T_c$. In the case of FeBS, the situation is somewhat more complicated. Typical data for some 1111 and 122 systems are shown in figure~\ref{fig:nmr}(a). Apparently, there is no peak below $T_c$ and the temperature dependence does not follow the same simple power or exponential law in all systems. However, simple arguments  can enable us to understand the main features found in experiments.

In case of a weakly coupled clean two-band superconductor below $T_c$,  assuming that the main contribution to $\mathrm{Im}\chi(\q,\omega)$ comes from  interband interactions, we have
\begin{equation}
\frac{1}{T_1 T} \propto \sum_{\k \k'} \left[1 + \frac{\Delta_{\k} \Delta_{\k'}}{E_{\k} E_{\k'}} \right] \left( -\frac{\partial f(E_{\k})}{\partial E_{\k}} \right) \delta\left( E_{\k} - E_{\k'} \right),
\end{equation}
where $\k$ and $\k'$ lie on hole and electron Fermi sheets, respectively, and $E_{\k}$ is the quasiparticle energy in the superconducting state. This is a straightforward generalization of the textbook expression \cite{schrieffer}. As in the spin resonance case, the coherence factor in square brackets gives rise to an important distinction between different symmetries of the gap. In the NMR $T_1^{-1}$ case, we see that the internal sign is different from that which occurs in   Equation~(\ref{eq:cohfac}) for the neutron spin resonance effect. Assuming first an isotropic $s_{++}$-wave gap with $\Delta_{\k} = \Delta_{\k'} = \Delta$, one finds
\begin{equation}
\frac{1}{T_1} \propto \int\limits_{\Delta(T)}^{\infty} dE \frac{E^2 + \Delta^2}{E^2 - \Delta^2} ~ \mathrm{sech}^2\left(\frac{E}{2T}\right).
\end{equation}
The denominator gives rise to a peak just below $T_c$, which is the famous
Hebel-Slichter peak. As pointed out earlier in \cite{Mazin_etal_splusminus}, it is
suppressed for the $s_{\pm}$ state. Indeed, if $\Delta_{\k} = -\Delta_{\k'} = \Delta$,
\begin{equation}
\frac{1}{T_1} \propto \int\limits_{\Delta(T)}^{\infty} dE \frac{E^2 - \Delta^2}{E^2 - \Delta^2} ~ \mathrm{sech}^2\left(\frac{E}{2T}\right) = \int\limits_{\Delta(T)}^{\infty} dE ~ \mathrm{sech}^2\left(\frac{E}{2T}\right),
\label{eq:1T1spm}
\end{equation}
which is just the Yoshida function, which decreases monotonically as temperature is decreased below $T_c$.  The same can be shown for a more general $s_\pm$ case of $|\Delta_{\k}| \neq |\Delta_{\k'}|$ \cite{Parker2008}.

It is well known that pair-breaking impurity scattering dramatically increases the subgap density of states just below $T_c$, and even weak magnetic scattering can broaden and eliminate the Hebel-Slichter peak in conventional superconductors. In FeBS, the same effect is present due to the nonmagnetic interband scattering \cite{Golubov}. Since the Hebel-Slichter peak is not present in this scenario even in a clean sample, see Equation~(\ref{eq:1T1spm}), the pair-breaking effect is more subtle: it changes exponential behavior below $T_c$ to a more power-law like one.  If the impurity-induced bound state lies at the Fermi level (Section~\ref{subsec:disorder}), the relaxation rate acquires a low-temperature linear-$T$ Korringa-like term over a range of temperatures corresponding to the
impurity bandwidth \cite{PJHconsequences}.

%

Qualitative arguments suggest that neither pure Born nor pure unitary limits with a simple isotropic $s_\pm$ state are well suited for explaining the observed $1/T_1$ behavior: the former leads to an exponential behavior at low temperatures in a relatively clean system, the latter to Korringa behavior.  Various early data on the 1111 systems appeared to be between these two limits.  On the other hand, early theory by Parker \textit{et al} studying the intermediate scattering regime seemed to be rather promising in this respect. Figure~\ref{fig:nmr}(b) shows various experimental data for 1111 systems \cite{Matano,Grafe,Nakai} together with a calculation of $T_1^{-1}$ for the simple $s_{\pm}$ gap \cite{Parker2008}.  We observe that the $s_{\pm}$ state result exhibits no coherence peak and as opposed to the Born and unitary limits, intermediate-$\sigma$ scattering is capable of reproducing the experimental behavior \cite{Parker2008, Chubukovdisorder, SengaKontani, Bang_disorder}.  It is not clear that these results, taken alone, should be taken as evidence for an isotropic $s_\pm$ state, since strong gap anisotropy is probably present in some of these systems, and will also lead to a higher
density of quasiparticles contributing at intermediate temperatures.  From the data, one can say with certainty  only that the K-doped Ba-122 system appears to have a large full gap, while the 1111 systems show a much higher density of low-energy excitations.

Regarding other systems, data obtained on \BFAP~shows a linear-$T$ term in $T_1^{-1}$ for an optimally doped sample, crossing over to something roughly approximating $T^3$ above $\sim 0.1T_c$ \cite{NakaiPdoped,NakaiPdoped2}, consistent with reports of nodes in this material from other probes.  Low-temperature data on Co-doped \BFCA~is not available at this writing. In Ba$_{0.68}$K$_{0.32}$Fe$_2$As$_2$, $1/T_1$ shows an exponential decrease below $T \approx 0.45T_c$ consistent with a full $s_\pm$ gap \cite{LiBa122}. Finally, consistent with other measurements, NMR in the LiFeAs system also shows a full gap \cite{Li111}.


\subsubsection{Electronic Raman scattering.}
Because the momentum and polarization of incoming and outgoing photons can be controlled in a Raman scattering measurement, this technique is useful to probe selectively different parts of the Fermi surface. The nonresonant electronic Raman intensity can to a good approximation be represented as an electron-hole bubble with Raman vertices $\gamma_{n\k}=\varepsilon_\alpha^i(\partial \epsilon_{n\k}/\partial k_\alpha\partial k_\beta)\varepsilon_\beta^f$, where the $\hat \varepsilon$'s are the incident and scattered (final) photon polarizations, $\epsilon_{n\k}$ is the electronic dispersion and $n$ is the band index. Muschler \textit{et al} \cite{Muschler} measured Raman scattering on an optimally doped sample of \BFCA~and presented a simple approximation to these vertices
which suggested that the  $A_{1g}$ polarization (symmetric configuration of $\hat\varepsilon^{i,f}$) intensity is maximal near the BZ center and thus probes the hole Fermi sheets; similarly, $B_{1g}$ probes the electron pockets, and $B_{2g}$ is maximal near $(\pi/2, \pi/2)$ points where there is no Fermi surface (we use the notation of the 1-Fe zone here, i.e. $B_{1g}=\widetilde{B_{2g}}$).  Within this interpretation, the large $B_{1g}$ peak observed corresponds to twice the maximum gap in the system, yielding a value of $\Delta_{max}\simeq$70cm$^{-1}$ on
the electron sheets.  Furthermore, Muschler \textit{et al} showed that in this polarization excitations were present down to the lowest measurement frequency, indicating nodes or deep gap minima less than their resolution of order 10 cm$^{-1}$. Strong in-plane anisotropy of the $B_{1g}$ peak was also reported across a wider range of dopings in \cite{Chauviere}.

The theory of electronic Raman scattering in the superconducting state has been reviewed by Devereaux and Hackl \cite{DevereauxRamanReview}.
 An early discussion of the intensities to be expected in a two-band isotropic $s_\pm$ state \cite{ChubukovRaman} predicted a peak at $2\Delta_0$ and a resonance below $2\Delta_0$---analogous to the neutron spin resonance (Section \ref{subsec:resonance})---in the $A_{1g}$ channel. No peak or resonance was observed later in Muschler \textit{et al}. It was then pointed out by Boyd \textit{et al} \cite{BoydRaman1} that Coulomb backflow effects in the doped multiband system, which did not occur in \cite{ChubukovRaman}, would strongly suppress the $2\Delta_0$ peak. Boyd \textit{et al}, however, did not consider vertex corrections due to short-range interactions which are important for a formation of the resonance below $2\Delta_0$ \cite{ChubukovRaman} and a subgap resonance for $A_{1g}$ polarization may therefore still be possible.
The experimental situation in this channel is still controversial with the reported observation of a weak peak in BaFe$_{1.84}$Co$_{0.16}$As$_2$ \cite{Sugai}.

Analysis of the $B_{1g}$ channel by Muschler \textit{et al} \cite{Muschler} and Boyd \textit{et al} \cite{BoydRaman1} in terms of highly anisotropic $s_\pm$ states provides internally consistent evidence for order parameter nodes or deep minima on the electron pockets in the electron doped 122 system. But it was also argued that the gap on the electron pocket observed in the $B_{1g}$ channel can be strongly affected by disorder \cite{BoydRaman2}, which not only broadens the $2\Delta_0$ peaks but can lift the nodes, as apparently observed by Muschler \textit{et al} upon doping.

More recently, a more detailed analysis of the correct Raman vertices for these systems based on DFT was attempted by Mazin \textit{et al} \cite{MazinDevereaux}, who concluded that the earlier approximation for the $B_{1g}$ vertex (which weighted the entire electron pocket essentially equally) was too crude, and used the Muschler \textit{et al} data to argue that gap nodes or deep minima had to be present in the form of loops on the electron barrels circling the $\Gamma-\mathrm{X}$ axis. It is interesting to note that this identification is consistent with that of Yamashita \textit{et al} \cite{Yamashita} from angle-dependent magnetic field thermal conductivity measurements on the \BFAP~system, supporting the notion that the gap minima in the \BFCA~might deepen and evolve into nodes in the \BFAP~system.

\subsection{Alkali-intercalated iron selenide}
\label{subsubsec:KFS}
 As this review was being finalized, a new intriguing
FeBS material was discovered, challenging both theory and experiment with
its novel properties. As of now, this is still work-in-progress, and the
field remains very controversial. Some would argue that the subject is not
ripe for a review yet, and indeed it is too early to pass any judgement on the
superconducting mechanism, superconducting symmetry, or even physical
properties of this system. Nevertheless, the authors of this review are of
the opinion that it is worth to review here the preliminary results, both on
the experimental and theoretical sides, as a matter of a status report,
rather than an analytical review along the lines of the previous Sections.

\begin{figure}[th]
\begin{indented}
\item[]
(a)~\includegraphics[width=0.32\columnwidth]{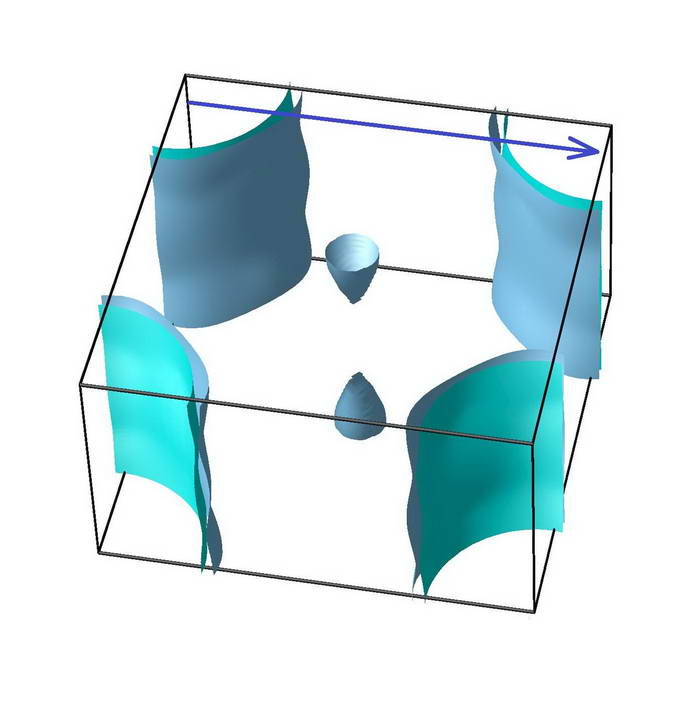}\\
(b)~\includegraphics[width=0.32\columnwidth]{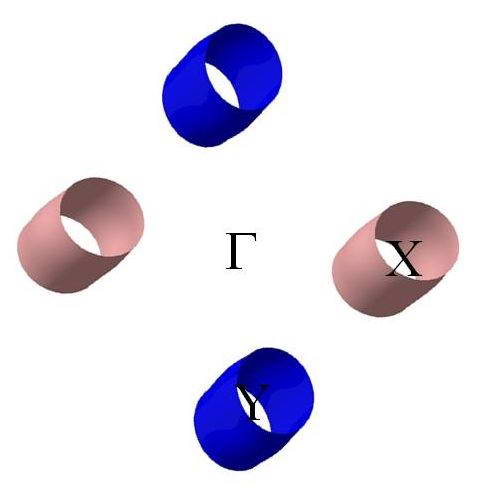}
(c)~\includegraphics[width=0.32\columnwidth]{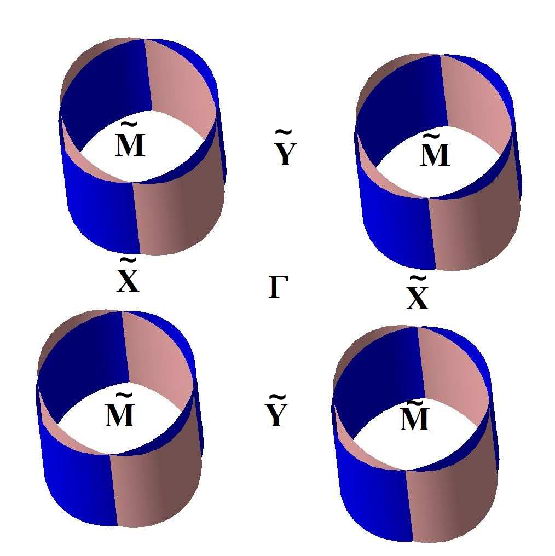}
\caption{(a) Fermi surface of K$_{0.8}$Fe$_{2}$Se$_{2}$ (from \protect\cite{trend}). (b) Cartoon showing a generic 3D Fermi surface for an AFe$_{2}$Se$_{2}$ material in the unfolded (one Fe/cell) Brillouin zone. Different colors show the signs of the order parameter in a nodeless $d$-wave state, allowed in the unfolded zone. The $\Gamma $ point is in the center (no Fermi surface pockets around $\Gamma $), and the electron pockets are around the $\mathrm{X}$, $\mathrm{Y}$ points. (c) Same as (b), but assuming a finite ellipticity (still zero $k_{z}$ dispersion). Different colors show the signs of the order parameter in a $d$-wave state. Wherever the two colors meet, turning on hybridization due to the Se potential creates nodes in the order parameter.}
\label{fig:fig2lbl}
\end{indented}
\end{figure}

In November 2010 a new superconductor, believed at that time to have the
chemical formula of K$_{0.8}$Fe$_{2}$Se$_{2}$, was reported, with a maximum
critical temperature of 33K. The formal electron count makes this compound
electron-doped at the level of 0.4 $e$ per Fe, the same as Ba(Fe$_{0.6}$Co$%
_{0.4}$)$_{2}$, which is far beyond the superconducting dome of the Co-doped
system, and significantly past the level of doping at which the hole pockets completely
sink under the Fermi level. The calculated band structure (figure~\ref{fig:fig2lbl}(a)) shows no hole pockets at all, but rather a large electron pocket at the corner of the Brillouin zone ($\tilde{M}$) and a small
electron pocket at the center ($\Gamma )$. Several ARPES measurements were
reported within a few months \cite{firstreports}, all agreeing among
themselves on the Fermi surface shown in figure~\ref{fig:fig2lbl}(a). In
addition, some reported a uniform nodeless gap around the large FS pocket.
This, by itself, is rather inspiring. Indeed, in the absence of the central
hole pocket the basis for the $s_{\pm }$ pairing is essentially lost, and
the whole theory seems to be in need of revisiting. Such revisiting, which
we will discuss in more detail later, did come nearly immediately from
numerous groups \cite{GraserKFeSe, Lee, Balat, Saito}.

However, the next wave of experiments, probing the bulk of the samples, came
to different conclusions, not readily compatible with the results of
photoemission. It appears that the most accurate methods for determining the
exact composition in the bulk, such as neutron scattering, invariably yield
the so-called charge-balanced compositions, namely K$_{2x}$Fe$_{2-x}$Se$_{2}$%
, where each extra electron brought in by alkaline intercalation is
compensated by holes introduced by Fe vacancies \cite{charge_balance}.
Obviously, if that is the case, and one neglects all effects of vacancy
ordering, this compositions brings us back to the parent FeSe material, with
equal hole and electron FSs, favorable for the $s_{\pm }$ model and
completely inconsistent with the ARPES data. One can, as usual, ascribe the
difference to surface effects, but that would require the surface to be
stoichiometric in Fe, with 20\% Fe vacancies in the bulk, which seems unlikely.

\begin{figure}[th]
\includegraphics[width=0.9\columnwidth]{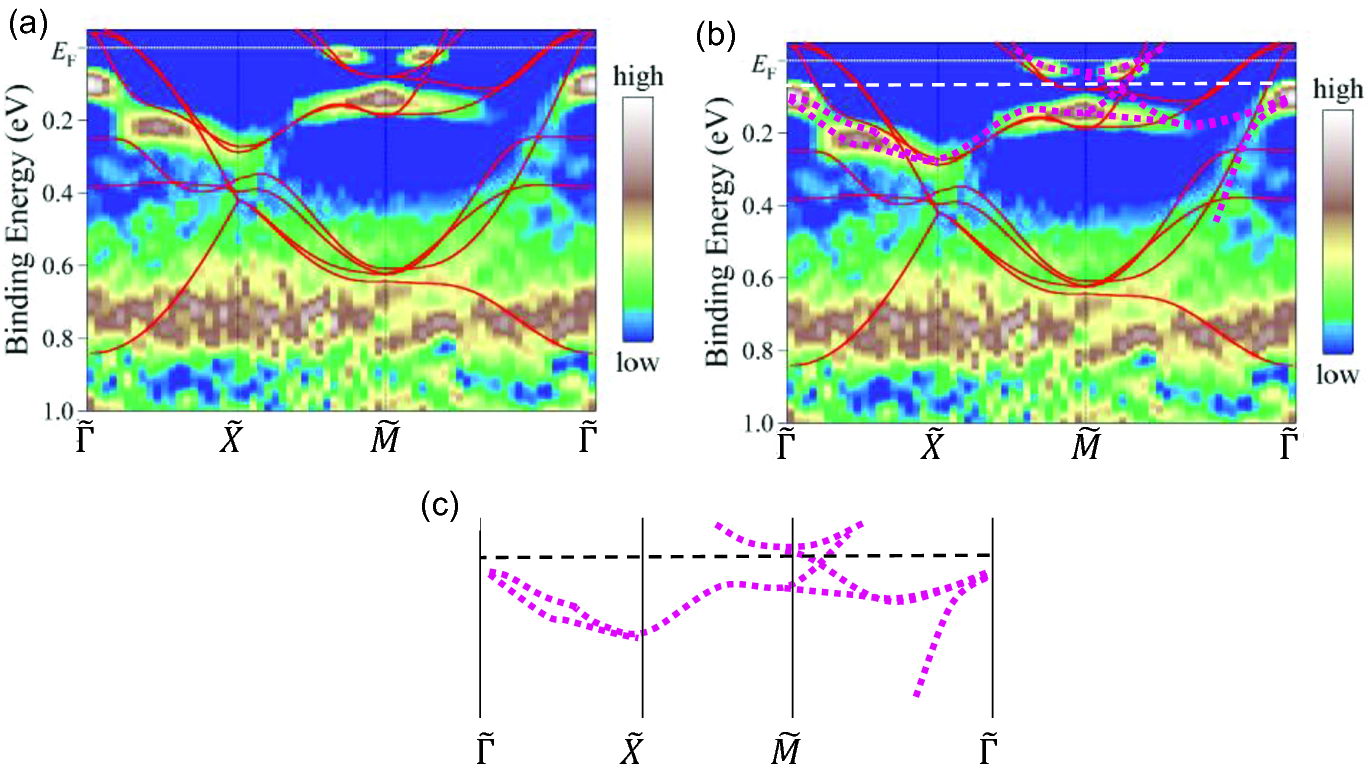}
\caption{Measured and calculated band structure of K$_{0}.8$Fe$_{1}.7$Se$_{2}
$. (a) ARPES data and DFT calculations from \protect\cite{HongDing}. (b) A
modification of the DFT bands according to the suggestion in \protect\cite%
{HongDing}. (c) Modified DFT bands alone. Note absence of a band gap between
the electron states at $\tilde{M}$ and hole states at $\Gamma $ points.}
\label{fig:ARPES_KFS}
\end{figure}

It was also suggested \cite{HongDing} that the ARPES-measured band
structures differs from DFT calculations in the sense that the hole bands
are strongly shifted down, and the electron bands up, so that the bottom of
the latter is above the top of the former. The authors of \cite{HongDing} claim that their actual composition is K$_{0.8}$Fe$_{1.7}$Se$_{2}$, corresponding
to 0.1 $e$/Fe doping, close to optimal doping in Ba(Fe,Co)$_{2}$As$_{2}$.
However, because of the assumed band shifts, they argue, the Fermi level
only crosses the electron bands. This explanation, however, is only valid
for this particular composition, and it does not seem very plausible that
this composition is special in any other way. Indeed, figure~\ref{fig:ARPES_KFS}(a) shows the data from \cite{HongDing}, overlaid with their own band structure calculations. The suggested band shifts are illustrated in figure~\ref{fig:ARPES_KFS}(b). As one can see, particularly in figure~\ref{fig:ARPES_KFS}(c) (where only the bands modified according to \cite{HongDing}'s suggestion are shown), if the composition is reduced to no
electron doping (K$_{0.8}$Fe$_{1.6}$Se$_{2}$), the Fermi level will shift
down and will cross a completely different band, neither the familiar hole
pockets near $\Gamma $, not the familiar electron pockets near $\tilde{M}$.

The surprises do not end there. Particularly stable appears the composition
with $x=0.4,$ K$_{0.8}$Fe$_{1.6}$Se$_{2},$ which can be also written as K$%
_{2}$Fe$_{4}$Se$_{5}$, suggesting a particular superstructure with a 5-fold
unit cell. Indeed such a superstructure was found \cite{vac},
and corresponds to Fe vacancies forming a $\sqrt{5}\times\sqrt{5}$
structure. The formula unit contains one vacancy and four Fe ions, forming a
square plaquette. Each plaquette is ordered ferromagnetically, forming a rigid ferromagnetic cluster. First principles calculations \cite{band_ins,CaoDai} have independently arrived at the same picture. Inside the plaquette the calculated Fe-Fe bonds are noticeably shorter than the average Fe-Fe distance \cite{band_ins}, also in agreement with the experiment \cite{vac}. Additionally, Yan \textit{et al} \cite{band_ins} have shown that shrinking of Fe-Fe bonds is not a magnetic effect, even though the ferromagnetic ordering benefits from such shrinking, but a covalent effect existing apart from Fe spin polarization \cite{band_ins}.
Moreover, both experiment and theory suggest an extremely large ordered magnetic moment on Fe, up to 3.3 $\mu_{B}$ in the experiment and even larger in the calculations. Note that this corresponds to a plaquette with a supermoment
of at least 13$\mu_B$. Several groups claim, nevertheless, that this ordered
magnetic state coexists microscopically with superconductivity.

Several considerations are in order here. This vacancy-ordered structure
corresponds to a lattice parameter of the order of 6\AA. The coherence
length has been measured \cite{Hc2} to be less than 60\AA. One can estimate
\cite{trend} that a net misalignment of the moments of the order of 0.05
degree will result in a net exchange field larger than $H_{c2}$.
It is hard to see how one can avoid this in real samples  with strain, grain boundaries, etc.. Furthermore,
calculations unambiguously show that the ideally ordered stoichiometric K$_{2}$Fe$_{4}$Se$_{5}$ is a band insulator with a large gap \cite{band_ins,CaoDai}. Experimental samples with this claimed composition are metallic, but an insulating phase has been found (and initially identified as a Mott insulator, although in view of the most recent data (see, e.g., \cite{optK122}) a band insulator appears more likely) nearby in the phase diagram; the metallicity of the ordered stoichiometric composition, in fact, may be an experimental artifact.

So now we face not only the fact that ARPES and bulk probes suggest
different valence states for Fe, we also seem to have high-temperature
superconductivity in a strongly magnetic insulator. An interesting, and
possibly correct explanation of this controversy was suggested recently in
\cite{haihu}. 
They suggest that although superconductivity and magnetism occur in the same
sample, and each involves nearly 100\% of carriers, they never occur
simultaneously. In short, the statement is that whenever vacancies are
disordered, the sample is superconducting, and whenever they order it
becomes antiferromagnetic, but not superconducting. 
The authors' claim that the vacancies reversibly order upon heating and disorder upon cooling sounds  counterintuitive. On the other hand, this picture was recently lent support by Li et al\cite{selenideSTM}, who found, using STM,
that areas of a K-intercalated Fe selenide with ordered Fe vacancies were
not superconducting, but areas closer to stoichiometric KFe$_2$Se$_2$ were.

One can say that in this contradictory experimental situation any
speculations are out of place. Yet, it is interesting to discuss what can
possibly happen in the superoverdoped regime that ARPES suggests. Indeed
this intriguing Fermi surface topology was already discussed in one of the
first theoretical papers, by Kuroki \textit{et al} \cite{k_kuroki_08} who
pointed out that in the absence of the hole pockets \textit{in the unfolded
BZ} this band structure is an ideal 2D representation of the
\textquotedblleft Agterberg-Barzykin-Gor'kov\textquotedblright\ gapless $d$-wave superconductivity \cite{ABG} [figure~\ref{fig:fig2lbl}(b)]. Indeed, in this case the quasi-nesting between the hole and electron pockets is
supplanted by the quasi-nesting between the electron pockets, resulting in a
$d$-wave state with alternating signs of the order parameters, while the
symmetry-required nodal lines fall between the Fermi surfaces. After the
first ARPES experiments on the Se-based 122 systems appeared, several
theoretical groups have revisited this idea \cite{GraserKFeSe,Lee}.

However, the original Agterberg \textit{et al} paper unambiguously identifies
possible locations of the zone-corner pockets allowing for a gapless $d$-wave superconductor. These are: $(\pi,0)$ and equivalent points in a
tetragonal symmetry, and $(\pi,0,0)$ and equivalent in a cubic one. This is
the case in the unfolded BZ, but in the folded zone the pockets are at the $(\pm \pi, \pm \pi)$ points, connected among themselves by reciprocal
lattice vector. The pockets in question are formed by shifting the unfolded
FSs by $(\pi, \pi, \pi)$, and overlapping the resulting pockets
(figure~\ref{fig:fig2lbl}). As discussed in details in \cite{Parker,MazinSe}, this unavoidably leads to node formation. At the same time, not only ARPES
data, but also various bulk probes \cite{Se122_SH,Se122_NMR} indicate that
superconductivity in K$_{x}$Fe$_{y}$Se$_{2}$ is nodeless.
Since the nodes in this case are driven by the hybridization of the two
unfolded FSs, the phase space for quasiparticle excitations associated with
such nodes is less than generic $d$ wave nodes if the hybridization gap is
small (\textit{cf.} \cite{Parker}). Roughly, the phase space affected is
reduced compared to $d$ wave as $V/\Delta E$, where $V$ is the matrix
element of the symmetry-breaking Se potential and $\Delta E\simeq
v_{F}\delta k$, where $\delta k$ is the ellipticity of the electron FS
pocket. While in actual DFT calculations it appears that $V\approx \Delta E$, and ARPES seems to show small ellipticity as well, one cannot exclude at
this stage a possibility that $V/\Delta E$ is considerably smaller than one,
in which case spectroscopic signatures of a $d$ state with no nodes enforced
by tetragonal symmetry but weak ones imposed by hybridization of the two
electron bands will be much harder to detect.

\begin{figure}[th]
\begin{indented}
\item[]
\includegraphics[width=0.83\columnwidth]{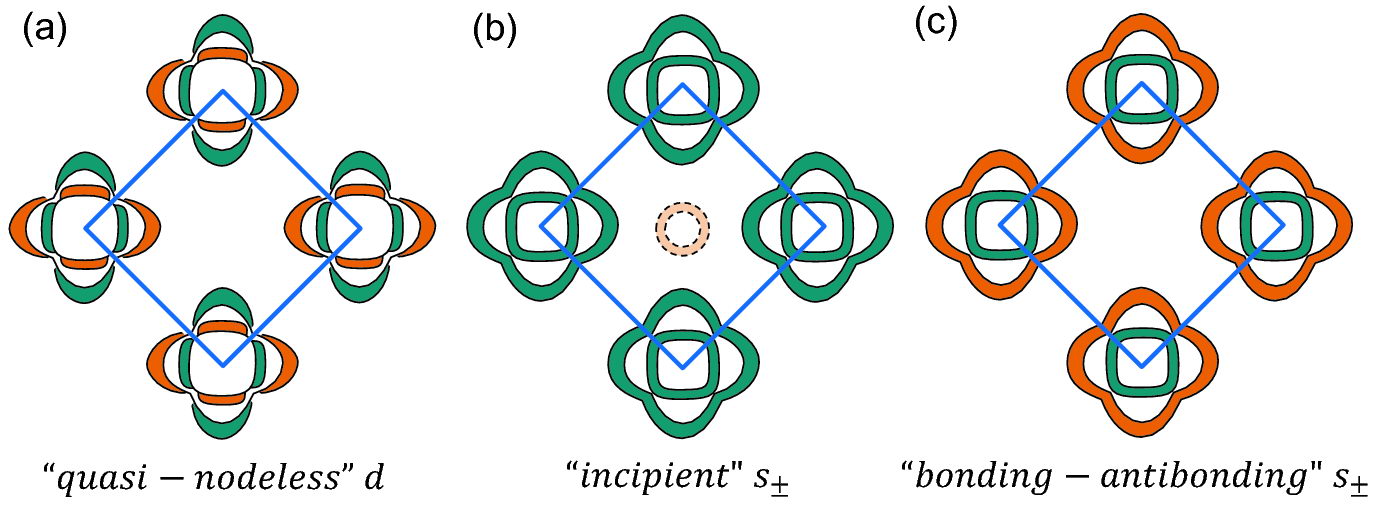}
\caption{This cartoon shows three proposed pairing states for K$_x$Fe$_{2-y}$Se$_2$ in the 2-Fe Brillouin zone. As suggested by the first
principles calculations, a finite gap between the inner and the
outer Fermi surface sheets is introduced. (a) $d$-wave state, including small parts of the Fermi surface where the gap is small; (b) the ``incipient'' $s_\pm$ state, with hole bands in proximity of the Fermi level, but not crossing it; (c) the ``bonding-antibonding'' $s_\pm$ state. Note that (a) and (c), but not (b), can give rise to a spin resonance at $(\pi,\pi)$ (in the unfolded Brillouin zone).}
\label{fig:new_FeSe}
\end{indented}
\end{figure}

At the present writing, at least three qualitatively different proposals exist. One is, as mentioned, a ``quasi-nodeless'' $d$-wave (nodes only induced by hybridization) (figure \ref{fig:fig2lbl}). The second is, essentially, the same $s_\pm$ as in other materials, but with the ``minus'' pairs formed not on the Fermi surface but on the hole band 50-100 meV below the Fermi level (the problem here is that superconductivity appears only in higher orders in the coupling constant) \cite{Lee,Chubukov-unpublished}; such a state would look for most practical purposes as regular same-sign s-wave, so it can be called ``incipient $s_{\pm }$''. The third state that allows for a sign-changing order parameter is an $s_\pm$ one, but entirely different form the one discussed throughout this review.   Rather, this is an $s_\pm$ state that was discussed more than a decade ago in connection with bilayer cuprates.  Here one refers to the fact that in a bilayer system every band is split in a bonding and an antibonding combination, and these may have different signs of the order parameter. Since DFT calculations predicts that the two electron-pockets in K$_{x}$Fe$_{y}$Se$_{2}$ are strongly hybridized, over most of the Fermi surface the calculations predict similar bonding-antibonding splitting and a possibility of a strictly-nodeless sign-changing s-wave superconductivity \cite{MazinSe}. The three different types of states are summarized in figure~\ref{fig:new_FeSe}.

One of the latest experimental developments relevant to the the order parameter in  K$_{x}$Fe$_{y}$Se$_{2}$ is a recent inelastic neutron scattering measurement \cite{InosovFeSe}. In agreement with ARPES-measured band structure, these authors did not find any peak around the $(\pi, 0)$ wave vector, indicating the absence of the conventional electron--hole nesting. In agreement with theoretical expectation \cite{GraserKFeSe}, there is not much  scattering at exactly $(\pi, \pi)$, even though this is the vector of nearly exact electron--electron nesting.  The reason is that the real part of the noninteracting spin-susceptibility is large when the Fermi velocities of the initial and the final states are opposite, and the real part controls the Stoner enhancement of the full susceptibility. Thus, a peak in susceptibility is expected when the FSs displaced by the given momentum just touch; if the radius of the electronic FSs in  K$_{x}$Fe$_{y}$Se$_{2}$ is $k_{F}$, then a peak in the neutron scattering is expected near $Q=(\pi/a, \pi/a)-(k_{F},k_{F})$. Actual calculations \cite{GraserKFeSe} show that due to the somewhat squarish shape of the FS the peak appears to be asymmetric and located at ($\pi, 0.625\pi)$ (for $0.1e$ doping). Experimentally, a peak is observed at ($\pi, \pi/2)$, not far from this predicted position, and found to be resonantly enhanced below $T_{c}$. The latter fact indicates that this wave vector connects two points on the FS, and these points have order parameters of the opposite signs, consistent, in principle, with the ``quasi-nodeless'' $d$-wave or with the bonding-antibonding $s_\pm$, but not with the ``incipient s$_{\pm }$''.

A second look, however, reveals that this straightforward interpretation may
be too naive. Indeed, the FS suggested by ARPES has by far too small
electron pockets to provide any states removed from each other by ($\pi,\pi/2)$. One either needs to assume that the Fe content is grossly
underestimated in \cite{InosovFeSe}, or that sizeable hole pockets are present (in which cases a feature at $q \approx (\pi,0)$ would be expected). Combined with the complex antiferromagnetic structure found in the same compound, and
possibly (even though unlikely) coexisting with superconductivity, one may
think that this antiferromagnetic structure, either present or incipient in
the superconducting state, may drastically change the physics of
superconductivity in these materials. In this regard, resolving the mystery
of interplay and possible coexistence between magnetism and superconductivity in these selenides is the most burning issue at this writing.

\subsection{Differences among materials: summary.}

\label{subsec:materials_summary}
In the above discussion we have focussed primarily on the heterovalently doped \BFA~materials, on which most experiments have been performed to date, and argued that thermodynamic properties tend to support fully gapped behavior near optimal doping, and increasing
anisotropy and eventual nodal behavior away from it, consistent with expectations from
a spin fluctuation picture.
The 122 family may indeed be representative of Fe-based superconductors, but it is worth noting that it is among the most weakly correlated families, and the most 3D as well.  Therefore it is important to review some of the other families, in the hope that new aspects of the physics may be gleaned from the comparison.  In Table~\ref{table:summary}, we have tried to
group materials into three categories, according to whether they display---in bulk thermodynamic properties alone---a large gap (at least several meV), deep gap minima or weak nodes,
or clear nodal behavior.    The middle category is clearly somewhat delicate, since it contains materials which have in some cases been explored extensively, whereas others have received less experimental attention.  In addition, the distinction between deep gap minima and
weak nodes (situations where the gap changes sign only over a very small part of the
Fermi surface) is not a clear one, since in A$_{1g}$ symmetry, a gap with deep minima   can be adiabatically transformed into a one with weak nodes, and vice versa,    with small
changes in electronic structure, disorder, etc.  Note we have deliberately omitted
AFe$_2$Se$_2$ from the Table due to its unclear materials properties and incomplete experimental information
on the superconducting state at this writing.

\begin{table}
\caption{\label{table:summary} Gap structures in Fe-based materials deduced from thermodynamic and transport measurements. OD=overdoped, OP=optimally doped, UD=underdoped.  Symbol * indicates possible evidence for ``$c$-axis nodes".}
\scriptsize
\begin{tabular}{@{}c|c|c|c}
\br
 Family & Full gap & Highly anisotropic & Strong nodal \\
\mr
 1111 & PrFeAsO$_{1-y}$[52K] \cite{HashimotoPrFeAsO} & LaFeAs(O,F)[26K] \cite{LOFFApendepth} & LaFePO[6K] \cite{j_fletcher_09,hicks09,YamashitaLaFePO}\\
 & SmFeAs(O,F)[55K] \cite{MaloneSmFeAsO} & NdFeAs(O,F) \cite{LOFFApendepth}& \\
\mr
 122 & (Ba,K)Fe$_2$As$_2$[40K] \cite{Yashima,x_luo_09,bobowski2010,Reid2011K-doped}  & Ba(Fe,Co)$_2$As$_2$ [OD] \cite{Tanatar2010,Reid}* & KFe$_2$As$_2$ [4K] \cite{kfe2as2,Dong_K122}\\
 & Ba(Fe,Co)$_2$As$_2$ [OP,23K] \cite{Tanatar2010,BFCA_full_gap_pendepth} & Ba(Fe,Ni)$_2$As$_2$ \cite{MartinNidoped122}* & BaFe$_2$(As,P)$_2$[OP,31K] \cite{k_hashimoto_09,NakaiPdoped} \\
 && Ba(Fe,Co)$_2$As$_2$ [UD] \cite{Reid}* & (Ba,K)Fe$_2$As$_2$ [UD] \cite{Reid2011K-doped}\\
\mr
 111 & LiFeAs [18K] \cite{ProzorovLiFeAs,tanatar_2011} & & LiFeP [6K] \cite{Shibauchi_LiFeP_2011} \\
\mr
 11 &  & Fe(Se,Te) [27K] \cite{Zeng,FeSe_kappa} & \\
\br
\end{tabular}
\normalsize
\end{table}

In our discussion of other materials, we first focus on the isovalently doped \BFAP~system, which displays clear nodal behavior.  This material is interesting in several regards, including its remarkable ``quantum critical" normal state properties and its phase diagram vs P concentration remarkably similar to \BFCA~despite the isovalent nature of the substitution of P for As \cite{BFAP_normalstate}.  It has the highest critical temperature (31K at optimal doping) of the confirmed nodal superconductors.  ($T_c$'s for LaFePO
and \KFA~are 6K and 4K, respectively).  The fact that the three P-doped superconductors \BFAP, \LFPO, and \LFP~are nodal whereas their As-doped counterparts \BFKA, \BFCA, \LOFA, and \LFA~are either fully gapped or anisotropic is also striking. In the case of the \LOFA/\LFPO~comparison, the claim has been made by Kuroki \textit{et al} \cite{Kuroki_pnictogen_ht_2009} and Wang \textit{et al} \cite{FaWangLAFP} that the distinction arises from the lack of a third hole pocket in the \LFPO~band structure.    This does not appear to provide an explanation
in the case of \LFA~and \LFP, where both materials have a third hole ($\gamma$) pocket. On the other hand, the third  pocket is important because it allows $(\pi,0)$ processes to couple to the $d_{xy}$ states on the electron pockets, as discussed in  Section \ref{subsec:sf}. Hashimoto \textit{et al} \cite{Shibauchi_LiFeP_2011} pointed out that a preponderance of $d_{xy}$ orbital character on the  $\alpha_2$ hole pocket of LiFeAs (negligible in the LiFeP case) may play the same role, and  induces a more isotropic state. Other possibilities to explain differences between As- and P-based compounds include the effective local interactions on Fe orbitals, which might be different due to the local polarizabilities of the ligand As or P \cite{Sawatzky_09}.  Such effects should be accounted for in DFT calculations; however these do not show significant differences in static local interactions $U$,$U'$,$J$, and $J'$ for As- and P-based systems \cite{Imada}.

As emphasized in Table~\ref{table:summary}, the 1111 materials  based on As
are both gapped and possibly nodal; it was this discrepancy among measurements on
these early samples which led to the initial diversity of low-$T$ results and impression of nonuniversality of the superconducting state in these materials.  The substitution on
the rare earth site is known to affect the lattice constants, in particular the $c$-axis constant, which shrinks as $T_c$ grows.  The concomitant changes in electronic structure have been ascribed primarily to the pnictogen height by Kuroki \textit{et al} \cite{Kuroki_pnictogen_ht_2009}, who parameterized them in an attempt to explain the sequence of $T_c$'s in this family.  It
is clearly significant that the more isotropic materials have the highest $T_c$'s, as would generally be expected from the spin fluctuation theory argument, but better crystals need to be made and other experiments performed to confirm the assignments given here.

 In the past year, a great deal of progress has been made in preparing high quality single crystals of LiFeAs ($T_c$=17K).
Since these crystals cleave well at the nonpolar Li surface, one might expect that surfaces would be excellent, and
recent unpublished STM work with atomic resolution indeed shows beautiful surfaces \cite{STMLiFeAs}.  No surface states near the
Fermi level are expected on the basis of DFT calculations
\cite{HEschrigLiFeAs}. But the earliest ARPES experiment on this system, by Borisenko \textit{et al} \cite{borisenko_2010}, reported a Fermi surface very different from the rather conventional set of hole and electron pockets predicted by DFT, in particular less clear nesting of hole and electron pockets.  More recently, on the other hand, de Haas-van Alphen measurements \cite{Putzke} showed good agreement with bulk DFT.

A second controversy relates to the spin excitations of the material.  While early NMR work reported a strongly temperature dependent Knight shift and $1/T_1$ below $T_c$, consistent with $s$-wave pairing \cite{Li_etal_LiFeAs_NMR}, this has been challenged recently by Baek \textit{et al} \cite{Buechner_LiFeAsNMR}, who report a Knight shift in some magnetic field directions with no $T$-dependence and claim consistency with recent theoretical analysis proposing triplet pairing for this system \cite{vdBrink_triplet}.
While the existence of a spin resonance does not definitively exclude triplet pairing \cite{Morr}, the LiFeAs Fermi surface seems much more likely to support a resonance in an $s_\pm$ state with wave vector $(\pi,0)$ \cite{Taylor_neutron_LiFeAs}.
These debates clearly need to be settled before it can be decided if LiFeAs fits into the usual framework discussed above, with pairing driven by spin fluctuations.  What is fairly certain phenomenologically is that this system has a full gap, as determined by various superconducting state measurements \cite{LiFeAspendepth, borisenko_2010, tanatar_2011, STMLiFeAs, Shibauchi_LiFeP_2011}.

A final system which deserves special comment is the strongly hole-doped compound KFe$_2$As$_2$, with a $T_c$ of 4K,
which may be considered to be the end point of the \BFKA~series with Ba entirely replaced by K.  The Fermi surface
calculated by DFT shows no electron pockets, and experiments have reported large quasiparticle densities at low $T$
in the superconducting state, consistent with gap nodes \cite{kfe2as2,Dong_K122,HashimotoK122}.  From a theoretical
standpoint, this material is particularly interesting because it allows one to ask what the subleading spin-fluctuation
pairing channel is when the pair interaction between electrons and holes no longer drives superconductivity.  An fRG study
by Thomale \textit{et al} \cite{ThomaleK122} found $d$-wave pairing, which was discussed in terms of  scattering between
$\Gamma$-centered and $\mathrm{M}$-centered pockets (1-Fe zone) yielding strong spin fluctuations at $\q=(\pi,\pi)$. Maiti \textit{et al} \cite{Maiti}
discussed this system in a more general context, studying the evolution of pairing within the RPA, and noted the decay of the $s_\pm$ channel relative to the $d$ channel as the system was doped in either direction. They also found a leading $d$-wave pairing instability for KFe$_2$As$_2$.
However, a recent study \cite{Forgan} of the vortex lattices in KFe$_2$As$_2$ claimed the results inconsistent with an in-plane anisotropy of the order parameter (including a $d$-wave state), and invoking a \textit{horizontal} nodal line instead. Lacking more direct probes of the gap anisotropy, though, the controversy about the possible $d$-wave pairing in KFe$_2$As$_2$ remains.

\section{Conclusions.}
\label{sec:conclusions}


In the past year, a confusing variety of experiments on the superconducting state of FeBS indicating a much wider diversity of gap structure among materials in the same family than expected have, with the help of theory, been classified in such a way that most variations can be qualitatively understood. The general hypothesis, which still requires further experimental and theoretical underpinning, is that in a ``typical'' FeBS, a spin fluctuation interaction between like orbitals on nearly nested hole and electron Fermi surfaces leads to an $s$-wave state which changes sign between these sheets.  At optimal doping,  the condensation energy obtained from such an interaction is maximal and there is a true spectral gap; the extent to which the overall gap is still somewhat anisotropic depends on details of the system, including whether three or two hole pockets are present.  Hole-doped systems are, within this picture, generally expected to have higher $T_c$ and be more anisotropic relative to their electron-doped counterparts.   In both cases, however, as one dopes the system away from optimal, the relative importance of subdominant interactions, in particular the intraband Coulomb interaction and pair scattering between the electron-like Fermi surface sheets, increases and frustrates the isotropic $s_\pm$ interaction, leading to anisotropy and eventually nodes.  There is considerable experimental evidence to support this picture in the As-based 122 family.

The $s$-wave ($A_{1g}$) nature of the superconducting allows for any distortion of the gap structure consistent with invariance under the operations of the tetragonal group; in particular nodes can be created or lifted without any thermodynamic singularities (except weak ones at $T=0$). Thus it is perhaps not a surprise that on the one hand the P-doped Ba-122 system exhibits a phase diagram  which resembles strongly that of the Co-doped Ba-122 system, yet the gap remains nodal across the doping range.  We have reviewed evidence in favor of deep gap minima on the electron pocket in the Co-doped system, which can easily shift to create nodes. The microscopic origin of this tendency is still unknown, but we have pointed to the striking tendency of the P-doped materials to be more anisotropic than their FeAs analogs, representing an obvious challenge to microscopic theory.

With the above explanation of the ``nonuniversality'' of the gap structure, the obvious question is whether one can develop an intuitive understanding of what physical principles are at work controlling the degree of anisotropy and, of course, the size of $T_c$, without having to perform a full-blown theoretical calculation for each material. One point emphasized early on by Kuroki and co-workers was the importance of the third, $d_{xy}$-dominated hole pocket, whose influence we noted already above.  Their observation that this feature of the electronic structure is controlled by the pnictogen height worked well within the 1111 family,
but appears not to apply directly in the 122 systems.  This may be because of the stronger 3D character of these systems. While few fully 3D theoretical calculations have been attempted thus far due to technical limitations, it is clear that the strong $k_z$ dispersion of certain bands forces admixtures of new bands near the top of the Brillouin zone in the 122 systems, and that this in turn can force strong gap variations and possibly nodes.

One of the themes of this review is that new physics can be found with the addition of more bands and more orbitals. The recipe for the highest $T_c$ is not clear, but the intuition gained from extensive spin fluctuation calculations can be summarized as follows: at least in the general framework of FeBS-type band structure, it is advantageous to maximize the number of bands but minimize the number of orbitals present at the Fermi surface.  If variable orbital weight is present, it should be distributed such that large momentum intraorbital pair scattering processes should be available to the system.

One can also search for a general intuitive principle describing the effect of disorder on gap structure. While disorder will always reduce $T_c$, in complex multiband superconductors with gap anisotropy, we have discussed that disorder can have several different effects on gap structure--even diametrically opposite ones--depending on impurity scattering and band interaction parameters. We are therefore somewhat pessimistic that systematic disorder studies will in this case aid substantially in identifying order parameter symmetry and structure.

We close by recalling the ancient remark that our knowledge is but a white spot in a black vastness, and therefore the more we  learn, the larger is the circle that
separates us from the unknown, and correspondingly larger is also our awareness of the limits of our knowledge.  In this case much hard work by many experimentalists and theorists has gone into building up a plausible ``standard'' story of the FeBS with ``typical'' electronic structure. The recent discovery of the AFe$_2$Se$_2$ based materials has shown us that there is  probably further terra incognita for superconductivity yet to be explored even within this particular set of chemical elements and structures.  It is the grand challenge for theory to guide the search through this terrain.

\section*{Acknowledgments}

We would like to thank R. Arita, A.V. Chubukov, T.P. Devereaux, O. Dolgov, I. Eremin, R.M. Fernandes, S. Graser, H. Ikeda, A.F. Kemper, H. Kontani, K. Kuroki, T.A. Maier, Y. Matsuda, V. Mishra, K.A. Moler, R. Prozorov, D.J. Scalapino, T. Shibauchi, G.R. Stewart, I. Vekhter, W. Wang, Y. Wang, and H.-H. Wen for useful discussions. Partial support was provided by DOE DE-FG02-05ER46236 (PJH and MMFK) and NSF-DMR-1005625 (PJH). MMK acknowledges  support from RFBR (grant 09-02-00127), Presidium of RAS program ``Quantum physics of condensed matter'' N5.7, FCP Scientific and Research-and-Educational Personnel of Innovative Russia for 2009-2013 (GK P891 and GK 16.740.12.0731), and President of Russia (grant MK-1683.2010.2). PJH and MMK are grateful for the support of the Kavli Institute for Theoretical Physics and the Stanford Institute for Materials \& Energy Science during the writing of this work.


\section*{References}

\begin{thebibliography}{1}

\bibitem{Hosono} Y. Kamihara, T. Watanabe, M. Hirano, and H. Hosono, J. Am. Chem. Soc. \textbf{130}, 3296 (2008).

\bibitem{SadovskiiReview} M.V. Sadovskii, Physics -- Uspekhi \textbf{51}, 1201 (2008).
\bibitem{IvanovskiiReview} A.L. Ivanovskii, Physics -- Uspekhi \textbf{51}, 1229 (2008).
\bibitem{IzyumovReview} Yu.A. Izyumov, E.Z. Kurmaev, Physics -- Uspekhi \textbf{51}, 1261 (2008).
\bibitem{IshidaReview} K. Ishida, Y. Nakai, and H. Hosono, J. Phys. Soc. Jpn. \textbf{78}, 062001 (2009).
\bibitem{JohnstonReview} D.C. Johnston, Advances in Physics \textbf{59}, 803 (2010).
\bibitem{PaglioneReview} J. Paglione and R.L. Greene, Nature Physics \textbf{6}, 645, (2010).
\bibitem{LumsdenReview} M.D. Lumsden and A.D. Christianson, J. Phys.: Condens. Matter \textbf{22}, 203203 (2010).
\bibitem{WenReview} H.-H. Wen and S. Li, Annu. Rev. Condens. Matter Phys. \textbf{2}, 121 (2011).
\bibitem{StewartReview} G.R. Stewart, arXiv:1106.1618.

\bibitem{Sakakibara} H. Sakakibara, H. Usui, K. Kuroki, R. Arita, and H. Aoki, Phys. Rev. Lett. \textbf{105}, 057003 (2010).

\bibitem{GolubovResistivity} A.A. Golubov, O.V. Dolgov, A.V. Boris, A. Charnukha, D.L. Sun, C.T. Lin, A.F. Shevchun, A.V. Korobenko, M.R. Trunin, and V.N. Zverev, Pis'ma v ZhETF \textbf{94}, 357 (2011); arXiv:1011.1900.

\bibitem{Sawatzky_09} G. A. Sawatzky, I. S. Elfimov, J. van den Brink, and J. Zaanen, Europhys. Lett. \textbf{86}, 17006 (2009).
\bibitem{Nakamura} K. Nakamura, R. Arita, and H. Ikeda, Phys. Rev. B \textbf{83}, 144512 (2011).

\bibitem{Brouet} V. Brouet, M. Marsi, B. Mansart, A. Nicolaou, A. Taleb-Ibrahimi, P. Le F\'evre, F. Bertran, F. Rullier-Albenque, A. Forget, and D. Colson, Phys. Rev. B \textbf{80}, 165115 (2009).

\bibitem{Kuroki_pnictogen_ht_2009} K. Kuroki, H. Usui, S. Onari, R. Arita, and H. Aoki, Phys. Rev. B \textbf{79}, 224511 (2009).
\bibitem{Mizuguhci} Y. Mizuguhci, Y. Hara, K. Deguchi, S. Tsuda, T. Yamaguchi, K. Takeda, H. Kotegawa, H. Tou, and Y. Takano, Supercond. Sci. Technol. \textbf{23} 054013 (2010).
\bibitem{Kuchinskii} E.Z. Kuchinskii, I.A. Nekrasov, and M.V. Sadovskii, JETP Letters \textbf{91}, 518 (2010).

\bibitem{Koshelev} A. Glatz and A. E. Koshelev, Phys. Rev. B \textbf{82}, 012507 (2010).

\bibitem{Tanatar} E. Boaknin, M.A. Tanatar, J. Paglione, D. Hawthorn, F. Ronning, R.W. Hill, M. Sutherland, L. Taillefer, J. Sonier, S.M. Hayden, and J.W. Brill, Phys. Rev. Lett. \textbf{90}, 117003 (2003).

\bibitem{Mazin_Andersen} I.I.Mazin, O.K. Andersen, O. Jepsen, O.V. Dolgov, J. Kortus, A.A. Golubov, A.B. Kuz'menko, D. van der Marel, Phys. Rev. Lett \textbf{89}, 107002 (2002).

\bibitem{CarringtonKortus}A. Carrington, J.D. Fletcher, J.R.  Cooper, O.J. Taylor, L. Balicas, N.D. Zhigadlo, S.M. Kazakov, J. Karpinski, J.P.H. Charmant and J. Kortus, Phys. Rev. B \textbf{72}, 060507(2005).


\bibitem{Terashima_dHvA}  T. Terashima, N. Kurita, M. Tomita, K. Kihou, C.-H. Lee, Y. Tomioka, T. Ito, A. Iyo, H. Eisaki, T. Liang, M. Nakajima, S. Ishida, S.-i. Uchida, H. Harima, and S. Uji, arXiv:1103.3329.

\bibitem{Coldea_Carrington_dHvA} H. Shishido, A.F. Bangura, A.I. Coldea, S. Tonegawa, K. Hashimoto, S. Kasahara, P.M.C. Rourke, H. Ikeda, T. Terashima, R. Settai, Y. \={O}nuki, D. Vignolles, C. Proust, B. Vignolle, A. McCollam, Y. Matsuda, T. Shibauchi, and A. Carrington, Phys. Rev. Lett. \textbf{104}, 057008 (2010).

\bibitem{Ortenzi} L. Ortenzi, E. Cappelluti, L. Benfatto, and L. Pietronero, Phys. Rev. Lett. \textbf{103}, 046404 (2009).

\bibitem{IIMazin_Nature} I.I. Mazin, Nature \textbf{464}, 183 (2010).

\bibitem{SigristUeda} M. Sigrist and K. Ueda, Rev. Mod. Phys. \textbf{63}, 239 (1991).

\bibitem{SinghDu} D.J. Singh and M.-H. Du, Phys. Rev. Lett. \textbf{100}, 237003 (2008).
\bibitem{Mazin_etal_splusminus} I.I. Mazin, D.J. Singh, M.D.  Johannes, and M.-H. Du,  Phys. Rev. Lett. \textbf{101}, 057003 (2008).


\bibitem{QCP}  ``Density Functional Calculations near Ferromagnetic Quantum Critical Points", I. I. Mazin, D.J. Singh, and A. Aguayo, in \textit{Proceedings of the NATO ARW on Physics of Spin in Solids: Materials, Methods and Applications}, ed. S. Halilov, Kluwer, 2003. Also at arXiv:cond-mat/0401563.

\bibitem{Anisimov} V.I. Anisimov, E.Z. Kurmaev, A. Moewes, and I.A. Izyumov, Physica C \textbf{469}, 442 (2009).
\bibitem{Georges}  M. Aichhorn, L. Pourovskii, V. Vildosola, M. Ferrero, O. Parcollet, T. Miyake, A. Georges, and S. Biermann, Phys. Rev. B \textbf{80}, 085101 (2009); M. Aichhorn, S. Biermann, T. Miyake, A. Georges, and M. Imada, Phys. Rev. B \textbf{82}, 064504 (2010).

\bibitem{Haule} Z.P. Yin, K. Haule, and G. Kotliar, Nature Materials
    (2011) doi:10.1038/nmat3120; arXiv:1104.3454.

\bibitem{ChubukovReview} A.V. Chubukov, Physica C \textbf{469}, 640 (2009).

\bibitem{Eremin2010} I. Eremin and A. V. Chubukov, Phys. Rev. B \textbf{81}, 024511 (2010).

\bibitem{Korshunov2008} M.M. Korshunov and I. Eremin, Phys. Rev. B \textbf{78}, 140509(R) (2008).

\bibitem{Knolle2010} J. Knolle, I. Eremin, A.V. Chubukov, and R. Moessner, Phys. Rev. B \textbf{81}, 140506(R) (2010).

\bibitem{Raghu} S. Raghu, X.-L. Qi, C.-X. Liu, D.J. Scalapino, and S.-C. Zhang, Phys. Rev. B \textbf{77}, 220503(R) (2008).

\bibitem{PALee} P.A. Lee and X.-G. Wen, Phys. Rev. B \textbf{78}, 144517 (2008).
\bibitem{s_graser_08} S. Graser, T.A. Maier, P.J. Hirschfeld, and D.J. Scalapino, New. J. Phys. \textbf{11}, 025016 (2009).
\bibitem{Maier_anisotropy}  T. Maier, S. Graser, P.J. Hirschfeld, and D.J. Scalapino, Phys. Rev. B \textbf{79}, 224510 (2009).
\bibitem{KemperSelfEnergy}  A.F. Kemper, M.M. Korshunov, T.P. Devereaux, J.N. Fry, H-P. Cheng, and P.J. Hirschfeld, Phys. Rev. B \textbf{83}, 184516 (2011).
\bibitem{Daghofer} M. Daghofer, A. Nicholson, A. Moreo, and E. Dagotto, Phys. Rev. B \textbf{81}, 014511 (2010).

\bibitem{Brydon}  P.M.R. Brydon, M. Daghofer, and C. Timm, J. Phys.: Condens. Matter \textbf{23}, 246001 (2011).

\bibitem{k_kuroki_08} K. Kuroki, S. Onari, R. Arita, H. Usui, Y. Tanaka, H. Kontani, and H. Aoki, Phys. Rev. Lett. \textbf{101}, 087004 (2008).

\bibitem{Cao} C. Cao, P.J. Hirschfeld, and H.-P. Cheng, Phys. Rev. B \textbf{77}, 220506(R) (2008).

\bibitem{Kemper_sensitivity} A. Kemper, T. Maier, S. Graser, H.-P. Cheng, P.J. Hirschfeld, and D.J. Scalapino, New J. Phys. \textbf{12}, 073030 (2010).

\bibitem{Graser3D} S. Graser, A.F. Kemper, T.A. Maier, H.-P. Cheng, P.J. Hirschfeld, and D.J. Scalapino, Phys. Rev. B \textbf{81}, 214503 (2010).
\bibitem{Kuroki3D} K. Suzuki, H. Usui, and K. Kuroki, J. Phys. Soc. Jpn. \textbf{80}, 013710 (2011).


\bibitem{BerkSchrieffer} N.P. Berk and J.R. Schrieffer, Phys. Rev. Lett. \textbf{17}, 433 (1966).

\bibitem{ScalapinoSFhistory} D.J. Scalapino, J. Low Temp. Phys. \textbf{117}, 179 (1999).

\bibitem{ABG} D.F. Agterberg, V. Barzykin, and L.P. Gor'kov, Phys. Rev. B \textbf{60}, 14868 (1999).

\bibitem{BulutScalapino}N. Bulut, D.J. Scalapino, and R.T. Scalettar, Physical Review B 45, 5577 (1992)

\bibitem{Mazin_bilayer} A.I. Liechtenstein, I.I. Mazin and O.K. Andersen, Phys. Rev. Lett. 74, 2303 (1995).

\bibitem{AS} A.G. Aronov and E.B. Sonin, Zh. Eksp. Teor. Fiz. \textbf{63}, 1059 (1972) [Sov. Phys.-JETP \textbf{36}, 556 (1973)].

\bibitem{Emery} V.J. Emery, Ann. Phys., NY, \textbf{28}, 1 (1964).


\bibitem{ScalapinoHF}  D.J. Scalapino, E. Loh, Jr., and J.E. Hirsch, Phys. Rev. B \textbf{34}, 8190 (1986).
\bibitem{VarmaHF}  K. Miyake, S. Schmitt-Rink, and C.M. Varma, Phys. Rev. B \textbf{34}, 6554 (1986).

\bibitem{ScalapinoPhysRep} D.J. Scalapino, Phys. Rep. \textbf{250}, 329 (1995).
\bibitem{Nakajima}  S. Nakajima, Prog. Theor. Phys. \textbf{50}, 1101 (1973).

\bibitem{Seoetal} K. Seo, B. Andrei Bernevig, and J. Hu, Phys. Rev. Lett. \textbf{101}, 206404 (2008).

\bibitem{Ueda_etal} T. Takimoto, T. Hotta, and K. Ueda, Phys. Rev. B \textbf{69}, 104504 (2004).
\bibitem{Kubo} K. Kubo, Phys. Rev. B \textbf{75}, 224509 (2007).
\bibitem{n_bickers_89} N.E. Bickers and D.J. Scalapino, Ann. Phys. (N.Y.) \textbf{193}, 206 (1989).

\bibitem{Chubukov_nodal-gapped} A.V. Chubukov, M.G. Vavilov, and A.B. Vorontsov, Phys. Rev. B \textbf{80}, 140515 (2009).
\bibitem{Thomale_nodal-gapped} R. Thomale, C. Platt, J. Hu, C. Honerkamp, and B.A. Bernevig, Phys. Rev. B \textbf{80}, 180505 (2009).
\bibitem{Thomale_nodal-gapped2} R. Thomale, C. Platt, W. Hanke, and B.A. Bernevig, Phys. Rev. Lett. \textbf{106}, 187003 (2011).

\bibitem{FaWangLAFP} F. Wang, H. Zhai, and D.-H. Lee, Phys. Rev. B \textbf{81}, 184512 (2010).

\bibitem{Ikeda_2008} H. Ikeda, J. Phys. Soc. Jpn. \textbf{77}, 123707 (2008).
\bibitem{Ikeda_2009} R. Arita and H. Ikeda, J. Phys. Soc. Jpn. \textbf{78}, 113707 (2009).
\bibitem{FernandesFLEX} J. Zhang, R. Sknepnek, R.M. Fernandes, and J. Schmalian, Phys. Rev. B \textbf{79}, 220502(R) (2009).
\bibitem{Ikeda_2010A} H. Ikeda, R. Arita, and J. Kunes, Phys. Rev. B \textbf{81}, 054502 (2010).
\bibitem{Ikeda_2010B} H. Ikeda, R. Arita, and J. Kunes, Phys. Rev. B \textbf{82}, 024508 (2010).

\bibitem{Shankar} R. Shankar, Rev. Mod. Phys. \textbf{66}, 129 (1994).
\bibitem{DHLee_2008} F. Wang, H. Zhai, Y. Ran, A. Vishwanath, and D.-H. Lee, Phys. Rev. Lett. \textbf{102}, 047005 (2009).
\bibitem{Honerkamp}  C. Honerkamp, M. Salmhofer, N. Furukawa, and T.M. Rice, Phys. Rev. B. \textbf{63}, 035109 (2001).
\bibitem{DHLee_2009} F. Wang, H. Zhai, and D.-H. Lee, Europhys. Lett. \textbf{85}, 37005 (2009).
\bibitem{Thomale_K122} R. Thomale, C. Platt, W. Hanke, J. Hu, and B.A. Bernevig, Phys. Rev. Lett. \textbf{107}, 117001 (2011).
\bibitem{Chubukovetal} A.V. Chubukov, D.V. Efremov, and I. Eremin, Phys. Rev. B \textbf{78}, 134512 (2008).
\bibitem{Tesanovic_2008} V. Cvetkovic and Z. Tesanovic, Phys. Rev. B \textbf{80}, 024512 (2009).
\bibitem{Maiti} S. Maiti, M.M. Korshunov, T.A. Maier, P.J. Hirschfeld, and A.V. Chubukov, Phys. Rev. Lett. \textbf{107}, 147002 (2011).

\bibitem{SiAbrahams} Q. Si and E. Abrahams, Phys. Rev. Lett. \textbf{101}, 076401 (2008).
\bibitem{CFang} C. Fang, H. Yao, W.-F. Tsai, J. Hu, and S.A. Kivelson, Phys. Rev. B \textbf{77}, 224509 (2008).

\bibitem{Sachdev_orborder}  C. Xu, M. Muller, S. Sachdev, Phys. Rev. B \textbf{78}, 020501(2008).
\bibitem{Goswami} P. Goswami, P. Nikolic, and Q. Si, Europhys. Lett. \textbf{91}, 37006 (2010).

\bibitem{Daghofer1} M. Daghofer, A. Moreo, J.A. Riera, E. Arrigoni, D.J. Scalapino, and  E. Dagotto, Phys. Rev. Lett. \textbf{101}, 237004 (2008).

\bibitem{Mazin2008} I.I. Mazin, M.D. Johannes, L. Boeri, K. Koepernik, and D.J. Singh, Phys. Rev. B \textbf{78}, 085104 (2008).

\bibitem{Yaresko} A.N. Yaresko, G.-Q. Liu, V.N. Antonov, and O.K. Andersen, Phys. Rev. B \textbf{79}, 144421 (2009).

\bibitem{Antropov} A.L. Wysocki, K.D. Belashchenko, and V.P. Antropov, Nature Physics \textbf{7}, 485 (2011).

\bibitem{Hund} I.I. Mazin and M.D. Johannes, Phys. Rev. B \textbf{79}, 220510(R) (2009); K. Haule and G. Kotliar, New J. Phys. \textbf{11} 025021 (2009).

\bibitem{Gunnarsson} O. Gunnarsson, Rev. Mod. Phys. \textbf{69}, 575 (1997).

\bibitem{MazinCohen}  I.I. Mazin and R.E. Cohen, Ferroelectrics \textbf{164}, 263 (1997).

\bibitem{Ginzburg_Kirzhnits}  V.L. Ginzburg and D.A. Kirzhnits (Eds) \textit{Problema Vysokotemperaturnoi Sverkhprovodimosti} (\textit{The Problem of High-Temperature Superconductivity}) (Moscow: Nauka, 1977) [Translated into English: \textit{High-Temperature Superconductivity} (New-York: Consultants Bureau, 1982)].

\bibitem{Little} W. Little, Phys. Rev. \textbf{134}, A1416 (1964).

\bibitem{Ginzburg} V. Ginzburg, Phys. Lett. \textbf{13}, 101 (1964).
\bibitem{Sawatzky_pairing}  M. Berciu, I. Elfimov, and G. A. Sawatzky, Phys. Rev. B \textbf{79}, 214507 (2009).

\bibitem{Kontani_Hubb_Holst}  H. Kontani and S. Onari, Phys. Rev. Lett. \textbf{104}, 157001 (2010).

\bibitem{Kontani_neutron}  S. Onari, H. Kontani, and M. Sato, Phys. Rev. B \textbf{81}, 060504(R) (2010).
\bibitem{Kontani_bond_angle} T. Saito, S. Onari, and H. Kontani, Phys. Rev. B \textbf{82}, 144510 (2010).

\bibitem{Zaanen_orborder}  F. Kruger, S. Kumar, J. Zaanen, and J. van den Brink, Phys. Rev. B \textbf{79}, 054504 (2009).
\bibitem{WeiKu_orborder}  C.-C. Lee, W.-G. Yin, and W. Ku, Phys. Rev. Lett. \textbf{103}, 267001 (2009).
\bibitem{Barzykin_orborder} V. Barzykin and L.P. Gor'kov, Phys. Rev. B \textbf{79}, 134510 (2009).
\bibitem{Imada} T. Miyake, K. Nakamura, R. Arita, and M. Imada, J. Phys. Soc. Jpn. \textbf{79}, 044705 (2010).

\bibitem{Yanagi}  Y. Yanagi, Y. Yamakawa, and Y. \={O}no, Phys. Rev. B \textbf{81}, 054518 (2010).
\bibitem{Matteo}  L. Boeri, M. Calandra, I.I. Mazin, O.V. Dolgov, and F. Mauri, Phys. Rev. B \textbf{82}, 020506(R) (2010).

\bibitem{Reznik} D. Reznik, K. Lokshin, D.C. Mitchell, D. Parshall, W. Dmowski, D. Lamago, R. Heid, K.-P. Bohnen, A.S. Sefat, M.A. McGuire, B.C. Sales, D.G. Mandrus, A. Subedi, D.J. Singh, A. Alatas, M.H. Upton, A.H. Said, A. Cunsolo, Yu. Shvydko, and T. Egami, Phys. Rev. B \textbf{80}, 214534 (2009).

\bibitem{Onari} S. Onari and H. Kontani, Phys. Rev. Lett. \textbf{103}, 177001 (2009).

\bibitem{Leeplot} C.-H. Lee, A. Iyo, H. Eisaki, H. Kito, M.T. Fernandez-Diaz, T. Ito, K. Kihou, H. Matsuhata, M. Braden, and K. Yamada, J. Phys. Soc. Jpn. \textbf{77}, 083704 (2008).


\bibitem{Suhl}  H. Suhl, B. T. Matthias, and L. R. Walker, Phys. Rev. Lett. \textbf{3}, 552 (1959).

\bibitem{Moskal}  V.A. Moskalenko, Fiz. Met. Metallov \textbf{8}, 503 (1959).

\bibitem{Kawabata} A. Kawabata, S.C. Lee, T. Moyoshi, Y. Kobayashi, and M. Sato, J. Phys. Soc. Jpn. \textbf{77}, 103704 (2008).
\bibitem{Sato} M. Sato, Y. Kobayashi, S.C. Lee, H. Takahashi, E. Satomi, and Y. Miura, J. Phys. Soc. Jpn. \textbf{79}, 014710 (2010).
\bibitem{SCLee} S.C. Lee, E. Satomi, Y. Kobayashi, and M. Sato, J. Phys. Soc. Jpn. \textbf{79}, 023702 (2010).
\bibitem{Tarantini} C. Tarantini, M. Putti, A. Gurevich, Y. Shen, R.K. Singh, J.M. Rowell, N. Newman, D.C. Larbalestier, P. Cheng, Y. Jia, H.-H. Wen, 	 Phys. Rev. Lett. \textbf{104}, 087002 (2010).

\bibitem{GraserTc} S. Graser, P.J. Hirschfeld, T. Dahm and L.-Y. Zhu, Phys. Rev. B \textbf{76}, 054516 (2007).

\bibitem{AG} A. Abrikosov and L. P. Gor'kov, Zh. Eksp. Teor. Fiz. \textbf{39}, 1781 (1960); Sov. Phys. JETP \textbf{12}, 1243 (1961).


\bibitem{Muzikar}  G. Preosti and P. Muzikar, Phys. Rev. B \textbf{54}, 3489 (1996).
\bibitem{Golubov} A.A. Golubov and I.I. Mazin, Phys. Rev. B \textbf{55}, 15146 (1997).
\bibitem{kulic} M.L. Kuli\'{c} and O.V. Dolgov, Phys. Rev. B \textbf{60}, 13062(1999).

\bibitem{Parker2008} D. Parker, O.V. Dolgov, M.M. Korshunov, A.A. Golubov, and I.I. Mazin, Phys. Rev. B \textbf{78}, 134524 {2008}.
\bibitem{Chubukovdisorder} A.V. Chubukov, D. Efremov, and I. Eremin, Phys. Rev. B \textbf{78}, 134512 (2008).

\bibitem{SengaKontani} Y. Senga and H. Kontani, J. Phys. Soc. Jpn. \textbf{77}, 113710 (2008); Y. Senga and H. Kontani, New J. Phys. \textbf{11}, 035005 (2009).
\bibitem{Bang_disorder} Y. Bang, H.-Y. Choi, and H. Won, Phys. Rev. B \textbf{79}, 054529 (2009).

\bibitem{JLTPreview}  P.J. Hirschfeld and W.A. Atkinson, J. Low Temp. Phys. \textbf{126}, 881 (2002).

\bibitem{Efremovdisorder} D.V. Efremov, M.M. Korshunov, O.V. Dolgov, A.A. Golubov, and P.J. Hirschfeld, arXiv:1104.3840.
\bibitem{Mishra_09} V. Mishra, G. Boyd, S. Graser, T.A. Maier, P.J. Hirschfeld, and D.J. Scalapino, Phys. Rev. B \textbf{79}, 094512 (2009).
\bibitem{Markowitz} D. Markowitz and L. P. Kadanoff, Phys. Rev. \textbf{131}, 563 (1963).
\bibitem{Balatskyreview} A.V. Balatsky, I. Vekhter, and J.-X. Zhu, Rev. Mod. Phys. \textbf{78}, 373 (2006).


\bibitem{Matsumoto2009} M. Matsumoto, M. Koga, and H. Kusunose: J. Phys. Soc. Jpn. \textbf{78}, 084718 (2009).
\bibitem{Tsai2009} W.-F. Tsai, Y.-Y. Zhang, C. Fang, and J. Hu, Phys. Rev. B \textbf{80}, 064513 (2009).
\bibitem{Ng2009} T. Ng and Y. Avishai, Phys. Rev. B \textbf{80}, 104504 (2009).
\bibitem{LiWang2009} J. Li and Y. Wang, Europhys. Lett. \textbf{88}, 17009 (2009).
\bibitem{DZhang2009} D. Zhang, Phys. Rev. Lett. \textbf{103}, 186402 (2009).
\bibitem{Akbari2010} A. Akbari, I. Eremin, and P. Thalmeier, Phys. Rev. B \textbf{81}, 014524 (2010).
\bibitem{TZhou2009} T. Zhou, X. Hu, J.-X. Zhu, and C.S. Ting, arXiv:0904.4273.
\bibitem{Ogata_1imp} T. Kariyado and M. Ogata, J. Phys. Soc. Jpn \textbf{79}, 083704 (2010).
\bibitem{Kemper_Co} A.F. Kemper, C. Cao, P.J. Hirschfeld, and H-P. Cheng, Phys. Rev. B \textbf{80}, 104511 (2009).
\bibitem{IkedaArita_imp} K. Nakamura, R. Arita, and H. Ikeda. Phys. Rev. B \textbf{83}, 144512 (2011).
\bibitem{Akbari_magnetic_imp} A. Akbari, P. Thalmaier, and I. Eremin, J. Supercond. Nov. Magn. \textbf{24}, 1173(2011).

\bibitem{MazinSchmalian} I.I. Mazin and J. Schmalian, Physica C \textbf{469}, 614 (2009).
\bibitem{Ole} O.K. Andersen and L. Boeri, Ann. der Phys. \textbf{523}, 8 (2011).

\bibitem{Ning} F. Ning, K. Ahilan, T. Imai, A.S. Sefat, R. Jin, M.A. McGuire, B.C. Sales, and D. Mandrus, J. Phys. Soc. Jpn. \textbf{77}, 103705 (2008).
\bibitem{Grafe} H.-J. Grafe, D. Paar, G. Lang, N.J. Curro, G. Behr, J. Werner,
    J. Hamann-Borrero, C. Hess, N. Leps, R. Klingeler, and B. B\"{u}chner,
    Phys. Rev. Lett. \textbf{101}, 047003 (2008).
\bibitem{Matano} K. Matano, Z.A. Ren, X.L. Dong, L.L. Sun, Z. X. Zhao, and G.-q. Zheng, Europhys. Lett. \textbf{83}, 57001 (2008).
\bibitem{MatanoBKFA} K. Matano, Z. Li, G.L. Sun, D.L. Sun, C.T. Lin, M. Ichioka, and G.-q. Zheng, Europhys. Lett. \textbf{87}, 27012 (2009).
\bibitem{Yashima} M. Yashima, H. Nishimura, H. Mukuda, Y. Kitaoka, K. Miyazawa, P.M. Shirage, K. Kihou, H. Kito, H. Eisaki, and A. Iyo, J. Phys. Soc. Japan \textbf{78}, 103702 (2009).
\bibitem{Jeglic111} P. Jegli\v{c}, A. Poto\v{c}nik, M. Klanj\v{s}ek, M. Bobnar, M. Jagodi\v{c}, K. Koch, H. Rosner, S. Margadonna, B. Lv, A.M. Guloy, and D. Ar\v{c}on, Phys. Rev. B \textbf{81}, 140511(R) (2010).
\bibitem{Li111} Z. Li, Y. Ooe, X.-C. Wang, Q.-Q. Liu, C.-Q. Jin, M. Ichioka, and G.-q. Zheng, J. Phys. Soc. Jpn. \textbf{79}, 083702 (2010).
\bibitem{NakaiPdoped} Y. Nakai, T. Iye, S. Kitagawa, K. Ishida, S. Kasahara, T. Shibauchi, Y. Matsuda, and T. Terashima, Phys. Rev. B \textbf{81}, 020503(R) (2010).


\bibitem{Mackenzie} A.P. Mackenzie and Y. Maeno, Rev. Mod. Phys. \textbf{75}, 657 (2003).
\bibitem{Amato2009} A. Amato, R. Khasanov, L. Luetkens, and H.-H. Klauss, Physica C \textbf{469}, 606 (2009).
\bibitem{Greene_tunneling} Y.S. Oh, Y. Liu, L.Q. Yan, K.H. Kim, R.L. Greene, and I. Takeuchi, Phys. Rev. Lett. \textbf{102}, 147002 (2009).

\bibitem{Geshkenbein_Meissner} V.B. Geshkenbein and A.I. Larkin, JETP Lett. \textbf{43}, 395 (1986).
\bibitem{KAM} C.W. Hicks, T.M. Lippman, M.E. Huber, Z.A. Ren, J. Yang, Z.X. Zhao, and K.A. Moler, J. Phys. Soc. Japan \textbf{78}, 013708 (2009).
\bibitem{Maier} T.A. Maier and D.J. Scalapino, Phys. Rev. B \textbf{78}, 020514(R) (2008).
\bibitem{Maier2} T.A. Maier, S. Graser, D.J. Scalapino, and P.J. Hirschfeld, Phys. Rev. B \textbf{79}, 134520 (200).
\bibitem{schrieffer} J.R. Schrieffer, \textit{Theory of superconductivity}, (Reading: Perseus), 1999.
\bibitem{MonthouxScalapino} P. Monthoux and D.J. Scalapino, Phys. Rev. Lett. \textbf{72}, 1874 (1994).
\bibitem{Inosov} D.S. Inosov, J.T. Park, P. Bourges, D.L. Sun, Y. Sidis, A. Schneidewind, K. Hradil, D. Haug, C.T. Lin, B. Keimer, and V. Hinkov, Nature Physics \textbf{6}, 178 (2010).
\bibitem{rossat} J. Rossat-Mignod, L.P. Regnault, C. Vettier, P. Bourges, P. Burlet, and J. Bossy, Physica C \textbf{185-189}, 86 (1991).
\bibitem{sato} N.K. Sato, N. Aso, K. Miyake, R. Shiina, P. Thalmeier, G. Varelogiannis, C. Geibel, F. Steglich, P. Fulde, and T. Komatsubara, Nature \textbf{410}, 340 (2001).
\bibitem{broholm} C. Stock, C. Broholm, J. Hudis, H.J. Kang, and C. Petrovic, Phys. Rev. Lett. \textbf{100}, 087001 (2008).
\bibitem{Qiu} Y. Qiu, M. Kofu, W. Bao, S.-H. Lee, Q. Huang, T. Yildirim, J.R.D. Copley, J.W. Lynn, T. Wu, G. Wu, and X.H. Chen, Phys. Rev. B \textbf{78}, 052508 (2008).
\bibitem{ChristiansonBKFA}  A.D. Christianson, E.A. Goremychkin, R. Osborn, S. Rosenkranz, M.D. Lumsden, C.D. Malliakas, I.S. Todorov, H. Claus, D.Y. Chung, M.G. Kanatzidis, R.I. Bewley, and T. Guidi, Nature \textbf{456}, 930 (2008).

\bibitem{Lumsden} M.D. Lumsden, A.D. Christianson, D. Parshall, M.B. Stone, S.E. Nagler, G.J. MacDougall, H.A. Mook, K. Lokshin, T. Egami, D.L. Abernathy, E.A. Goremychkin, R. Osborn, M.A. McGuire, A.S. Sefat, R. Jin, B.C. Sales, and D. Mandrus, Phys. Rev. Lett. \textbf{102}, 107005 (2009).
\bibitem{ChristiansonBFCA} A.D. Christianson, M.D. Lumsden, S.E. Nagler, G.J. MacDougall, M.A. McGuire, A.S. Sefat, R. Jin, B.C. Sales, and D. Mandrus, Phys. Rev. Lett. \textbf{103}, 087002 (2009).
\bibitem{Park} J.T. Park, D.S. Inosov, A. Yaresko, S. Graser, D.L. Sun, Ph. Bourges, Y. Sidis, Y. Li, J.-H. Kim, D. Haug, A. Ivanov, K. Hradil, A. Schneidewind, P. Link, E. Faulhaber, I. Glavatskyy, C.T. Lin, B. Keimer, and V. Hinkov, Phys. Rev. B \textbf{82} 134503 (2010).
\bibitem{Argyriou} D.N. Argyriou, A. Hiess, A. Akbari, I. Eremin, M.M. Korshunov, J. Hu, B. Qian, Z. Mao, Y. Qiu, C. Broholm, and W. Bao, Phys. Rev. B \textbf{81}, 220503(R) (2010).

\bibitem{Osborn} J.-P. Castellan, S. Rosenkranz, E.A. Goremychkin, D.Y. Chung, I.S. Todorov, M.G. Kanatzidis, I. Eremin, J. Knolle, A.V. Chubukov, S. Maiti, M.R. Norman, F. Weber, H. Claus, T. Guidi, R.I. Bewley, and R. Osborn, arXiv:1106.0771.

\bibitem{Lipscombe} O.J. Lipscombe, L.W. Harriger, P.G. Freeman, M. Enderle, C. Zhang, M. Wang, T. Egami, J. Hu, T. Xiang, M.R. Norman, and P. Dai, Phys. Rev. B \textbf{82}, 064515 (2010).

\bibitem{LiBKFA} Z. Li, D.L. Sun, C.T. Lin, Y.H. Su, J.P. Hu, and G.-q. Zheng, Phys. Rev. B \textbf{83}, 140506(R) (2011).

\bibitem{Eremin2002} I. Eremin, D. Manske, and K.H. Bennemann, Phys. Rev. B \textbf{65}, 220502(R) (2002).

\bibitem{Kuroki_recent_neutron} Y. Nagai and K. Kuroki, Phys. Rev. B \textbf{83}, 220516(R) (2011).

\bibitem{Kontani_recent_neutron} S. Onari and H. Kontani, arXiv:1105.6233.
\bibitem{QiuFeSeTe} Y. Qiu, W. Bao,Y. Zhao, C. Broholm, V. Stanev, Z. Tesanovic, Y.C. Gasparovic, S. Chang, J. Hu, B. Qian, M. Fang, and Z. Mao, Phys. Rev. Lett. \textbf{103}, 067008 (2009).
\bibitem{Babkevich} P. Babkevich, M. Bendele, A.T. Boothroyd, K. Conder, S.N. Gvasaliya, R. Khasanov, E. Pomjakushina, and B. Roessli, J. Phys.: Condens. Matter \textbf{22}, 142202 (2010).

\bibitem{InosovDeltaTc}  D.S. Inosov, J.T. Park, A. Charnukha, Y. Li, A.V. Boris, B. Keimer, and V. Hinkov, Phys. Rev. B \textbf{83}, 214520 (2011).
\bibitem{Yu} G. Yu, Y. Li, E.M. Motoyama, and M. Greven, Nature Phys. \textbf{5}, 873 (2009).
\bibitem{ParkerPRL} D. Parker and I.I. Mazin, Phys. Rev. Lett. \textbf{102}, 227007 (2009).
\bibitem{Wu} J. Wu and P. Phillips, Phys. Rev. B \textbf{79}, 092502 (2009).
\bibitem{Chen} C.-T. Chen, C.C. Tsuei, M.B. Ketchen, Z.-A. Ren, and Z.X. Zhao, Nature Physics \textbf{6}, 260 (2010).


\bibitem{WangLeeQPI} Q.H. Wang and D.-H. Lee, Phys. Rev. B \textbf{67}, 020511(R) (2003).
\bibitem{Capriotti} L. Capriotti, R.D. Sedgewick, and D.J. Scalapino, Phys. Rev. B \textbf{68}, 14508 (2003).
\bibitem{ZhuQPI} L. Zhu, W.A. Atkinson, and P. J. Hirschfeld, Phys. Rev. B \textbf{69}, 060503 (2004).
\bibitem{NunnerQPI} T.S. Nunner, W. Chen, B.M. Andersen, A. Melikyan, and P.J. Hirschfeld, Phys. Rev. B \textbf{73}, 104511 (2006).
\bibitem{HanaguriColeman} T. Hanaguri, Y. Kohsaka, M. Ono, M. Maltseva, P. Coleman, I. Yamada, M. Azuma, M. Takano, K. Ohishi, and H. Takagi, Science \textbf{323}, 923 (2009).
\bibitem{PeregBarneaFranz} T. Pereg-Barnea and M. Franz, Phys. Rev. B \textbf{78}, 020509 (2008). See also M. Maltseva and P. Coleman, Phys. Rev. B \textbf{80}, 144514 (2009).
\bibitem{early_QPI} Y.-Y. Zhang, C. Fang, X. Zhou, K. Seo, W.-F. Tsai, B.A. Bernevig, and J.P. Hu, Phys. Rev. B \textbf{80}, 094528 (2009).
\bibitem{Knolle} J. Knolle, I. Eremin, A. Akbari, and R. Moessner, Phys. Rev. Lett. \textbf{104}, 257001 (2010).
\bibitem{Akbari} A. Akbari, J. Knolle and I. Eremin, Phys. Rev. B \textbf{82}, 224506, (2010).
\bibitem{Davis_Ca122} T.-M. Chuang, M.P. Allan, J. Lee, Y. Xie, N. Ni, S.L. Bud'ko, G.S. Boebinger, P.C. Canfield, and J.C. Davis, Science \textbf{327}, 181 (2010).
\bibitem{MazinArgyriou} I.I. Mazin, S.A.J. Kimber, and D.N. Argyriou, Phys. Rev. B \textbf{83}, 052501 (2011).


\bibitem{Hanaguri2} T. Hanaguri, S. Niitaka, K. Kuroki, and H. Takagi, Science \textbf{328}, 474 (2010).

\bibitem{Mazin_Singh} I.I. Mazin and D.J. Singh, arXiv:1007.0047.

\bibitem{HanaguriReply} T. Hanaguri, S. Niitaka, K. Kuroki, H. Takagi, arXiv:1007.0307.

\bibitem{Bobroff} Y. Laplace, J. Bobroff, F. Rullier-Albenque, D. Colson, and A. Forget, Phys. Rev. B \textbf{80}, 140501(R) (2009).

\bibitem{Fernandes} R.M. Fernandes and J. Schmalian, Phys. Rev. B \textbf{82}, 014521 (2010).

\bibitem{Vorontsov} A.B. Vorontsov, M.G. Vavilov, and A.V. Chubukov, Phys. Rev. B \textbf{79}, 060508(R) (2009).

\bibitem{Bulaevskii} L.N. Bulaevskii, A.I. Rusinov, and M.L. Kulic, J. Low Temp. Phys. \textbf{39}, 256 (1980).

\bibitem{Overhauser} L.L. Daemon and A.W. Overhauser, Phys. Rev. B \textbf{39}, 6431 (1989).

\bibitem{Parker} D. Parker, M.G. Vavilov, A.V. Chubukov, and I.I. Mazin, Phys. Rev. B \textbf{80}, 100508(R) (2009).

\bibitem{Grossetal} F. Gross, B.S. Chandrasekhar, D. Einzel, P.J. Hirschfeld, K. Andres, H. R. Ott, J. Beuers, Z. Fisk, and J.L. Smith, Z. Physik B \textbf{64}, 175 (1986).
\bibitem{j_fletcher_09} J.D. Fletcher, A. Serafin, L. Malone, J.G. Analytis, J.-H. Chu, A.S. Erickson, I.R. Fisher, and A. Carrington, Phys. Rev. Lett. \textbf{102}, 147001 (2009).
\bibitem{hicks09} C.W. Hicks, T.M. Lippman, M.E. Huber, J.G. Analytis, J.-H. Chu, A.S. Erickson, I.R. Fisher, and K.A. Moler, Phys. Rev. Lett. \textbf{103}, 127003 (2009).

\bibitem{k_hashimoto_09} K. Hashimoto, M. Yamashita, S. Kasahara, Y. Senshu, N. Nakata, S. Tonegawa, K. Ikada, A. Serafin, A. Carrington, T. Terashima, H. Ikeda, T. Shibauchi, and Y. Matsuda, Phys. Rev. B \textbf{81}, 220501 (2010).
\bibitem{r_gordon_08} R.T. Gordon, C. Martin, H. Kim, N. Ni, M. A. Tanatar, J. Schmalian, I.I. Mazin, S. L. Bud'ko, P. C. Canfield, and R. Prozorov, Phys. Rev. B \textbf{79}, 100506(R) (2009).
\bibitem{CMartin_2010} C. Martin, H. Kim, R.T. Gordon, N. Ni, V.G. Kogan, S.L. Bud'ko, P.C. Canfield, M.A. Tanatar, and R. Prozorov, Phys. Rev. B \textbf{81}, 060505 (2010).
\bibitem{BFCA_full_gap_pendepth} L. Luan, T.M. Lippman, C.W. Hicks, J.A. Bert, O.M. Auslaender, J.-H. Chu, J.G. Analytis, I.R. Fisher, and K.A. Moler, Phys. Rev. Lett. \textbf{106}, 067001 (2011).

\bibitem{KoganMartinProzorov2009} V.G. Kogan, C. Martin, and R. Prozorov, Phys. Rev. B \textbf{80}, 014507 (2009).

\bibitem{Hashimoto2009BKFA} K. Hashimoto, T. Shibauchi, S. Kasahara, K. Ikada, S. Tonegawa, T. Kato, R. Okazaki, C. J. van der Beek, M. Konczykowski, H. Takeya, K. Hirata, T. Terashima, and Y. Matsuda, Phys. Rev. Lett. \textbf{102}, 207001 (2009).
\bibitem{kfe2as2} C. Martin, R.T. Gordon, M.A. Tanatar, H. Kim, N. Ni, S.L. Bud'ko, P.C. Canfield, H. Luo, H.-H. Wen, Z. Wang, A.B. Vorontsov, V.G. Kogan, and R. Prozorov, Phys. Rev. B \textbf{80}, 020501 (2009).
\bibitem{LiFeAspendepth} H. Kim, M.A. Tanatar, Y.J. Song, Y.S. Kwon, and R. Prozorov, Phys. Rev. B \textbf{83}, 100502(R) (2011).

\bibitem{SmFeAsOFpendepth} L. Malone, J.D. Fletcher, A. Serafin, A. Carrington, N.D. Zhigadlo, Z. Bukowski, S. Katrych, and J. Karpinski, Phys. Rev. B \textbf{79}, 140501(R) (2009).
\bibitem{LOFFApendepth} C. Martin, M.E. Tillman, H. Kim, M.A. Tanatar, S.K. Kim, A. Kreyssig, R.T. Gordon, M.D. Vannette, S. Nandi, V.G. Kogan, S.L. Bud'ko, P.C. Canfield, A.I. Goldman, and R. Prozorov, Phys. Rev. Lett. \textbf{102}, 247002 (2009).

\bibitem{Proz_Voront_disorder} H. Kim, R.T. Gordon, M.A. Tanatar, J. Hua, U. Welp, W.K. Kwok, N. Ni, S.L. Bud'ko, P.C. Canfield, A.B. Vorontsov, and R. Prozorov Phys. Rev. B \textbf{82}, 060518 (2010).
\bibitem{Moler} K.A. Moler, D.J. Baar, J.S. Urbach, R. Liang, W.N. Hardy, and A. Kapitulnik, Phys. Rev. Lett. \textbf{73}, 2744 (1994).

\bibitem{Kogan2009} V.G. Kogan, Phys. Rev. B \textbf{80}, 214532 (2009).

\bibitem{Volovik} G.E. Volovik, JETP Lett. \textbf{58}, 469 (1993).
\bibitem{Kuebert} C. K\"ubert and P.J. Hirschfeld, Sol. St. Commun. \textbf{105}, 459 (1998).
\bibitem{Mu2009} G. Mu, H. Luo, Z. Wang, L. Shan, C. Ren, and H.-H. Wen, Phys. Rev. B \textbf{79}, 174501 (2009).
\bibitem{Mu2008} G. Mu, X.Y. Zhu, L. Fang, L. Shan, C. Ren, and H.-H. Wen, Chin. Phys. Lett. \textbf{25}, 2221 (2008).
\bibitem{Gofryk} K. Gofryk, A.S. Sefat, E.D. Bauer, M.A. McGuire, B.C. Sales, D. Mandrus, J.D. Thompson, F. Ronning,	 New J. Phys. \textbf{12}, 023006 (2010).
\bibitem{StewartPdoped1} J.S. Kim, P.J. Hirschfeld, G.R. Stewart, S. Kasahara, T. Shibauchi, T. Terashima, and Y. Matsuda, Phys. Rev. B \textbf{81}, 214507 (2010).
\bibitem{StewartPdoped2} Y. Wang, J.S. Kim, G.R. Stewart, P.J. Hirschfeld, S. Graser, S. Kasahara, T. Terashima, Y. Matsuda, T. Shibauchi, and I. Vekhter, arXiv:1109.0554.
\bibitem{Gofryk2011} K. Gofryk, A.B. Vorontsov, I. Vekhter, A.S. Sefat, T. Imai, E.D. Bauer, J.D. Thompson, and F. Ronning, Phys. Rev. B \textbf{83}, 064513 (2011)
\bibitem{Vekhter1999} I. Vekhter, P.J. Hirschfeld, J.P. Carbotte, and E.J. Nicol,   Phys. Rev. B \textbf{59}, R9023 (1999).
\bibitem{VekhterVorontsov} A.B. Vorontsov and I. Vekhter, Phys. Rev. Lett. \textbf{96}, 237001 (2006); A.B. Vorontsov and I. Vekhter, Phys. Rev. B \textbf{75}, 224501 (2007).
\bibitem{Boyd} G.R. Boyd, A. Vorontsov, P.J. Hirschfeld, and I. Vekhter, Phys. Rev.  B \textbf{79}, 064525 (2009).
\bibitem{Tsuei} C.C. Tsuei and J.R. Kirtley, Rev. Mod. Phys. \textbf{72}, 969 (2000).
\bibitem{Graser_spht} S. Graser, G.R. Boyd, C. Cao, H.-P. Cheng, P.J. Hirschfeld, and D.J. Scalapino, Phys. Rev. B \textbf{77}, 180514 (2008).
\bibitem{Zeng} B. Zeng, G. Mu, H.Q. Luo, T. Xiang, I.I. Mazin, H. Yang, L. Shan, C. Ren, P.C. Dai, and H.-H. Wen, Nat. Commun. \textbf{1:112}, doi:10.1038/ncomms1115 (2010).
\bibitem{Vekhter_FeSe} A.B.Vorontsov and I.Vekhter, Phys. Rev. Lett. \textbf{105}, 187004 (2010).
\bibitem{Chubukov_FeSe} A.V. Chubukov and I. Eremin, Phys. Rev. B \textbf{82}, 060504(R) (2010).
\bibitem{TailleferReview} H. Shakeripour, C. Petrovic, and L. Taillefer, New. J. Phys. \textbf{11}, 055065 (2009).
\bibitem{Kuebert_kappa} C. K\"ubert and P.J. Hirschfeld, Phys. Rev. Lett. \textbf{80}, 4963 (1998).
\bibitem{x_luo_09} X.G. Luo, M.A. Tanatar, J.-Ph. Reid, H. Shakeripour, N. Doiron-Leyraud, N. Ni, S.L. Bud'ko, P.C. Canfield, H. Luo, Z. Wang, H.-H. Wen, R. Prozorov, and L. Taillefer, Phys. Rev. B \textbf{80}, 140503(R) (2009).
\bibitem{Ding2009} L. Ding, J.K. Dong, S.Y. Zhou, T.Y. Guan, X. Qiu, C. Zhang, L.J. Li, X. Lin, G.H. Cao, Z.A. Xu, and S.Y. Li, New J. Phys. \textbf{11}, 093018 (2009).
\bibitem{Tanatar2010} M.A. Tanatar, J.P. Reid, H. Shakeripour, X.G. Luo, N. Doiron-Leyraud, N. Ni, S.L. Bud'ko, P.C. Canfield, R. Prozorov, L. Taillefer, Phys. Rev. Lett. \textbf{104}, 067002 (2010).
\bibitem{v_mishra_09} V. Mishra, A. Vorontsov, P.J. Hirschfeld, and I. Vekhter,  Phys. Rev. B \textbf{80}, 224525 (2009).
\bibitem{ybang} Y. Bang, Phys. Rev. Lett. \textbf{104}, 217001 (2010).
\bibitem{Reid} J.-Ph. Reid, M.A. Tanatar, X. G. Luo, H. Shakeripour, N. Doiron-Leyraud, N. Ni, S.L. Bud'ko, P.C. Canfield, R. Prozorov, L. Taillefer, Phys. Rev. B \textbf{82}, 064501 (2010).
\bibitem{Reid2011K-doped} J.-Ph. Reid, M.A. Tanatar, X.G. Luo, H. Shakeripour, S. Ren\'e de Cotret, N. Doiron-Leyraud, J. Chang, B. Shen, H.-H. Wen, H. Kim, R. Prozorov, and L. Taillefer, arXiv:1105.2232.
\bibitem{Mishra2011} V. Mishra, S. Graser, and P.J. Hirschfeld, Phys. Rev. B \textbf{84}, 014524 (2011).
\bibitem{MazinDevereaux} I.I. Mazin, T.P. Devereaux, J.G. Analytis, J.-H. Chu, I.R. Fisher, B. Muschler, and R. Hackl, Phys. Rev. B \textbf{82}, 180502(R) (2010).
\bibitem{VekhterReview} Y. Matsuda, K. Izawa, and I. Vekhter, J. Phys.: Condens. Matter \textbf{18}, R705 (2006).
\bibitem{FeSe_kappa} J.K. Dong, T.Y. Guan, S.Y.  Zhou, X.  Qiu, L. Ding, C. Zhang, U. Patel, Z.L. Xiao, S.Y. Li, Phys. Rev. B \textbf{80}, 024518 (2009).
\bibitem{Yamashita} M. Yamashita, Y. Senshu, T. Shibauchi, S. Kasahara, K. Hashimoto, D. Watanabe, H. Ikeda, T. Terashima, I. Vekhter, A.B. Vorontsov, Y. Matsuda, Phys. Rev. B \textbf{84}, 060507(R) (2011).
\bibitem{l_zhao_08} L. Zhao, H. Liu, W. Zhang, J. Meng, X. Jia, G. Liu, X. Dong, G. F. Chen, J.L. Luo, N.L. Wang, G. Wang, Y. Zhou, Y. Zhu, X. Wang, Z. Zhao, Z. Xu, C. Chen, and X.J. Zhou, Chin. Phys. Lett. \textbf{25}, 4402 (2008).
\bibitem{h_ding_08} H. Ding, P. Richard, K. Nakayama, T. Sugawara, T. Arakane, Y. Sekiba, A. Takayama, S. Souma, T. Sato, T. Takahashi, Z. Wang, X. Dai, Z. Fang, G. Chen, J. Luo, and N. Wang, Europhys. Lett. \textbf{83}, 47001 (2008).
\bibitem{t_kondo_08} T. Kondo, A. Santander-Syro, O. Copie, C. Liu, M. Tillman, E. Mun, J. Schmalian, S. Bud'ko, M. Tanatar, P. Canfield, and A. Kaminski, Phys. Rev. Lett. \textbf{101}, 147003 (2008).
\bibitem{d_evtushinsky_09} D. Evtushinsky, D. Inosov, V. Zabolotnyy, A. Koitzsch, M. Knupfer, B. Buchner, G. Sun, V. Hinkov, A. Boris, C. Lin, B. Keimer, A. Varykhalov, A. Kordyuk, and S. Borisenko, Phys. Rev. B \textbf{79}, 054517 (2009).

\bibitem{k_nakayama_09} K. Nakayama, T. Sato, P. Richard, Y.-M. Xu, Y. Sekiba, S. Souma, G.F. Chen, J.L. Luo, N.L. Wang, H. Ding, and T. Takahashi, Europhys. Lett. \textbf{85}, 67002 (2009).
\bibitem{l_wray_08} L. Wray, D. Qian, D. Hsieh, Y. Xia, L. Li, J. Checkelsky, A. Pasupathy, K. Gomes, C. Parker, A. Fedorov, G. Chen, J. Luo, A. Yazdani, N. Ong, N. Wang, and M. Hasan, Phys. Rev. B \textbf{78}, 184508 (2008).
\bibitem{borisenko_2010} S.V. Borisenko, V.B. Zabolotnyy, D.V. Evtushinsky, T.K. Kim, I.V. Morozov, A.N. Yaresko, A.A. Kordyuk, G. Behr, A. Vasiliev, R. Follath, an B. B\"uchner	 Phys. Rev. Lett. \textbf{105}, 067002 (2010).


\bibitem{Shin2011} T. Shimojima, F. Sakaguchi, K. Ishizaka, Y. Ishida, T. Kiss, M. Okawa, T. Togashi, C.-T. Chen, S. Watanabe, M. Arita, K. Shimada, H. Namatame, M. Taniguchi, K. Ohgushi, S. Kasahara, T. Terashima, T. Shibauchi, Y. Matsuda, A. Chainani, and S. Shin, Science \textbf{332}, 564 (2011).
\bibitem{Damascelli} A. Damascelli, Z.-X. Shen, Z.Hussain, Rev. Mod. Phys. \textbf{75}, 473 (2003).
\bibitem{theory_LiFeAs} C.Platt, R. Thomale, and W. Hanke, arXiv:1103.2101.

\bibitem{tanatar_2011}  M.A. Tanatar, J.-Ph. Reid, S. Rene de Cotret, N. Doiron-Leyraud, F. Laliberte, E. Hassinger, J. Chang, H. Kim, K. Cho, Y.J. Song, Y.S. Kwon, R. Prozorov, and L. Taillefer, arXiv:1104.2209.
\bibitem{STMLiFeAs} T. Hanaguri, private communication (2011).
\bibitem{stockert_11} U. Stockert, M. Abdel-Hafiez, D.V. Evtushinsky, V.B. Zabolotnyy, A.U.B. Wolter, S. Wurmehl, I. Morozov, R. Klingeler, S.V. Borisenko, and B. B\"uchner, arXiv:1011.4246.
\bibitem{PJHconsequences} P.J. Hirschfeld, P. W\"olfle and D. Einzel, Phys. Rev. B \textbf{37}, 83 (1988).
\bibitem{Nakai} Y. Nakai, K. Ishida, Y. Kamihara, M. Hirano, and H. Hosono, J.
    Phys. Soc. Japan \textbf{77}, 073701 (2008).
\bibitem{NakaiPdoped2}  Y. Nakai, T. Iye, S. Kitagawa, K. Ishida, H. Ikeda, S. Kasahara, H. Shishido, T. Shibauchi, Y. Matsuda, and T. Terashima, Phys. Rev. Lett. \textbf{105}, 107003 (2010).
\bibitem{LiBa122} Z. Li, D.L. Sun, C.T. Lin, Y.H. Su, J.P. Hu, and G.-q. Zheng, Phys. Rev. B \textbf{83}, 140506(R) (2011).
\bibitem{Muschler} B. Muschler, W. Prestel, R. Hackl, T.P. Devereaux, J.G. Analytis, J.-H. Chu, and I.R. Fisher, Phys. Rev. B \textbf{80}, 180510(R) (2009).
\bibitem{Chauviere} L. Chauvi\`ere, Y. Gallais, M. Cazayous, M.A. M\'easson, A. Sacuto, D. Colson, and A. Forget, Phys. Rev. B \textbf{82}, 180521(R) (2010).

\bibitem{DevereauxRamanReview}  T.P. Devereaux and R. Hackl, Rev. Mod. Phys. \textbf{79}, 175 (2007).

\bibitem{ChubukovRaman} A.V. Chubukov, I. Eremin, and M.M. Korshunov, Phys. Rev. B \textbf{79}, 220501(R) (2009).

\bibitem{BoydRaman1}  G. Boyd, T.P. Devereaux, P.J. Hirschfeld, V.Mishra, and D.J. Scalapino, Phys. Rev. B \textbf{79}, 174521 (2009).

\bibitem{Sugai} S. Sugai, Y. Mizuno, K. Kiho, M. Nakajima, C.H. Lee, A. Iyo, H. Eisaki, and S. Uchida, Phys. Rev. B \textbf{82}, 140504(R) (2010); \textit{ibid} \textbf{83}, 019903(E) (2011).

\bibitem{BoydRaman2} G.R. Boyd, P.J. Hirschfeld, and T.P. Devereaux, Phys. Rev. B \textbf{82}, 134506 (2010).

\bibitem{firstreports} J. Guo, S. Jin, G. Wang, S. Wang, K. Zhu, T. Zhou, M. He, and X. Chen, Phys. Rev. B \textbf{82}, 180520 (2010).

\bibitem{trend} I.I. Mazin, Physics \textbf{4}, 26 (2011).

\bibitem{GraserKFeSe} T.A. Maier, S. Graser, P.J. Hirschfeld, D.J. Scalapino, Phys. Rev. B \textbf{83}, 100515(R) (2011).

\bibitem{Lee} F. Wang, F. Yang, M. Gao, Z.-Y. Lu, T. Xiang, D.-H. Lee, Europhy. Lett. \textbf{93}, 57003 (2011).

\bibitem{Chubukov-unpublished} A.V. Chubukov, unpublished (2011).

\bibitem{InosovFeSe} J.T. Park, G. Friemel, Y. Li, J.-H. Kim, V. Tsurkan, J. Deisenhofer, H.-A. Krug von Nidda, A. Loidl, A. Ivanov, B. Keimer, and D.S. Inosov, arXiv:1107.1703.

\bibitem{Balat} T. Das and A.V. Balatsky, Phys. Rev. B \textbf{84}, 060507(R) (2011).

\bibitem{Saito} T. Saito, S. Onari, and H. Kontani, Phys. Rev. B \textbf{83}, 140512(R) (2011).

\bibitem{charge_balance} W. Bao, G.N. Li, Q. Huang, G.F. Chen, J.B. He, M.A. Green, Y. Qiu, D.M. Wang, J.L. Luo, and M.M. Wu, arXiv:1102.3674.

\bibitem{HongDing} T. Qian, X.-P. Wang, W.-C. Jin, P. Zhang, P. Richard, G. Xu, X. Dai, Z. Fang, J.-G. Guo, X.-L. Chen, and H. Ding, Phys. Rev. Lett. \textbf{106}, 187001 (2011).

\bibitem{vac}  W. Bao, Q. Huang, G.F. Chen, M.A. Green, D.M. Wang, J.B. He, X.Q. Wang, and Y. Qiu, Chinese Phys. Lett. \textbf{28}, 086104 (2011); V.Yu. Pomjakushin, E.V. Pomjakushina, A. Krzton-Maziopa, K. Conder, and Z. Shermadini, J. Phys.: Condens. Matter \textbf{23} (2011) 156003; F. Ye, S. Chi, W. Bao, X.F. Wang, J.J. Ying, X.H. Chen, H.D. Wang, C.H. Dong, and M. Fang, Phys. Rev. Lett. \textbf{107}, 137003 (2011); P. Zavalij, W. Bao, X.F. Wang, J.J. Ying, X.H. Chen, D.M. Wang, J.B. He, X.Q. Wang, G.F Chen, P.-Y. Hsieh, Q. Huang, and M.A. Green, Phys. Rev. B \textbf{83}, 132509 (2011); J. Bacsa, A.Y. Ganin, Y. Takabayashi, K.E. Christensen, K. Prassides, M.J. Rosseinsky, and J.B. Claridge, Chem. Sci. \textbf{2}, 1054 (2011).

\bibitem{band_ins} X.W. Yan, M. Gao, Z.Y. Lu, and T. Xiang, Phys. Rev. Lett. \textbf{106}, 087005 (2011).

\bibitem{CaoDai} C. Cao and J. Dai, Phys. Rev. Lett. \textbf{107}, 056401 (2011).

\bibitem{optK122} Z.G. Chen, R.H. Yuan, T. Dong, G. Xu, Y.G. Shi, P. Zheng, J.L. Luo, J.G. Guo, X.L. Chen, N.L. Wang, Phys. Rev B \textbf{83}, 220507(R) (2011).

\bibitem{Hc2} M.I. Tsindlekht, I. Felner, M. Zhang, A.F. Wang, and X.H. Chen, arXiv:1105.1065.

\bibitem{haihu} F. Han, B. Shen, Z.-Y. Wang, and H.-H. Wen, arXiv:1103.1347.

\bibitem{order} J.Q. Li, Y.J. Song, H.X. Yang, Z. Wang, H.L. Shi, G.F. Chen, Z.W. Wang, Z. Chen, and H.F. Tian, arXiv:1104.5340.

\bibitem{selenideSTM} W. Li, H. Ding, P. Deng, K. Chang, C. Song, K. He, L. Wang, X. Ma, J. Hu, X. Chen, Q. Xue, arXiv:1108.0069.

\bibitem{MazinSe}  I.I. Mazin, Phys. Rev. B, in press, arXiv:1102.3655.


\bibitem{Se122_SH} B. Zeng, B. Shen, G.F. Chen, J.B. He, D.M. Wang, C.H. Li, and H.-H. Wen, Phys. Rev. B \textbf{83}, 144511 (2011).
\bibitem{Se122_NMR} W. Yu, L. Ma, J.B. He, D.M. Wang, T.-L. Xia, G.F. Chen, and W. Bao, Phys. Rev. Lett. \textbf{106}, 197001 (2011).
\bibitem{HashimotoPrFeAsO} K. Hashimoto, T. Shibauchi, T. Kato, K. Ikada, R. Okazaki, H. Shishido, M. Ishikado, H. Kito, A. Iyo, H. Eisaki, S. Shamoto, and Y. Matsuda, Phys. Rev. Lett. \textbf{102}, 017002 (2009).
\bibitem{YamashitaLaFePO} M. Yamashita, N. Nakata, Y. Senshu, S. Tonegawa, K. Ikada, K. Hashimoto, H. Sugawara, T. Shibauchi, and Y. Matsuda, Phys. Rev. B \textbf{80}, 220509(R) (2009).
\bibitem{MaloneSmFeAsO} L. Malone, J.D. Fletcher, A. Serafin, A. Carrington, N.D. Zhigadlo, Z. Bukowski, S. Katrych, J. Karpinski, Phys. Rev. B \textbf{79}, 140501(R) (2009).
\bibitem{bobowski2010} J.S. Bobowski, J.C. Baglo, J. Day, P. Dosanjh, R. Ofer, B.J. Ramshaw, R. Liang, D.A. Bonn, W.N. Hardy, H. Luo, Z.-S. Wang, L. Fang, H.-H. Wen, Phys. Rev. B \textbf{82}, 094520 (2010).
\bibitem{MartinNidoped122} C. Martin, H. Kim, R.T. Gordon, N. Ni, V.G. Kogan, S.L. Bud'ko, P.C. Canfield, M.A. Tanatar, and R. Prozorov, Phys. Rev. B \textbf{81}, 060505 (2010).
\bibitem{ProzorovLiFeAs} H. Kim, M.A. Tanatar, Y.J. Song, Y.S. Kwon, and R. Prozorov, Phys. Rev. B \textbf{83}, 100502 (2011).
\bibitem{Shibauchi_LiFeP_2011} K. Hashimoto, S. Kasahara, R. Katsumata, Y. Mizukami, M. Yamashita, H. Ikeda, T. Terashima, A. Carrington, Y. Matsuda, and T. Shibauchi, arXiv:1107.4505.

\bibitem{BFAP_normalstate} S. Kasahara, T. Shibauchi, K. Hashimoto, K. Ikada, S. Tonegawa, H. Ikeda, H. Takeya, K. Hirata, T. Terashima, and Y. Matsuda, Phys. Rev. B \textbf{81}, 184519 (2010).

\bibitem{HEschrigLiFeAs} A. Lankau, K. Koepernik, S. Borisenko, V. Zabolotnyy, B. B\"uchner, J. van den Brink, and H. Eschrig, Phys. Rev. B \textbf{82}, 184518 (2010).

\bibitem{Putzke} C. Putzke, A.I. Coldea, I. Guillamon, D. Vignolles, A. McCollam, D. LeBoeuf, M.D. Watson, I.I. Mazin, S. Kasahara, T. Terashima, T. Shibauchi, Y. Matsuda, A. Carrington, arXiv:1107.4375.

\bibitem{Boriskenkocomment} S.V. Borisenko, V. B. Zabolotnyy, D.V. Evtushinsky, T.K. Kim, I.V. Morozov, A.A. Kordyuk, B. B\"uchner, arXiv:1108.1159.

\bibitem{Li_etal_LiFeAs_NMR} Z. Li, Y. Ooe, Z.-C. Wang, Q.-Q. Liu, C.-Q. Jin, M. Ichioka, and G. Zheng, J. Phys. Soc. Jpn \textbf{79}, 083702 (2010).

\bibitem{Buechner_LiFeAsNMR} S.-H. Baek, H.-J. Grafe, F. Hammerath, M. Fuchs, C. Rudisch, L. Harnagea, S. Aswartham, S. Wurmehl, J. van den Brink, and B. B\"uchner, arXiv:1108.2592.

\bibitem{vdBrink_triplet} P.M.R. Brydon, M. Daghofer, C. Timm, J. van den Brink, Phys. Rev. B \textbf{83}, 060501 (2011).

\bibitem{Morr} D.K. Morr, P.F. Trautman, and M.J. Graf, Phys. Rev. Lett. \textbf{86}, 5978 (2001).

\bibitem{Taylor_neutron_LiFeAs} A.E. Taylor, M.J. Pitcher, R.A. Ewings, T.G. Perring, S.J. Clarke, A.T. Boothroyd, Phys. Rev. B \textbf{83}, 220514(R) (2011).

\bibitem{Dong_K122} J.K. Dong, S.Y. Zhou, T.Y. Guan, H. Zhang, Y.F. Dai, X. Qiu, X.F. Wang, Y. He, X.H. Chen, S.Y. Li, Phys. Rev. Lett. \textbf{104}, 087005 (2010).

\bibitem{HashimotoK122} K. Hashimoto, A. Serafin, S. Tonegawa, R. Katsumata, R. Okazaki, T. Saito, H. Fukazawa, Y. Kohori, K. Kihou, C.H. Lee, A. Iyo, H. Eisaki, H. Ikeda, Y. Matsuda, A. Carrington, T. Shibauchi, Phys. Rev. B \textbf{82}, 014526 (2010).

\bibitem{ThomaleK122} R. Thomale, C. Platt, W. Hanke, J. Hu, and B.A. Bernevig, Phys. Rev. Lett. \textbf{107}, 117001 (2011).

\bibitem{Forgan} H. Kawano-Furukawa, C. J. Bowell, J. S. White, R. W. Heslop, A. S. Cameron, E. M. Forgan, K. Kihou, C. H. Lee, A. Iyo, H. Eisaki, T. Saito, H. Fukazawa, Y. Kohori, R. Cubitt, C. D. Dewhurst, J. L. Gavilano, and M. Zolliker, Phys. Rev. B 84, 024507 (2011)

\end{thebibliography}
\end{document}